\documentclass[11pt,a4paper]{article}
\pdfoutput=1

\usepackage[nosort]{cite}
\usepackage{graphicx}
\usepackage[british]{babel} 
\usepackage{amsthm}
\usepackage{epstopdf} 
\usepackage{amsmath,amssymb,amsthm}
\usepackage{mathrsfs}
\usepackage{bm}
\usepackage{slashed}
\usepackage{framed}
\usepackage{color}
\usepackage[textwidth = 430 pt, textheight = 630 pt]{geometry}
\definecolor{MyDarkBlue}{rgb}{0.15,0.25,0.45}
\usepackage[linktocpage=true,hypertexnames=false]{hyperref}
\hypersetup{
colorlinks=true,
citecolor=MyDarkBlue,
linkcolor=MyDarkBlue,
urlcolor=MyDarkBlue,
breaklinks=true
}

\let\fn\footnote
\renewcommand{\footnote}[1]{\linespread{1.1}\fn{#1}\linespread{1.29}}

\makeatletter\renewcommand{\section}{\@startsection
{section}{1}{\z@}{-3.5ex plus -1ex minus
    -.2ex}{2.3ex plus .3ex}{\Large\bf\mathversion{bold} }}

\renewcommand\tableofcontents{%
    \section*{\large\contentsname
        \@mkboth{%
          \MakeUppercase\contentsname}{\MakeUppercase\contentsname}}%
       {\baselineskip=15pt plus 2pt minus 1pt
    \@starttoc{toc}}%
}

\renewenvironment{thebibliography}[1]
     {\baselineskip=16pt plus 2pt minus 1pt
      \section*{\large\refname
        \@mkboth{\MakeUppercase\refname}{\MakeUppercase\refname}}%
     \list{\@biblabel{\@arabic\c@enumiv}}%
           {\settowidth\labelwidth{\@biblabel{#1}}%
            \leftmargin\labelwidth
            \advance\leftmargin\labelsep
            \@openbib@code
            \usecounter{enumiv}%
            \let\p@enumiv\@empty
            \renewcommand\theenumiv{\@arabic\c@enumiv}}%
      \sloppy
      \clubpenalty4000
      \@clubpenalty \clubpenalty
      \widowpenalty4000%
      \sfcode`\.\@m
 \catcode`\^^M=10%
}

\newcommand{\appendices}{
\section*{Appendix}\label{appendices}\setcounter{subsection}{0}
\addcontentsline{toc}{section}{Appendix}
\setcounter{equation}{0}
\makeatletter
\renewcommand{\theequation}{\Alph{subsection}.\arabic{equation}}
\renewcommand{\thesubsection}{\Alph{subsection}}
\@addtoreset{equation}{subsection}
\makeatother
}
\numberwithin{equation}{section}

\begin{document}

\begin{titlepage}

\setcounter{page}{0}

\vspace{1cm}

\begin{center}

\textbf{\Large\mathversion{bold} Lectures on S-matrices and Integrability}

\vspace{1cm}

{\large Diego Bombardelli} 

\vspace{1cm}

{\it Dipartimento di Fisica and INFN, Universit\`a di Torino\\
via Pietro Giuria 1, 10125 Torino, Italy\\
and\\
Dipartimento di Fisica e Astronomia and INFN, Universit\`a di Bologna\\
via Irnerio 46, 40126 Bologna, Italy} \\
\href{mailto:diegobombardelli@gmail.com}{\ttfamily diegobombardelli@gmail.com}

\vspace{1cm}

\end{center}

\begin{abstract}
In these notes we review the S-matrix theory in (1+1)-dimensional integrable models,
focusing mainly on the relativistic case. Once the main definitions and physical properties 
are introduced, we discuss the factorization of scattering processes due to
integrability. We then focus on the analytic properties of the two-particle scattering amplitude and illustrate the derivation of the S-matrices
for all the possible bound states using the so-called bootstrap principle. General algebraic structures underlying the S-matrix
theory and its relation with the form factors axioms are briefly mentioned.
Finally, we discuss the S-matrices of sine-Gordon and
$SU(2)$, $SU(3)$ chiral Gross-Neveu models.

This is part of a collection of lecture notes for the
\emph{Young Researchers Integrability School}, organized by the GATIS network at Durham University on 6-10 July 2015.
\end{abstract}
\vfill

\emph{In loving memory of Lilia Grandi}
\setcounter{footnote}{0}\renewcommand{\thefootnote}{\arabic{thefootnote}}

\end{titlepage}

\tableofcontents

\section{Introduction}

The \emph{S-matrix} program is a non-perturbative analytic approach to the scattering problem in quantum field theory (QFT), whose origins date back to works by Wheeler \cite{Wheeler} and Heisenberg \cite{Heisenberg}. The main purpose of the program was to overcome the problems of QFT related on one hand to the divergences emerging from standard 
perturbative methods, and on the other to the discovery, in the 1950s and 60s, of many hadronic resonances with high spin.

The idea, further developed by Chew \cite{Chew}, Mandelstam \cite{Mandelstam:1958xc} and many others, was to compute scattering amplitudes and mass spectra without the use of a Lagrangian formulation, by imposing analytic constraints on the S-matrix, that is the operator relating initial and final states in a scattering process, and by giving a physical interpretation of all its singularities. Moreover, higher spin particles were treated on the same footing as the fundamental ones. This latter aspect will be illustrated in these notes when the so-called \emph{bootstrap principle} is discussed.

Unfortunately, after the initial successes, not many quantitative results were obtained in real-world particle physics. Moreover, the search for exact S-matrix models was finally discouraged by the Coleman-Mandula theorem \cite{Coleman:1967ad}, stating that QFT models in $d>2$ space-time dimensions, with higher-order conserved charges, can only have trivial S-matrices.
However, between the 1970s and 80s the program was given a new boost in the context of $d=2$ \emph{integrable} theories, whose S-matrices are non trivial and can be uniquely fixed. Furthermore, as we will see, knowing the amplitudes for the scattering
between two-particle states is sufficient, at least in principle, to reconstruct the correlation functions of the theory, through the \emph{form factor} program.

The S-matrix plays an essential role also in calculating the spectrum of integrable theories. Both in the large volume approximation, through the derivation of the asymptotic Bethe ansatz, reviewed in \cite{Fedor}, and at finite volume,
being the key ingredient of techniques like the L\"uscher formulas \cite{Luscher:1983rk} and the thermodynamic Bethe ansatz (TBA), reviewed in \cite{Stjin}. 

We believe that this is one of the strongest reasons to study the integrable S-matrix theory, which is the subject of these lectures. 
In particular, we will try to describe in a pedagogical way some of the fundamental concepts developed in this research field, assuming
that the reader is familiar with the basics of quantum mechanics and special relativity.  
In order to give a deeper understanding of a few technical aspects, calculations will be described in full detail in some simple cases only. These tools and ideas can be then adapted to the study of much more 
complicated systems and the interested reader may find more complete and advanced discussions on the many important applications of such techniques in the quoted references. Since we could not cover the enormous literature on the subject, we selected a few reviews and original papers concerning the models discussed in these notes. 

In particular, the lectures focused mostly on \emph{relativistic} cases and were built mainly on the book by Mussardo \cite{Mussardobook}, the lectures by Dorey \cite{Dorey} and the paper by Zamolodchikov
and Zamolodchikov \cite{ZZ}. For the important non-relativistic case of $AdS_5/CFT_4$, the reading of \cite{AF}, especially chapter 3, is suggested, as well as the
reviews \cite{Ahn:2010ka, Vieira:2010kb} and the seminal papers \cite{Staudacher:2004tk, Beisert:2005tm, Janik:2006dc, Arutyunov:2004vx, Arutyunov:2006yd}, and \cite{Beisert:2010jr} for an overview of many other recent developments in the context of gauge/gravity dualities.

The outline of the lectures is the following: after the introduction of the necessary definitions, with a brief description of the S-matrix physical properties, the main ideas underlying the demonstration of the 
factorization property for integrable S-matrices will be explained.
Then we will focus on two-particle S-matrices, including those for the processes involving bound states, and on their analytic and algebraic properties.

A few examples, regarding two-particle S-matrices of the sine-Gordon and chiral Gross-Neveu models, will be given.

The latter theories will be used also to explain the links between S-matrices and correlation functions, through a very introductory discussion on the \emph{form factor} program in integrable models.
This part is built mainly on the paper \cite{Karowski:1978vz} and the review \cite{Babujian:2006km} (see also \cite{Yurov:1990kv, Babujian:1998uw}, the book \cite{Smirnovbook} and the recent review \cite{Delfino:2015ria}, that includes a discussion of the S-matrix in $d=2$ theories and applications to critical phenomena). 

Finally, we conclude with a short guide to the literature about recent developments on S-matrices in $AdS/CFT$ correspondences. 

\section{Asymptotic states and the S-matrix}

\subsection{Definitions}

It is well known that in quantum mechanics the time evolution of a system can be defined through an unitary operator $U(t,t_{0})$, which generates the state $|\psi(t)\rangle$ by acting on a state $|\psi(t_{0})\rangle$:
\begin{equation}
|\psi(t)\rangle=U(t,t_{0})|\psi(t_{0})\rangle\,.
\end{equation}
In order to study a scattering process, actually, it is not necessary to know $U(t,t_{0})$ at any values of $t, t_{0}$, but it is enough to know it at $t_{0}\rightarrow -\infty$ and $t\rightarrow +\infty$.
Indeed, if we assume that interactions among particles occur in a very small region of the space-time, then, very far away from the interaction region, we can treat them as free particles.
Thus we need to define in a formal way these quantum states of free excitations introducing the so-called \emph{asymptotic states}
\begin{equation}
 \left|p_1, p_2,\dots, p_n\right\rangle_{a_1a_2\dots a_n}^{in/out}\,,
\label{asymst}
\end{equation}
where $n$ is the number of particles, $p_i$ are their momenta and indices $a_i$ label their flavors. Essentially, the asymptotic states describe wave packets with approximate
positions at given times: in particular, $n$ free particles at time $t\rightarrow-\infty$ for the \emph{in} states and at $t\rightarrow+\infty$ for the \emph{out} ones.
We choose the order of momenta to be $p_1>p_2>\dots>p_n$.
Any intermediate state can equivalently be expanded on the $in$ or $out$ bases.

The S-matrix is defined as the linear operator that maps final asymptotic states into initial asymptotic states (or vice versa, depending on the convention
adopted, related to the inversion of such operator):
\begin{equation}
|\dots\rangle_{in}=S|\dots\rangle_{out}\,.
\label{eq:Sdef}
\end{equation}
Written in components, this reads
\begin{eqnarray}
\label{Sdef2}
 &&\hspace{-0.5cm}\left|p_1,p_2,\dots,p_n\right\rangle^{in}_{a_1,a_2,\dots,a_n}\\
 &&\hspace{-0.5cm}=\sum_{m=2}^{\infty}\mathop{\sum_{p_1'>\dots>p_m'}}_{b_1,\dots,b_m}S_{a_1,\dots,a_n}^{b_1,\dots,b_m}(p_1,\dots,
 p_n;p_1',\dots,p_m')\left|p_1',p_2'\dots, p_m'\right\rangle^{out}_{b_1,b_2,\dots,b_m}\,,\nonumber
\end{eqnarray}
where the second line actually involves integrals in $p_1',p_2'\dots, p_m'$.

Hence $S$ is the time evolution operator from $t=-\infty$ to $t=+\infty$: 
\begin{equation}
S=\mathop{\lim_{t_{0}\rightarrow-\infty}}_{t\rightarrow\infty}U(t,t_{0})\,.
\end{equation}
If the system has an Hamiltonian
\begin{equation}
H=H_{0}+H_{I}\,,
\end{equation}
where $H_{0}$ is the Hamiltonian of the free system and $H_{I}=H_I(t)$ is the interaction part in the interaction (Dirac) picture\footnote{In this representation, both states and operators depend on time, then a generic physical state is defined as $|s_{I}(t)\rangle=e^{\mathrm{i}H_{0}t}|s_{S}(t)\rangle$, where $|s_{S}(t)\rangle$ is the corresponding state in the Schr\"odinger picture. Then a generic operator in interaction picture is given in terms of the operator in Schr\"odinger representation by $O_{I}=O_{I}(t)=e^{\mathrm{i}H_{0}t}O_{S}e^{-\mathrm{i}H_{0}t}$.}, then $S$ can be expressed as 
\begin{equation}
S=\mathcal{T}\exp\left[-\mathrm{i}\int_{-\infty}^{+\infty}dt H_{I}(t)\right]\,, 
\label{linop}
\end{equation}
where $\mathcal{T}$ denotes the time-ordering for the series expansion of the exponential in (\ref{linop}).

\subsection{General properties}
\label{sec:genprop}

In this section we discuss some general assumptions motivated by physical properties fulfilled by usual QFTs.
As previously mentioned, \emph{interactions} among particles are assumed to occur only at \emph{short range}. 
Another obvious assumption is the validity of the 
\emph{QM superposition principle}, meaning that asymptotic states form a complete basis for initial and final states and any \emph{in} state can be expanded in the basis of \emph{out} 
states and vice versa, through the time evolution linear operator $S$, as expressed by (\ref{eq:Sdef}).
Moreover, \emph{probability conservation} implies that
\begin{equation}
1=\sum_m|\langle m|S|\psi\rangle|^2\,,
 \end{equation}
where $|\psi\rangle=\sum_n a_n|n\rangle$ and $|m\rangle$, $|n\rangle$ are orthogonal, complete basis vectors generating the Hilbert space of the asymptotic states. Then one can show that 
 \begin{equation}
  \hspace{-0.5cm}1=\sum_m|\langle m|S|\psi\rangle|^2=\sum_m\langle \psi|S^{\dagger}|m\rangle\langle m|S|\psi\rangle=\langle \psi|S^{\dagger}S|\psi\rangle=\sum_{n,m}a_n^*a_m
  \langle n|S^{\dagger}S|m\rangle\,,
 \end{equation}
meaning that the S-matrix has to be \emph{unitary}: $S^{\dagger}S=1$. 
We will refer to this property also as \emph{physical unitarity}.
Working mainly with relativistic theories, we will be interested in the consequences of \emph{Lorentz invariance}. In particular, given a generic Lorentz transformation 
denoted by $L|m\rangle=|m'\rangle$, requiring invariance under such transformation at the level of the S-matrix is equivalent to 
\begin{equation}
\langle m'|S|n'\rangle=\langle m|S|n\rangle\,.
\end{equation}
In order to explain the consequences of this assumption, let us consider a two-to-two-particle scattering process, where the incoming (outgoing) particles have momenta $p_1,p_2$ ($p_3,p_4$).
In a relativistic (1+1)-dimensional theory, energies and momenta of the particles involved in such scattering process can be conveniently encoded in a set of relativistic
invariants, called Mandelstam variables \cite{Mandelstam:1958xc}:
\begin{equation}
 s=(p_1+p_2)^2\,,\ \ t=(p_1-p_3)^2\,,\ \ u=(p_1-p_4)^2\,,
 \label{mandel}
\end{equation}
where $p_i=(p_i^{(0)}=E_i,p_i^{(1)})$. Because of the conservation law $p_1+p_2=p_3+p_4$ and the on-shell condition $p_i^2=m_i^2$, then $s+t+u=\sum_{i=1}^4m_i^2$. Hence the amplitude depends only on these Lorentz-invariant combinations of momenta, and in particular, since they are not independent, on two Mandelstam variables only.

Now, momenta and energies can be parametrized respectively as $p_i=m_i\sinh\theta_i$ and $E_i=m_i\cosh\theta_i$ in terms of the rapidity variable $\theta$, while Mandelstam variables can be written as
\begin{eqnarray}
\label{stheta}
 &s=m_1^2+m_2^2+2m_1m_2\cosh(\theta_{12})\,,&\\
 &t=m_1^2+m_3^2-2m_1m_3\cosh(\theta_{13})\,,&\\
 &u=m_1^2+m_4^2-2m_1m_4\cosh(\theta_{14})\,,&
\end{eqnarray}
where we introduced the notation $\theta_{ij}=\theta_{i}-\theta_{j}$.
Then Lorentz invariance implies that the scattering phases depend only on the difference of the rapidities.

Another fundamental assumption is the so-called \emph{macrocausality}, that play a fundamental role in the factorization property discussed in the next section. 
Roughly speaking, macrocausality tells us that outgoing particles 
can propagate only once the interaction
among the incoming ones has happened, where ``macro'' means that this property can be violated on microscopic time scales.
Finally, we will assume the  \emph{analyticity} of the S-matrices, namely they will be assumed to be analytic functions in the $\theta$-plane with a minimal number of singularities dictated by specific physical processes.

\section{Conserved charges and factorization}

In a QFT, the notion of integrability is related to the existence of an infinite number of independent, conserved and mutually commuting charges $Q_{s}$. Then they can be diagonalized simultaneously:
\begin{equation}
Q_{s}|p\rangle_a = q_{s}^{(a)}(p)|p\rangle_a\,.
\end{equation} 
If they are local, \emph{i.e.} they can be expressed as integrals of local densities, then they are additive:
\begin{equation}
Q_{s}|p_{1},\dots,p_{n}\rangle_{a_1,\dots,a_n} = (q_{s}^{(a_{1})}(p_{1})+\dots+q_{s}^{(a_{n})}(p_{n}))|p_{1},\dots,p_{n}\rangle_{a_1,\dots,a_n}\,.
\end{equation} 
Integrability has dramatic consequences on the form of the S-matrix: in $d>2$ dimensions the Coleman-Mandula theorem \cite{Coleman:1967ad} states that, even with a
single charge being a second (or higher) order tensor, the theory has a trivial S-matrix: $S=1$.

In (1+1) dimensions, instead, S-matrices do not trivialize. However, integrability is still very constraining and in particular we show that it implies
\begin{enumerate}
 \item  \emph{no particle production};
\item \emph{final set of momenta = initial set of momenta};
\item \emph{factorization}.
\end{enumerate}
Points 1. and 2. can be understood as follows. If a charge $Q_{s}$ is conserved, then an initial eigenstate of $Q_{s}$ with a given eigenvalue must evolve
into a superposition of states sharing the same eigenvalue:
\begin{equation}
\sum_{i\in in}^{n}q_{s}^{(a_{i})}(p_{i})=\sum_{j\in out}^{m}q_{s}^{(b_{j})}(p_{j}')\,.
\end{equation}
Since we have an infinite sequence of such constraints, these imply that $n=m$ and $p_{i}=p_{j}'$ ($q_{a_{i}}^{(s)}=q_{b_{i}}^{(s)},i=1,\dots,n$),
namely the number of particles is the same before and after scattering and the initial and final sets of momenta are equal: in a word, the scattering is \emph{elastic}.

\subsection{Factorization and the Yang-Baxter equation}

In order to show point 3., that is the factorization of $n$-particle scattering into a product of two-particle events
\begin{equation}
S_{n}(p_1,\dots, p_{n})=\prod_{i=1}^{n-1}\prod_{j=i+1}^{n}S_{2}(p_{i},p_{j})\,,
\end{equation}
we begin by an heuristic argument due to Zamolodchikov and Zamolodchikov \cite{ZZ}.

Let us consider an $n$-particle configuration space ($\mathbb{R}^{n}$), with particles interacting at short range $R$.
Then it is possible to consider $n!$ disconnected domains where the particles, with a permutation $\sigma$ of ordered coordinates $x_{\sigma_1}<x_{\sigma_2}<\dots<x_{\sigma_n}$ and momenta $p_{\sigma_1}>p_{\sigma_2}>\dots>p_{\sigma_n}$, are very far apart  ($|x_{\sigma_{i+1}}-x_{\sigma_i}|\gg R$), so that they can be considered free.

Because of points 1. and 2., the wave function describing the particles in any single domain is a superposition of a finite number of $n$-particle plane waves:
\begin{equation}
\psi_{\sigma}(x_{1},\dots,x_{n})=\sum_{\sigma'}c(\sigma,\sigma')\exp[\mathrm{i}(p_{\sigma'_{1}}x_{\sigma_{1}}+\dots+p_{\sigma'_{n}}x_{\sigma_{n}})]\,,
\end{equation}
with $\sigma, \sigma'$ being permutations of $p_{1},\dots,p_{n}$ allowed by the conditions of no particle production and conservation of momenta: basically, the set of momenta can only be reshuffled by scattering.

Since we assumed the existence of an asymptotic region (of free motion) for any permutation of particles, then the scattering process can be thought as a plane wave propagating from one of these asymptotic regions to another 
by passing through boundary interaction regions. Thus the propagation path can always be chosen in a way such that it goes through interaction regions where only two particles are so close to 
interact. For example, let us take those two particles as particle 1 and 2, then such region is identified by
\begin{equation}
|x_{1}-x_{2}|\ll R\ ,\ \ |x_{1}-x_{j}|\gg R\ ,\ \ |x_{2}-x_{j}|\gg R\ ,\ \ |x_{i}-x_{j}|\gg R\ ,\ \ i,j=3,4,\dots\,.
\end{equation}
In this way only one particle at a time can overtake another, until all the particles 
starting from the configuration $(x_{1}, p_{1}),\dots,(x_{n}, p_{n})$ have overtaken each other, to reach the configuration $(x_{1}, p_{n}),\dots,(x_{n}, p_{1})$.
All the other possible choices of paths connecting the same initial and final configurations, passing also through boundary regions with more than two interacting particles, have to give the same final result for the 
total scattering amplitude.
 
Not completely satisfied by this heuristic proof, we want to discuss  a more rigorous argument, that dates back to \cite{Polyakov} and 
\cite{ShankarWitten}.
For the reader interested in the details of the demonstration we refer to those papers and to \cite{Parke}, while what follows is mainly inspired by the review \cite{Dorey}. 
Demonstrations based on different approaches are given in \cite{Iagolnitzer:1977sw} and, using non-local charges\footnote{See \cite{Florian} for a definition of those charges on the basis of \cite{Luscher:1977uq}.}, in \cite{Luscher:1977uq}.

Let us start by considering a wave packet
\begin{equation}
\psi(x)=\int_{-\infty}^{+\infty}dp \exp[-a^2(p-p_{0})^2]\exp[\mathrm{i}p(x-x_{0})]\,,
\end{equation}
with position and momentum centered around $x_{0}$ and $p_{0}$ respectively. 
We act on $\psi(x)$ with an operator $e^{\mathrm{i}cQ_{s}}$, where $c$ is an arbitrary constant and $Q_{s}$ is a conserved tensor of order $s$. The resulting wave function is given by
\begin{equation}
\tilde\psi(x)=\int_{-\infty}^{+\infty}dp \exp[-a^2(p-p_{0})^2]\exp[\mathrm{i}p(x-x_{0})]e^{\mathrm{i}cp^{s}}\,,
\end{equation}
\emph{i.e.} $e^{\mathrm{i}cQ_{s}}|p\rangle=e^{\mathrm{i}cp^{s}}|p\rangle$, since under a Lorentz transformation $Q_{s}$ transforms as $s$ copies of the total momentum $P=Q_{1}$.

Now, the wave packet is localized at a new position $x=x_{0}-scp_{0}^{s-1}$, that is where the new phase is stationary ($\phi'(p_{0})=0$ with $\phi(p)=-a^{2}(p-p_{0})^{2}+\mathrm{i}p(x-x_{0})+\mathrm{i}cp^{s}$).
Thus the charge with $s=1$, the total momentum, translates all the particles by the same amount $c$. In the case $s>1$, instead, particles with different momenta are displaced by different amounts.
In what follows, actually, we actually only need a couple of conserved charges $Q_{s}$, $Q_{-s}$, with $s>1$ \cite{Parke}.

Let us then consider a scattering process with two incoming and $m$ outgoing particles: the related scattering amplitude is
\begin{equation}
_{a_3,\dots,a_{m+2}}\langle p_{3},\dots,p_{m+2}|S|p_{1},p_{2}\rangle_{a_1,a_2}\,,
\end{equation}
where the momenta are ordered as $p_{1}>p_{2};p_{3}>p_{4}>\dots>p_{m+2}$.
Now, the assumption of \emph{macrocausality} for the S-matrix essentially tells us that the scattering amplitude is nonzero only if the outgoing particles are created after the 
incoming ones. In other words, the time $t_{12}$ when the incoming particle 1 collides with particle 2 has to be smaller than the time $t_{23}$ when the slowest 
incoming particle (particle 2) interacts with the fastest outgoing particle (particle 3): $t_{23}\ge t_{12}$.

Since the charge $Q_{s}$ commutes with the S-matrix, we can use it to rearrange initial and final configurations without changing the amplitude:
\begin{equation}
\hspace{-0.5cm}_{a_3,\dots,a_{m+2}}\hspace{-0.05cm}\langle p_{3},\dots,p_{m+2}|S|p_{1},p_{2}\rangle_{a_1,a_2}=_{~a_3,\dots,a_{m+2}}\hspace{-0.15cm}\langle p_{3},\dots,p_{m+2}|e^{-\mathrm{i}cQ_{s}}Se^{\mathrm{i}cQ_{s}}|p_{1},p_{2}\rangle_{a_1,a_2}\,.
\label{fact}
\end{equation}
This means that, with a suitable choice of $c$, $t_{23}$ can be made smaller than $t_{12}$, and, if any of the outgoing particles is different 
from the incoming ones, then the amplitude vanishes, following the macrocausality principle.

Therefore the only possibility is that one has just two outgoing particles with the same momenta $p_{1},p_{2}$ as the incoming ones.
With this we have shown that the scattering has to be \emph{elastic}.

\begin{figure}
\begin{centering}
\includegraphics[width=0.65\textwidth]{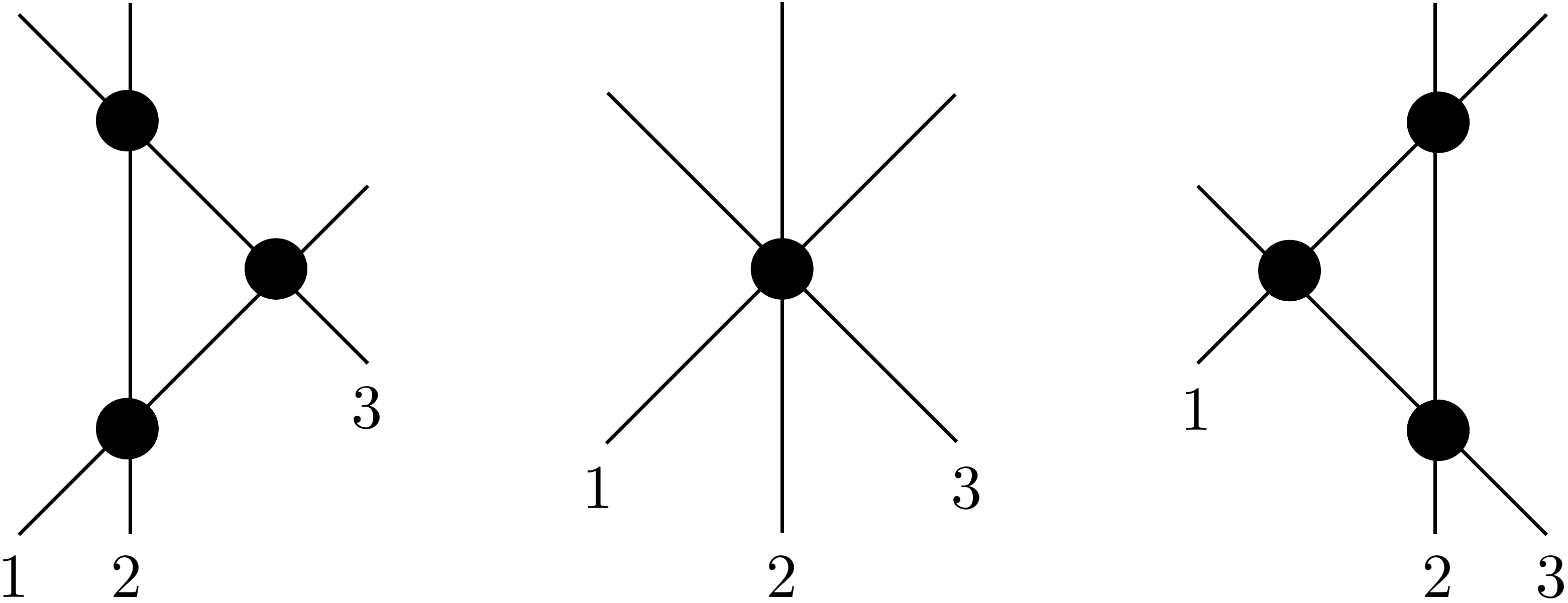}
\par\end{centering}
\caption{Three-particle scattering amplitudes.}
\label{fig:YBE}
\end{figure}

In order to prove the \emph{factorization}, we have to consider processes with more than two particles. In this case, we know now that, acting with a charge like in (\ref{fact}), 
we can separate the trajectories of the particles as much as we want without changing the resulting amplitude, and then also the points of interaction 
between couples of particles (which we know now can produce only couples of particles with momenta equal to the incoming ones): then the total scattering can happen as a sequence of two-particle interactions.

In other words, considering the three-particle example, the three types of possible collision shown in figure \ref{fig:YBE} can be obtained by suitable actions of $e^{icQ_{s}}$ with different values of $c$.
As all of these commute with the Hamiltonian and the S-matrix, then they have to give physically equivalent processes.

\begin{figure}
\begin{centering}
\includegraphics[width=0.65\textwidth]{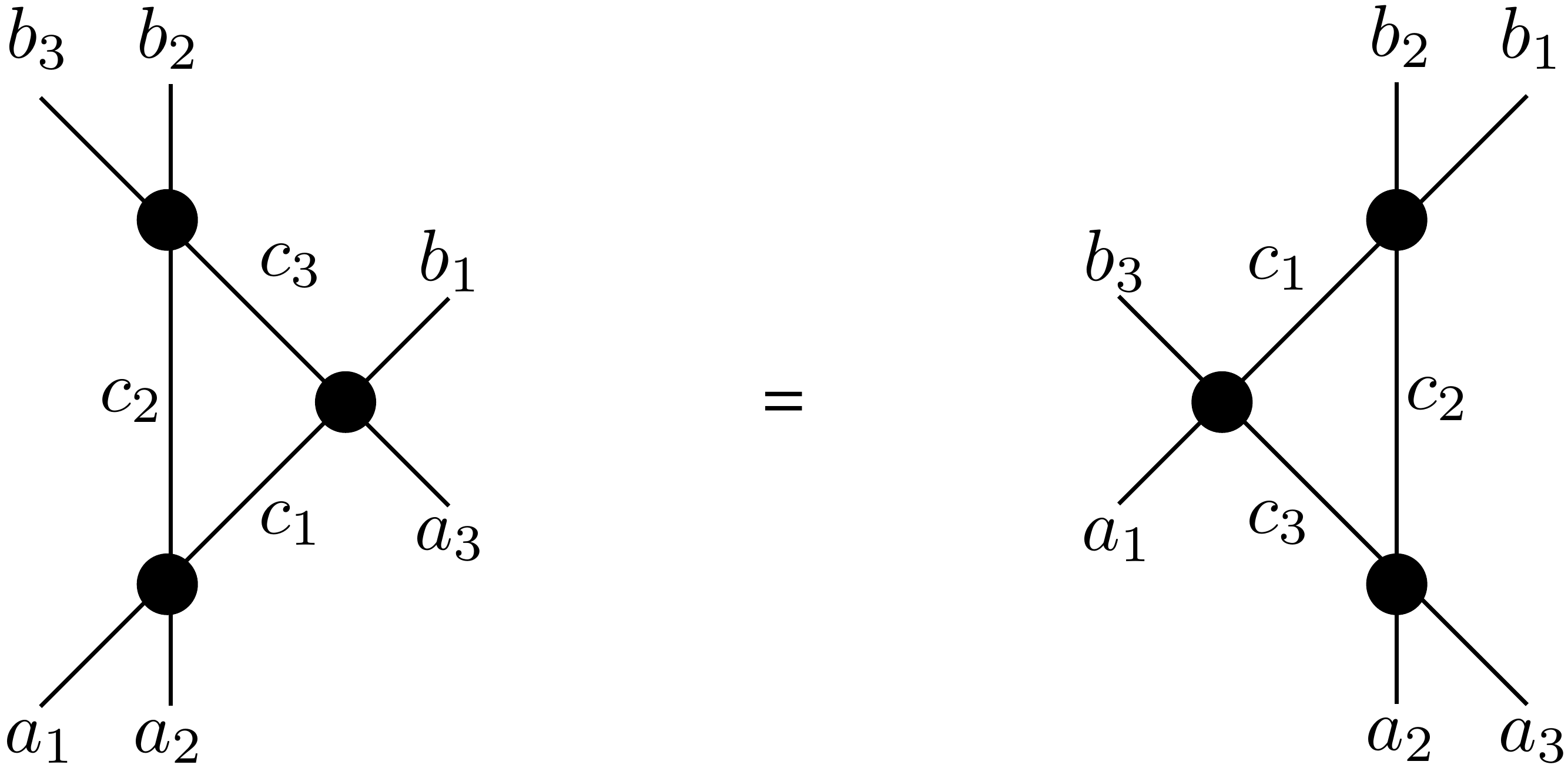}
\par\end{centering}
\caption{Yang-Baxter equation.}
\label{fig:YBEd}
\end{figure}

This equivalence is formalized in the famous \emph{Yang-Baxter equation} (YBE) \cite{Yang-Baxter}:
\begin{equation}
S_{23}S_{13}S_{12}=S_{12}S_{13}S_{23}\,,
\label{YBE}
\end{equation}
where for simplicity we labeled the S-matrices just by the labels of the particles of kind $1,2,3$ involved in a three-particle process.
We can write (\ref{YBE}) in components in order to show the matrix elements involved in a generic non-diagonal process, where exchanges of flavors among particles are possible, 
in the following way (see also figure \ref{fig:YBEd}):
\begin{equation}
 \sum_{c_1,c_2,c_3}S_{a_1a_2}^{c_1c_2}(\theta_{12})S_{c_1a_3}^{b_1c_3}(\theta_{13})S_{c_2c_3}^{b_2b_3}(\theta_{23})= \sum_{c_1,c_2,c_3}S_{a_2a_3}^{c_2c_3}(\theta_{23})S_{a_1c_3}^{c_1b_3}(\theta_{13})
 S_{c_1c_2}^{b_1b_2}(\theta_{12})\,.
 \label{YBEd}
\end{equation}
The generalization to $n$-particle is straightforward. A four-particle process can be always separated in a three-particle one, for which the YBE (\ref{YBEd}) is already shown, and three two-particle processes, by displacing a particle. Then the YBE is proven for four-particle processes. In the same way one decomposes a five-particle scattering in processes involving at most four particles, and so on.

Now we can understand better why in $d>2$ the S-matrix of an integrable theory must be trivial: essentially, in $d>2$ it is always possible to move the trajectories of the particles to create equivalent scattering processes where particles are not crossing each other.  

\section{Two-particle S-matrix}

From the discussion of the previous section, it turns out that any $n$-particle scattering process in integrable theories is completely determined by the knowledge of the 
two-particle S-matrix.
Therefore, in this section, we will focus on general physical properties and the analytic structure of two-particle S-matrices.

\subsection{Properties and analytic structure}
\label{proprel}

\begin{figure}
\begin{centering}
\includegraphics[width=0.45\textwidth]{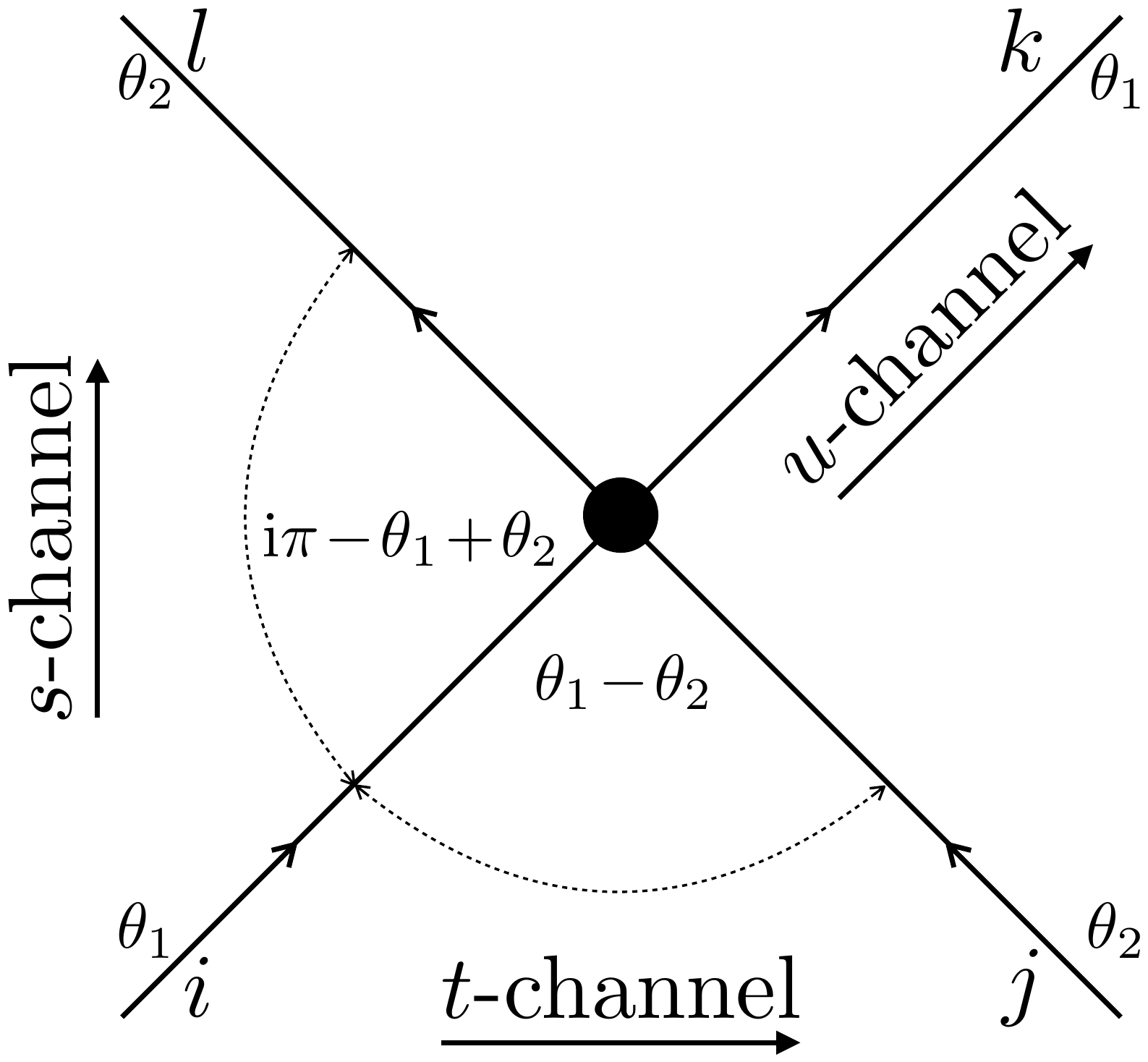}
\par\end{centering}
\caption{Two-particle elastic S-matrix element $S_{ij}^{kl}(\theta_{1}-\theta_{2})$.}
\label{Sdef}
\end{figure}

Following the general definition (\ref{Sdef2}), in the case of a two-particle \emph{elastic scattering} with incoming (outgoing) rapidities $\theta_1$, $\theta_2$ ($\theta_3$, $\theta_4$) we have $\theta_{1}=\theta_{4}, \theta_{2}=\theta_{3}$ and $S=S(\theta_1-\theta_2)$. A two-particle elastic relativistic S-matrix is then given by
\begin{equation}
\left|\theta_1,\theta_2\right\rangle_{i,j}^{in}=S_{ij}^{kl}(\theta_{1}-\theta_{2})\left|\theta_1,\theta_2\right\rangle_{k,l}^{out}\,,
\label{inSout}
\end{equation}
with $\theta_{1}>\theta_{2}$, and represented graphically in figure \ref{Sdef}.
In terms of Mandelstam variables,
$u=0$ and $t(\theta_{12})=s(\mathrm{i}\pi-\theta_{12})$, then the S-matrix depends only on one variable, say $S=S(s)$.

Now, we want to answer the question of how to determine the two-particle S-matrix elements.
Let us begin from the constraints given by \emph{discrete symmetries} usually respected by physical QFTs. If the theory is invariant under reflection of space coordinates, $i.e.$ under \emph{parity}, it means that looking at figure \ref{Sdef} from left to right or from right to left has to be equivalent. Namely, the particles $i$ and $k$ can be exchanged with $j$ and $l$ respectively, leaving the amplitudes unchanged:
\begin{equation}
S_{ij}^{kl}(\theta)=S_{ji}^{lk}(\theta)\,. 
\end{equation}
Analogously, the symmetry under \emph{time reversal} implies that the amplitude represented in figure \ref{Sdef} is the same if we look at it from bottom to top or vice versa, then by exchanging particles $i$ and $l$, $j$ and $k$:
\begin{equation}
S_{ij}^{kl}(\theta)=S_{lk}^{ji}(\theta)\,. 
\end{equation}
If a theory is invariant under \emph{charge conjugation}, then we require that the S-matrix does not change under conjugation of the particles involved in the scattering process:
\begin{equation}
 S_{\bar i\bar j}^{\bar k\bar l}(\theta)=S_{ij}^{kl}(\theta)\,,
 \label{cc}
\end{equation}
where we denoted the charge-conjugated particles by barred indices.

\begin{figure}
\begin{centering}
\includegraphics[width=0.7\textwidth]{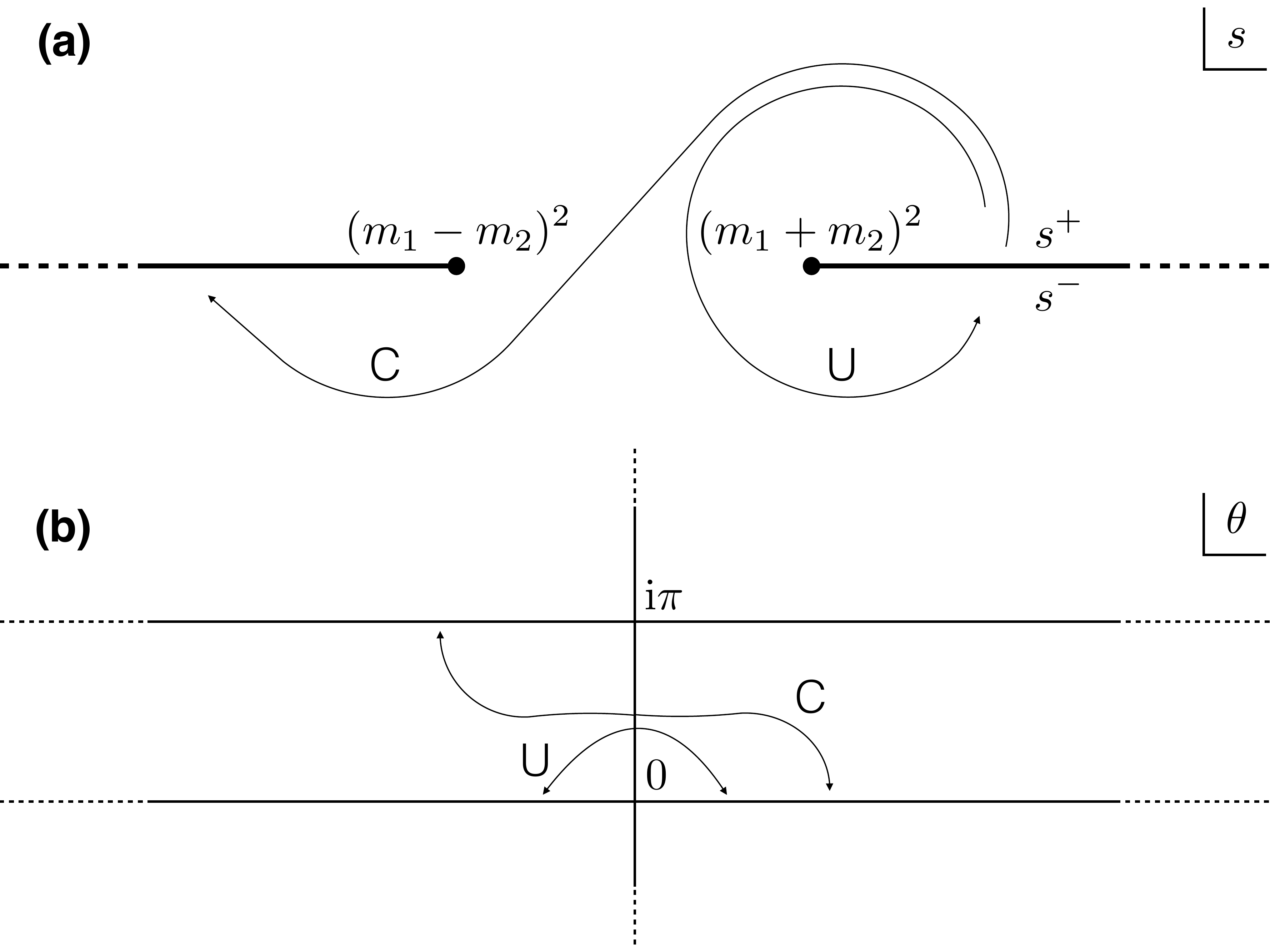}
\par\end{centering}
\caption{S-matrix analytic properties in the $s$-plane (a) and $\theta$-plane (b). $U$ and $C$ stand for unitarity and crossing transformations, respectively.}
\label{anprop}
\end{figure}

Now, in order to study the \emph{analytic properties} of the S-matrix, we recall the definitions (\ref{mandel}) of the Mandelstam variables. From their definitions (\ref{mandel}), it is easy to understand that $s, t$ and $u$ are 
the square of the center-of-mass energies in the channels defined by the process $i+j\rightarrow k+l$ ($s$-channel),  $i+\bar l\rightarrow k+\bar j$ ($t$-channel) and $i+\bar k\rightarrow l+\bar j$ ($u$-channel) respectively, as depicted in figure \ref{Sdef}.
In a physical process, $\theta_{i}-\theta_{j}$ has to be real, then $s$ has to be in the so-called \emph{physical region}, defined by $s^+=s+\mathrm{i}0$ and 
$s\ge (m_{i}+m_{j})^{2}$, \emph{i.e.} slightly above the right cut in the first of figure \ref{anprop} (a).

Then let us study the analytic continuation of $S(s)$ to the $s$-plane. We begin by imposing \emph{unitarity} in the physical region:
\begin{equation}
S_{ij}^{kl}(s^+)(S_{lk}^{mn})^{*}(s^+)=\delta_{i}^{n}\delta_{j}^{m}\,.
\label{physuni}
\end{equation}
According to the \emph{analyticity} assumption, the S-matrix in the physical region is the boundary value of a function that is analytic in the whole $s$-plane, then the unitarity property (\ref{physuni}) can be extended 
to the so-called \emph{hermitian analyticity}:
\begin{equation}
S_{ij}^{kl}(s^{*})=(S_{kl}^{ij})^{*}(s)\,.
\end{equation}
Adding to this property the time reversal symmetry, we get a stronger condition, that is the \emph{real analyticity}\footnote{See an interesting discussion on hermitian and real analyticity of the S-matrix, and their interplay, in 
\cite{Miramontes:1999gd} and references therein.}:
\begin{equation}
S_{ij}^{kl}(s^{*})=(S_{ij}^{kl})^{*}(s)\,,
\label{realan}
\end{equation} 
\emph{i.e.} the S-matrix is real for real values of $s$ and $(m_{i}-m_{j})^{2}\le s\le(m_{i}+m_{j})^{2}$. This means, in general, that real S-matrices do not describe physical processes.

Another fundamental property constraining relativistic S-matrices is the \emph{crossing symmetry}, meaning that the process in figure \ref{Sdef} has to be read equivalently along the $s$- and $t$-channels:
\begin{equation}
S_{ij}^{kl}(s)=S_{\bar li}^{\bar jk}(t)\,,
\end{equation}
where, as in (\ref{cc}), barred indices denote charge-conjugated particles or anti-particles, that can be considered also as particles propagating backwards in time.   
In terms of the rapidity, since $t(\theta)=s(\mathrm{i}\pi-\theta)$, crossing symmetry can be written as
\begin{equation}
S_{ij}^{kl}(\theta)=S_{\bar li}^{\bar jk}(\mathrm{i}\pi-\theta)\,.
\end{equation}
In particular, denoting by $\mathcal{C}$ the charge-conjugation operator, crossing symmetry can be written also as
\begin{equation}
S_{ij}^{kl}(\theta)=\mathcal{C}_{jn}S_{mi}^{nk}(\mathrm{i}\pi-\theta)\mathcal{C}^{ml}\,.
\label{crossym}
\end{equation}
Note that relation (\ref{crossym}) involves a dynamical transformation - in contrast to unitarity or discrete symmetries, where only the matrix form is involved - on the rapidity. As we will see in section \ref{gensol} and in the examples of section \ref{examples}, crossing symmetry plays a fundamental role in fixing the scalar factors of the S-matrices.
It is a property that profoundly reflects the relativistic invariance of the theory, since it uses the invariance of physical processes under exchange of 
space and time, {\it i.e.} under rotation of the $s$-channel to the $t$-channel. 

However, it is possible to generalize crossing symmetry to non-relativistic theories like $AdS/CFT$ thanks to its formulation in completely algebraic
ways \cite{Janik:2006dc, Arutyunov:2006yd}, that will be discussed in sections \ref{ZF} and \ref{Hopf}.

Turning back to the relativistic case, we notice that real analyticity (\ref{realan}) entails
\begin{equation}
S_{ij}^{kl}(s^{+})(S_{lk}^{mn})(s^{-})=\delta_{i}^{n}\delta_{l}^{m}\,,
\label{rean}
\end{equation}
where $s^{-}=s-\mathrm{i}0$. Equation (\ref{rean}) means that the S-matrix\footnote{Except for the trivial cases of $S=\pm1$.} has a branch cut between the regions where $s^+$ and $s^-$ are respectively defined, namely
there is a branch point in $s=(m_i+m_j)^2$. This is expected also since that point corresponds to the two-particle threshold, \emph{i.e.} it is a discontinuity point of the amplitude imaginary part (see more details on this aspect in \cite{Mussardobook}). Because of crossing symmetry, another branch cut starting from
$s=(m_i-m_j)^2$ towards $-\infty$ must exist, as depicted in Figure \ref{anprop} (a)\footnote{A different choice of the branch cuts is not equivalent, since $s=\pm\infty$ are branch points too.}. These are the only two branch cuts if the S-matrix is factorized, since particle production thresholds for more 
than two particles cannot appear.

Moreover, it is possible to show that the branch cut is of square root type, since unitarity gives
\begin{equation}
S(s^{+})S_{\gamma}(s^{+})=1\,,
\end{equation}
where $S_{\gamma}$ is the S-matrix analytically continued below the cut around the branch point $(m_{i}+m_{j})^{2}$, and then
\begin{equation}
S_{\gamma}(s^{-})=S^{-1}(s^{-})=S(s^{+})\,,
\end{equation}
where we used the real analyticity (\ref{rean}).
The last relation means basically that a double continuation around the branch point gives back the original S-matrix, \emph{i.e.} the branch cut is of square root type.

In order to show in a more concise way the analytic properties of the S-matrix, it is convenient to switch from the variable $s$ to the difference of rapidities, by inverting the relation (\ref{stheta}):
\begin{eqnarray}
\theta_{12}&=&\mbox{arccosh}\left(\frac{s-m_{1}^2-m_{2}^2}{2m_1m_2}\right)\nonumber\\
&=&\log\left(\frac{s-m_{1}^{2}-m_{2}^{2}+\sqrt{(s-(m_{1}+m_{2})^{2})(s-(m_{1}-m_{2})^{2})}}{2m_{1}m_{2}}\right)\,.
\end{eqnarray}
Then the \emph{physical sheet} maps to the strip $0\le\mbox{Im}(\theta_{12})\le\pi$, the second sheet corresponds to $\pi\le\mbox{Im}(\theta_{12})\le2\pi$ and so on, with periodicity $2\pi \mathrm{i}$.
Essentially, the branch cuts of the $s$-plane open up in such a way that all the Riemann sheets are mapped into strips $n\pi\le\mbox{Im}(\theta)\le (n+1)\pi$ and $S(\theta)$ is analytic at the images $\mathrm{i}n\pi$ of the branch points.
In conclusion, $S(\theta)$ is a meromorphic function of $\theta$ and its real analyticity implies that it is real on the imaginary axis of $\theta$. The main analytic properties of the two-particle relativistic
S-matrix can be represented in the $\theta$-plane as in figure \ref{anprop} (b).

\subsection{The Zamolodchikov-Faddeev algebra}
\label{ZF}

Having discussed the analytic properties of the two-particle integrable S-matrix, let us move to its algebraic features.
To do this, we introduce a purely algebraic setup, which is fully consistent with the properties studied in the previous section and will be also useful to extend some properties to the non-relativistic case, as explained in section \ref{sec:nonrel}.

Let us start by defining the creation and annihilation operators of excitations out of the vacuum state $|0\rangle$, that is left invariant by the symmetry algebra of the particular integrable quantum model under study:
\begin{equation}
\bar A^{a_j}(p_j)|0\rangle = 0=\langle 0|A_{a_j}(p_j)\,.
 \label{anndef}
\end{equation}
In particular, the particles created by $|p_{j}\rangle_{a_{j}}=A_{a_{j}}(p_{j})|0\rangle$ have momenta $p_{i}$ and transform in a linear 
irreducible representation of the symmetry algebra.

Then the asymptotic states (\ref{asymst}) can be written as
\begin{eqnarray}
\label{ZFasnonrel}
|p_1,p_2,\dots,p_n\rangle^{in}_{a_1,a_2,\dots,a_n}&=&A_{a_1}(p_1)A_{a_2}(p_2)\dots A_{a_n}(p_n)|0\rangle\,,\\
|p_1,p_2,\dots,p_n\rangle^{out}_{a_1,a_2,\dots,a_n}&=&A_{a_n}(p_n)\dots A_{a_2}(p_2)A_{a_1}(p_1)|0\rangle\,,
\end{eqnarray}
with $p_{1}>p_{2}>\dots>p_{n}$. 
On the other hand, the conjugated operators $\bar A_{a_j}(p_j)$ generate the dual states
\begin{eqnarray}
_{a_1,a_2,\dots,a_n}^{\ \ \ \ \ \ \ \ \ in}\langle p_1,p_2,\dots,p_n|&=&\langle0|\bar A^{a_1}(p_1)\bar A^{a_2}(p_2)\dots \bar A^{a_n}(p_n)\,,\\
_{\ a_1,a_2,\dots,a_n}^{\ \ \ \ \ \ \ \ \ out}\langle p_1,p_2,\dots,p_n|&=&\langle0|\bar A^{a_n}(p_n)\dots \bar A^{a_2}(p_2)\bar A^{a_1}(p_1)\,.
\label{ZFasnonreldual}
\end{eqnarray}
The operators $A_{a_j}(p_j), \bar A^{a_j}(p_j)$ are elements of an associative non-commutative algebra, the so-called Zamolodchikov-Faddeev (ZF) algebra
\cite{ZZ, Faddeev}.
In a relativistic case, (\ref{ZFasnonrel})-(\ref{ZFasnonreldual}) can be conveniently parametrized by the particles rapidities: 
\begin{eqnarray}
|\theta_1,\theta_2,\dots,\theta_n\rangle^{in}_{a_1,a_2,\dots,a_n}&=&A_{a_1}(\theta_1)A_{a_2}(\theta_2)\dots A_{a_n}(\theta_n)|0\rangle\,,
\label{ZFas1}\\
|\theta_1,\theta_2,\dots,\theta_n\rangle^{out}_{a_1,a_2,\dots,a_n}&=&A_{a_n}(\theta_n)\dots A_{a_2}(\theta_2)A_{a_1}(\theta_1)|0\rangle\,,
\label{ZFas2}\\
_{a_1,a_2,\dots,a_n}^{\ \ \ \ \ \ \ \ \ in}\langle \theta_1,\theta_2,\dots,\theta_n|&=&\langle0|\bar A^{a_1}(\theta_1)\bar A^{a_2}(\theta_2)\dots \bar A^{a_n}(\theta_n)\,,
\label{ZFas3}\\
^{\ \ \ \ \ \ \ \ out}_{a_1,a_2,\dots,a_n}\langle \theta_1,\theta_2,\dots,\theta_n|&=&\langle0|\bar A^{a_n}(\theta_n)\dots \bar A^{a_2}(\theta_2)\bar A^{a_1}(\theta_1)\,.
\label{ZFas4}
\end{eqnarray}
For simplicity of notation, all the following equations involving ZF operators will be understood as acting on $|0\rangle$.

Defining the asymptotic states in this way allows to interpret the scattering processes as simple reordering of ZF operators in the rapidity space.
Indeed, writing explicitly the asymptotic states of equation (\ref{inSout}) in terms of ZF generators as in (\ref{ZFas1}), (\ref{ZFas2}) and dropping the vacuum states, it becomes
\begin{equation}
A_{i}(\theta_{1})A_{j}(\theta_{2})=A_{l}(\theta_{2})A_{k}(\theta_{1})S_{ij}^{kl}(\theta_{1}-\theta_{2})\,,
\label{ZFcr}
\end{equation}
that is the commutation relation between the ZF algebra elements, and it can be interpreted as definition of the two-particle S-matrix.
The ZF algebra is completed by the commutation relations involving the annihilation operators (\ref{anndef}):
\begin{eqnarray}
\label{ZFhcr}
\bar A^{i}(\theta_{1})\bar A^{j}(\theta_{2})&=&S^{ij}_{kl}(\theta_{1}-\theta_{2})\bar A^{l}(\theta_{2})\bar A^{k}(\theta_{1})\,,\\
\label{ZFmcr}
\bar A^k(\theta_1)A_j(\theta_2)&=&A_l(\theta_2)S_{ij}^{kl}(\theta_2-\theta_1)\bar A^i(\theta_1)+\delta(\theta_1-\theta_2)\delta_j^k\,,
\end{eqnarray}
that generalize the usual bosonic and fermionic canonical commutation relations, corresponding to $S=1$ and $S=-1$ respectively. The $\delta$-function in the r.h.s of (\ref{ZFmcr}) is related to the normalization of the states, that is $_{i}\langle\theta_{1}|\theta_{2}\rangle_{j}=\delta(\theta_{1}-\theta_{2})\delta_{ij}$.
 
Now, writing the commutation relation for the elements labeled by $k$ and $l$
\begin{equation}
A_{l}(\theta_{2})A_{k}(\theta_{1})=S_{lk}^{mn}(\theta_{2}-\theta_{1})A_{n}(\theta_{1})A_{m}(\theta_{2})\,,
\label{ZFcrkl}
\end{equation}   
and plugging it into (\ref{ZFcr}), one can show
\begin{equation}
A_{i}(\theta_{1})A_{j}(\theta_{2})=S_{ij}^{kl}(\theta_{1}-\theta_{2})S_{lk}^{mn}(\theta_{2}-\theta_{1})A_{n}(\theta_{1})A_{m}(\theta_{2})\,,
\label{unishow}
\end{equation}
that is equivalent to
\begin{equation}
S_{ij}^{kl}(\theta_{1}-\theta_{2})S_{lk}^{mn}(\theta_{2}-\theta_{1})=\delta_{i}^{n}\delta_{j}^{m}\,.
\label{uni}
\end{equation}
This property is also called \emph{braiding unitarity}.

On the other hand, in order to get (\ref{physuni}), also referred as \emph{physical unitarity}, we have to take the hermitian conjugation of (\ref{ZFcr}):
\begin{equation}
\bar A^{j}(\theta_{2})\bar A^{i}(\theta_{1})=(S^{\dagger})^{ij}_{kl}(\theta_{1}-\theta_{2})\bar A^{k}(\theta_{1})\bar A^{l}(\theta_{2})\,.
\label{ZFhcr2}
\end{equation}
Thus, exchanging $\theta_{1}$ with $\theta_{2}$ and permuting the ZF operators, we get
\begin{equation}
\bar A^{i}(\theta_{1})\bar A^{j}(\theta_{2})=(S^{\dagger})^{ij}_{kl}(\theta_{2}-\theta_{1})\bar A^{l}(\theta_{2})\bar A^{k}(\theta_{1})\,.
\label{ZFhcrperm}
\end{equation}
But we also know that (\ref{ZFhcr}) holds, then $S_{ij}^{kl}(\theta_{1}-\theta_{2})=(S^{\dagger})_{ij}^{kl}(\theta_{2}-\theta_{1})$. Finally, using the braiding unitarity (\ref{uni}), we get
(\ref{physuni}): $SS^{\dagger}=1$.
\begin{framed}
\paragraph*{Exercises}
\begin{enumerate}
\item We leave as an exercise the derivation of CPT invariance using the ZF algebra and knowing that $A_{i}(\theta)\rightarrow A_{i}(-\theta)$ under parity and time reversal. The charge-conjugation symmetry, on the 
other hand, requires that the ZF algebra maps to itself under the transformations $A_{i}(\theta)\rightarrow \bar A_{i}^{t}(\mathrm{i}\pi+\theta)\mathcal{C}$, $\bar A_{i}(\theta)\rightarrow \mathcal{C}^{\dagger}A_{i}^{t}(\mathrm{i}\pi+\theta)$, 
where $\mathcal{C}$ is the charge-conjugation matrix defined by $A_i(\theta)=\mathcal{C}_{ij}\bar A^j$ and the superscript $t$ denotes the transposition. 

\item Prove that, if the charge conjugation acts only on one sector of the two-particle space, one gets the crossing symmetry relation (\ref{crossym}).

\item Show that the associativity property of the ZF algebra implies the YBE (\ref{YBE}).

\end{enumerate}
\end{framed}
\subsubsection{Non-relativistic case}
\label{sec:nonrel}

In a non-relativistic model, the S-matrix does not depend on the difference of rapidities, but separately on the momenta of the particles.
Therefore, the ZF algebra generalizes to
\begin{eqnarray}
\label{ZFcrnonrel1}
A_{i}(p_{1})A_{j}(p_{2})&=&A_{l}(p_{2})A_{k}(p_{1})S_{ij}^{kl}(p_{1},p_{2})\,,\\
\bar A^{i}(p_{1})\bar A^{j}(p_{2})&=&S^{ij}_{kl}(p_{1},p_{2})\bar A^{l}(p_{2})\bar A^{k}(p_{1})\,,\\
\bar A^k(p_1)A_j(p_2)&=&A_l(p_2)S_{ij}^{kl}(p_2,p_1)\bar A^i(p_1)+\delta(p_1-p_2)\delta_j^k\,.
\label{ZFcrnonrel3}
\end{eqnarray}
Analogously, the YBE (\ref{YBEd}) becomes
\begin{equation}
\hspace{-0.1cm}\sum_{c_1,c_2,c_3}\hspace{-0.1cm}S_{a_1a_2}^{c_1c_2}(p_{1},p_{2})S_{c_1a_3}^{b_1c_3}(p_{1},p_{3})S_{c_2c_3}^{b_2b_3}(p_{2},p_{3})=\hspace{-0.1cm}\sum_{c_1,c_2,c_3}\hspace{-0.1cm}S_{a_2a_3}^{c_2c_3}(p_{2},p_{3})S_{a_1c_3}^{c_1b_3}(p_{1},p_{3})
S_{c_1c_2}^{b_1b_2}(p_{1},p_{2})\,,
\label{YBEnonrel}
\end{equation}
and the physical properties discussed in section \ref{proprel} can be derived using the properties of the ZF algebra. For example, relations similar to (\ref{ZFcrkl}) and (\ref{unishow}) lead to
the braiding unitarity condition
\begin{equation}
S_{ij}^{kl}(p_{1},p_{2})S_{lk}^{mn}(p_{2},p_{1})=\delta_{i}^{n}\delta_{j}^{m}\,.
\label{nonrelbu}
\end{equation} 
Together with relations analogous to (\ref{ZFhcr2}) and (\ref{ZFhcrperm}), (\ref{nonrelbu}) gives the physical unitarity condition
\begin{equation}
(S^{\dagger})_{ij}^{kl}(p_{1},p_{2})S_{kl}^{mn}(p_{1},p_{2})=\delta_{i}^{m}\delta_{j}^{n}\,.
\end{equation} 
Furthermore, the properties of the asymptotic states under transformations of parity and time reversal, respectively denoted by $\mathcal{P}$ and $\mathcal{T}$,
\begin{eqnarray}
\mathcal{P}|p_{1},p_{2},\dots,p_{n}\rangle^{(in)}_{i_{1},\dots,i_{n}}&=&|-p_{1},-p_{2},\dots,-p_{n}\rangle^{(in)}_{i_{1},\dots,i_{n}}\,,\\
\mathcal{T}|p_{1},p_{2},\dots,p_{n}\rangle^{(in)}_{i_{1},\dots,i_{n}}&=&|-p_{1},-p_{2},\dots,-p_{n}\rangle^{(out)}_{i_{1},\dots,i_{n}}\,,
\end{eqnarray}
written in terms of ZF operators as
\begin{eqnarray}
\mathcal{P}A_{i_{1}}(p_{1})\dots A_{i_{n}}(p_{n})|0\rangle&=&A_{i_{n}}(-p_{n})\dots A_{i_{1}}(-p_{1})|0\rangle\,,\\
\mathcal{T}A_{i_{1}}(p_{1})\dots A_{i_{n}}(p_{n})|0\rangle&=&A_{i_{1}}(-p_{1})\dots A_{i_{n}}(-p_{n})|0\rangle\,,
\end{eqnarray}
allow us to generalize the discrete symmetries listed in section \ref{proprel} for the relativistic case in the following way (see Chapter 3 of \cite{AF} for further details on the derivation): 
\begin{itemize}
 \item \emph{parity}: $S_{ij}^{kl}(p_1,p_2)=S_{ji}^{lk}(-p_2,-p_1)$,
 \item \emph{time reversal}: $S_{ij}^{kl}(p_1,p_2)=S_{lk}^{ji}(-p_2,-p_1)$,
\end{itemize}
while the symmetry under \emph{charge conjugation} translates trivially to
the condition
\begin{equation}
S_{ij}^{kl}(p_1,p_2)=S_{\bar i\bar j}^{\bar k\bar l}(p_1,p_2)\,,
\end{equation}
or, using the charge-conjugation operator $\mathcal{C}$,
\begin{equation}
S_{ij}^{kl}(p_1,p_2)=\mathcal{C}_{ir}\mathcal{C}_{js}S_{mn}^{rs}(p_1,p_2)\mathcal{C}^{mk}\mathcal{C}^{nl}\,,
\end{equation}
where $\mathcal{C}_{ij}\mathcal{C}^{jk}=\delta_i^k$.

Although \emph{crossing symmetry} is a property that emerges naturally in the context of relativistic scattering theories and at a first approach its generalization to systems where time and space cannot be exchanged  might seem impossible, it can be recovered, as all the other properties discussed above, from an additional requirement on the ZF algebra \cite{AF, Arutyunov:2006yd}.
Basically, we recall that in the relativistic case the crossing transformation entails an exchanging of a particle with an anti-particle and a kinematic map $\theta\rightarrow \mathrm{i}\pi+\theta$ on the rapidity of the conjugated particle. This translates to the maps $p\rightarrow -p$ and $E\rightarrow -E$ on the momentum and energy of a non-relativistic particle. 

Then the ZF generators must transform as
\begin{equation}
A_{i}(p)\rightarrow A_{\bar i}(-p)=A_{j}(-p)\mathcal{C}^{ji}\ ;\quad \bar A^{i}(p)\rightarrow \bar A^{\bar i}(-p)=\mathcal{C}_{ij}A^{j}(-p)\,.
\end{equation}
Requiring that the commutation relations (\ref{ZFcrnonrel1})-(\ref{ZFcrnonrel3}) are invariant under this transformation implies 
\begin{eqnarray}
S_{ij}^{kl}(p_1,-p_2)&=&\mathcal{C}_{jn}S_{mi}^{nk}(p_1,p_2)\mathcal{C}^{ml}\,,\\
S_{ij}^{kl}(-p_1,p_2)&=&\mathcal{C}_{in}S_{jm}^{ln}(p_1,p_2)\mathcal{C}^{mk}\,,
\label{crossnonrel}
\end{eqnarray}
which are the crossing symmetry relations for a non-relativistic two-particle S-matrix.

\subsection{General relativistic solutions}
\label{gensol}

Turning back to relativistic S-matrices, we want to show here how they can be completely determined using their analytic properties and symmetries.
First of all, the YBE can determine the ratios between S-matrix elements that belong to the same mass multiplet.
Thus, a general solution of the YBE can be written as
\begin{equation}
S_{ij}^{kl}(\theta)=\frac{1}{f(\theta)}R_{ij}^{kl}(\theta)\,,
\end{equation}
where $R$ is the matrix of the ratios between amplitudes fixed by the YBE, $f$ and $R_{ij}^{kl}$ are meromorphic functions of $\theta$.
\begin{framed}
\paragraph*{Exercises}
\begin{enumerate}
\item
Show that $R$ satisfies $R_{ij}^{kl}(0)=\delta_{i}^{l}\delta_{j}^{k}R_{0}$ (from the commutation relation of the ZF algebra).
\item
Using the previous relation and the YBE (\ref{YBEd}), show that
\begin{equation}
R_{ij}^{nm}(\theta)R_{nm}^{kl}(-\theta)=\delta_{i}^{k}\delta_{j}^{l}Q(\theta)\,,
\end{equation}
and that the braiding unitarity reduces to
\begin{equation}
f(\theta)f(-\theta)=Q(\theta)\,.
\label{unif}
\end{equation}
\end{enumerate}
\end{framed}
Rescaling the rapidity by an arbitrary constant $\lambda$ ($\theta\rightarrow\lambda\theta$), $R$ is still solution of the YBE and, for a suitable choice of $\lambda$, in all the known cases it satisfies
\begin{equation}
R_{ij}^{kl}(\theta)=R_{il}^{kj}(\mathrm{i}\pi-\theta)\,.
\end{equation}
Then the crossing symmetry reduces to
\begin{equation}
f(\theta)=f(\mathrm{i}\pi-\theta)\,.
\label{crossf}
\end{equation}

Therefore $f(\theta)$ is fixed by (\ref{unif}) and (\ref{crossf}) up to a function $\phi(\theta)$, called the CDD factor after Castillejo-Dalitz-Dyson \cite{CDD}, satisfying
\begin{eqnarray}
&&\phi(\theta)=\phi(\mathrm{i}\pi-\theta)\,,\\
&&\phi(\theta)\phi(-\theta)=1\,.
\label{cdd}
\end{eqnarray}
This ambiguity corresponds to the freedom to add zeros and poles with period $2\pi \mathrm{i}$ to $f(\theta)$, due to the infinite discrete set of solutions for $\phi$, in general. So, if we denote as $f_{min}(\theta)$ the solution of (\ref{unif}) and (\ref{crossf}) with minimal number of poles and zeros, then the general solution for $f$ is $f(\theta)=f_{min}(\theta)\phi(\theta)$.

Another fundamental restriction for generic S-matrix elements is the invariance under  the symmetry algebra of the model under study.
The corresponding constraints can be derived by acting with the symmetry generators $J^{a}$, where $a$ runs from 1 to the dimension of the symmetry algebra, on the ZF relations (\ref{ZFcr}):
\begin{equation}
J^{a}A_{i}(\theta_{1})A_{j}(\theta_{2})=S_{ij}^{kl}(\theta_{1}-\theta_{2})J^{a}A_{l}(\theta_{2})A_{k}(\theta_{1})\,.
\end{equation}
The action of $J^{a}$ on the two-particle states is given by
\begin{equation}
J^{a}A_{i}(\theta_{1})A_{j}(\theta_{2})=(J^{a})_{ij}^{kl}(\theta_{1},\theta_{2})A_{k}(\theta_{1})A_{l}(\theta_{2})\,,
\end{equation}
where $(J^{a})_{ij}^{kl}$ are the matrix elements of the two-particle generator $J^{a}_{12}$, that acts on the two-particle spaces as $J^{a}_{12}=J^{a}\otimes\mathbb{I}+\mathbb{I}\otimes J^{a}$.
Thus, the S-matrix invariance can be written in matrix form as
\begin{equation}
(J^{a}\otimes\mathbb{I}+\mathbb{I}\otimes J^{a})S=S(J^{a}\otimes\mathbb{I}+\mathbb{I}\otimes J^{a})\,.
\label{syminv}
\end{equation}
Summarizing, the steps necessary to compute the S-matrix in an integrable theory are the following:

\begin{itemize}
\item determine the structure of the S-matrix by imposing invariance under the symmetry generators (by solving the equations given by the condition (\ref{syminv}));
\item find the ratios between the remaining undetermined S-matrix elements by imposing the YBE (\ref{YBE});
\item fix the remaining (minimal) overall scalar factor, up to CDD factors, by imposing unitarity and crossing symmetry.
\end{itemize}

We will see some detailed application of this algorithm in few particular cases (sine-Gordon, $SU(2)$ and $SU(3)$ chiral Gross-Neveu models), which are discussed in section \ref{examples}.

In the non-relativistic example of $AdS_{5}/CFT_{4}$, for instance, the S-matrix for the fundamental excitations was determined in \cite{Beisert:2005tm}, up to a scalar factor, imposing invariance under two copies of centrally extended $SU(2|2)$ symmetry algebras. Such S-matrix turned out to satisfy identically the YBE, while the crossing symmetry condition, implemented in \cite{Arutyunov:2006yd} and \cite{Janik:2006dc} through the algebraic frameworks illustrated respectively in the previous section and in section \ref{Hopf}, led to an equation for the scalar factor, that was solved in \cite{Volin:2009uv} (see also the review \cite{Vieira:2010kb}).

\subsubsection{Purely elastic case}

While elastic scattering essentially means that the set of outgoing particles is identical to the incoming one, \emph{purely elastic scattering} is further constrained by not having reflection between particles.
So, particles can be just transmitted and the S-matrix is \emph{diagonal}:
\begin{equation}
S_{ij}^{kl}=\delta_{i}^{k}\delta_{j}^{l}S_{ij}\,.
\end{equation}
The YBE is identically satisfied and the system of equations of unitarity and crossing symmetry is solved by a function $S(\theta)$ with period $2\pi\mathrm{i}$, given by \cite{Klassen:1989ui}
\begin{equation}
S (\theta)= \prod_{\alpha} f_{\alpha}(\theta)\,;\quad f_{\alpha}(\theta)=\frac{\sinh\left(\frac{\theta+\mathrm{i}\pi\alpha}{2}\right)}{\sinh\left(\frac{\theta-\mathrm{i}\pi\alpha}{2}\right)}\,,
\end{equation}
with $\alpha$ belonging to a subset $A_{ij}$ of $\mathbb{C}$ invariant under complex conjugation. Indeed one can easily verify that
\begin{equation}
f_{\alpha}(\theta)f_{\bar\alpha}(\theta)=f_{\alpha}(2\pi\mathrm{i}+\theta)f_{\bar\alpha}(2\pi\mathrm{i}+\theta)\,,
\end{equation}
for any complex $\alpha$.
Periodicity implies that $\alpha$ can be chosen in the interval $-1\le\alpha\le1$. Poles are at $\mathrm{i}\pi\alpha$, while zeros are at $-\mathrm{i}\pi\alpha$, then they are contained in the strip $-\pi\le\mbox{Im}(\theta)\le\pi$.

In case of neutral particles (particle = antiparticle), then $S_{ij}(\theta)=S_{ij}(\mathrm{i}\pi-\theta)$ and the solution of unitarity and crossing is a product over arbitrary $\alpha$ of the functions
\begin{equation}
F_{\alpha}(\theta)=f_{\alpha}(\theta)f_{\alpha}(\mathrm{i}\pi-\theta)\,,
\end{equation}
with simple poles at $\theta=\mathrm{i}\pi\alpha, \mathrm{i}\pi(1-\alpha)$, zeros at $\theta=-\mathrm{i}\pi\alpha, -\mathrm{i}\pi(1-\alpha)$, related by crossing.
Anyway, unitarity and crossing are not sufficient to fix the sets of poles/zeros $A_{ij}$ of $\prod_{\alpha\in A_{ij}}f_{\alpha}$ or  $\prod_{\alpha\in A_{ij}}F_{\alpha}$.
We will see in the next section how it is actually possible to fix them.

\section{Poles structure and bootstrap principle}

Since in the region $(m_{i}-m_{j})^{2}\le s\le(m_{i}+m_{j})^{2}$ ($0\le\mbox{Im}(\theta_{ij})\le\pi$ in terms of rapidity) it is possible to create, from incoming particles of masses $m_{i}$ and $m_{j}$, only one-particle states with $m<m_{i}+m_{j}$, then simple poles of the S-matrix in that range of $s$ are generally expected to correspond to \emph{bound states}.
In order to clarify this correspondence with a simple example, let us consider the one-dimensional scattering problem associated to a quantum mechanical system with delta potential.

\subsection{Delta potential scattering problem}

We want to solve the Schr\"odinger problem corresponding to the Hamiltonian
\begin{equation}
H=\frac{p^2}{2m}+V(x)\,,
\label{ham}
\end{equation}
with potential $V(x)=-2g\delta(x)$ and $g>0$. The Schr\"odinger equation
\begin{equation}
\left(-\frac{\hbar^2}{2m}\frac{d^2}{dx^2}-2g\delta(x)\right)|\psi\rangle=E|\psi\rangle\,, 
\end{equation}
can be conveniently rewritten as
\begin{equation}
\left(-\frac{d^2}{dx^2}-2g\delta(x)\right)|\psi\rangle=k^2|\psi\rangle\,, 
\label{Schrprob}
\end{equation}
by rescaling $g\rightarrow \frac{\hbar^2}{2m}g$ and defining $k^2=E\frac{2m}{\hbar^2}$.
We look for solutions of (\ref{Schrprob}) in the following generic form
\begin{equation}
\psi(x)=\left\{
\begin{array}{ll}
 A e^{\mathrm{i}kx}+B e^{-\mathrm{i}kx}\,;\ \ x<0\,,\\
 C e^{\mathrm{i}kx}+D e^{-\mathrm{i}kx}\,;\ \ x>0\,.
\end{array}
\right.
\label{Schrsol}
\end{equation}
Since we want to consider the scattering of incident particles coming from the left and being reflected or transmitted by the $\delta$-potential barrier, then we have not incoming waves from the right, $i.e.$ $D=0$, and 
$A$, $B$, $C$ are the amplitudes of the incoming, reflected and transmitted wave packets respectively.
These coefficients can be found by solving the continuity condition of the wave function across the point $x=0$
\begin{equation}
\psi(0^-)=\psi(0^+)\,,
\label{cond1}
\end{equation}
and the discontinuity condition on the first derivative of $\psi(x)$ given by integrating equation (\ref{Schrprob}) between $\epsilon$ and $-\epsilon$, with $\epsilon\rightarrow 0$
\begin{equation}
\psi'(0^-)-\psi'(0^+)-2g\psi(0)=0\,.
\label{cond2}
\end{equation}
Condition (\ref{cond1}) gives
\begin{equation}
A+B=C\,, 
\end{equation}
while (\ref{cond2}) implies
\begin{equation}
\mathrm{i}kB+gC=0\,. 
\end{equation}
Thus, the transmission and reflection coefficient are, respectively
\begin{equation}
 T=\frac{C}{A}=\frac{k}{k-\mathrm{i}g}\,;\quad R=\frac{B}{A}=\frac{\mathrm{i}g}{k-\mathrm{i}g}\,.
\end{equation}
If $k$ is complex, its imaginary part contributes to the real parts of the exponentials in (\ref{Schrsol}).
Moreover, both the transmission and reflection coefficients in (\ref{Schrsol}) have a pole in $k=\mathrm{i}g$, but we can still normalize the incoming wave function by setting $A=0$.
Thus, at the value $k=\mathrm{i}g$, with $g>0$, (\ref{Schrsol}) gives a physically admissible solution, $i.e.$ decreasing to zero at large distances, with just outgoing waves and not incoming ones:
it corresponds to a bound state.

\begin{figure}
\begin{centering}
\includegraphics[width=15cm]{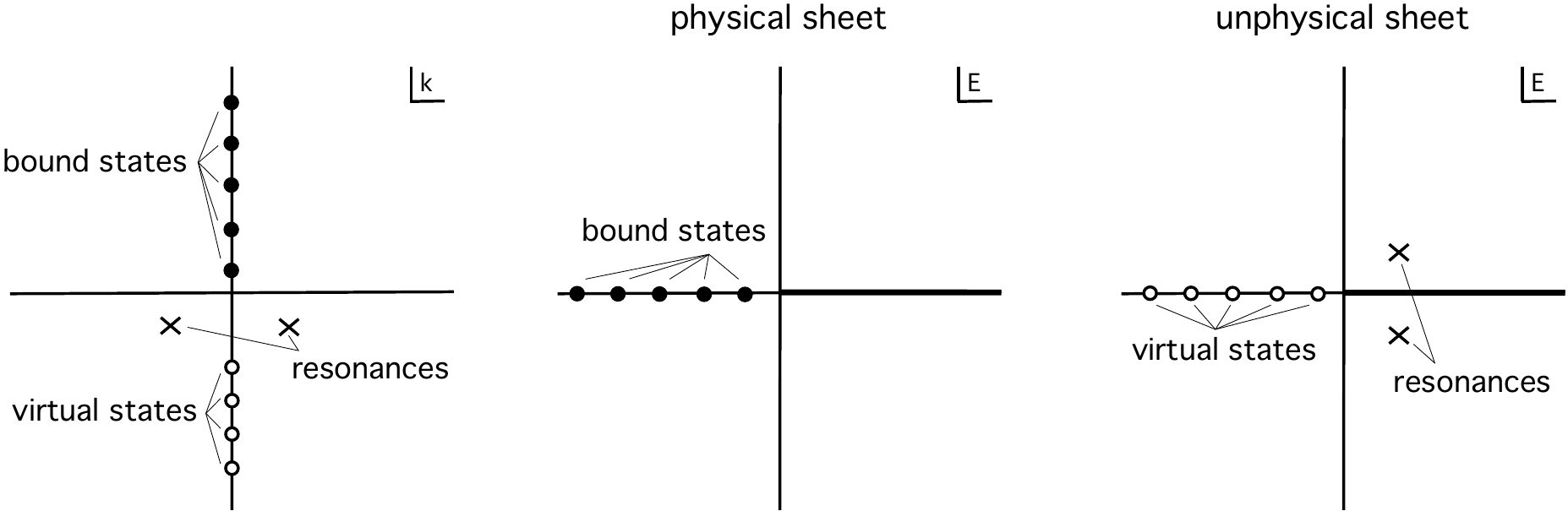}
\par\end{centering}
\caption{S-matrix poles on $k$- and $E$-plane.}
\label{fig:Eplane}
\end{figure}

Moreover, considering the time evolution of (\ref{Schrsol})
\begin{equation}
 \psi(x,t)=e^{-\mathrm{i}t\frac{\hbar k^2}{2m}}\psi(x)
 \label{timev}
\end{equation}
we see that no solutions can exist with $k=k_1+\mathrm{i}k_2$ ($k_{1,2}\in\mathbb{R}$), $k_1\neq0$ and $k_2>0$, since (\ref{timev}) would increase exponentially with time in some channel. This would contradict the conservation of probability, then there are no poles of the S-matrix with non-vanishing real part in the upper half plane of $k$. 

Poles of the S-matrix with negative imaginary part lead still to unphysical states, since the corresponding amplitude increase exponentially in a given channel, but such divergences at large distances are compensated by exponential decreasing amplitudes in another channel, giving an overall conservation of probability. 

In particular, purely imaginary negative poles, that can be realized in our $\delta$-potential case by considering $g<0$, take the name of \emph{virtual states}. 

With $k_1\neq0$, instead, we have a so-called \emph{resonance}, since it can be shown \cite{Mussardobook} that the corresponding cross section takes the typical shape of a Breit-Wigner distribution.

In summary, if we parametrize the S-matrix with the energy $E$, then $S(E)$ has a cut on the positive real axis and the region Im$(k)>0$ corresponds to the first (physical) sheet, while the region Im$(k)<0$ maps to the second or unphysical sheet. Moreover, poles on the negative real axis in the physical sheet correspond to bound states, resonances and virtual states are poles on the unphysical sheet, with the latter placed on the negative real
axis, as in figure \ref{fig:Eplane}.

\subsection{Bound states and bootstrap equations}

\begin{figure}
\begin{centering}
\hspace{-2cm}
\includegraphics[width=0.3\textwidth]{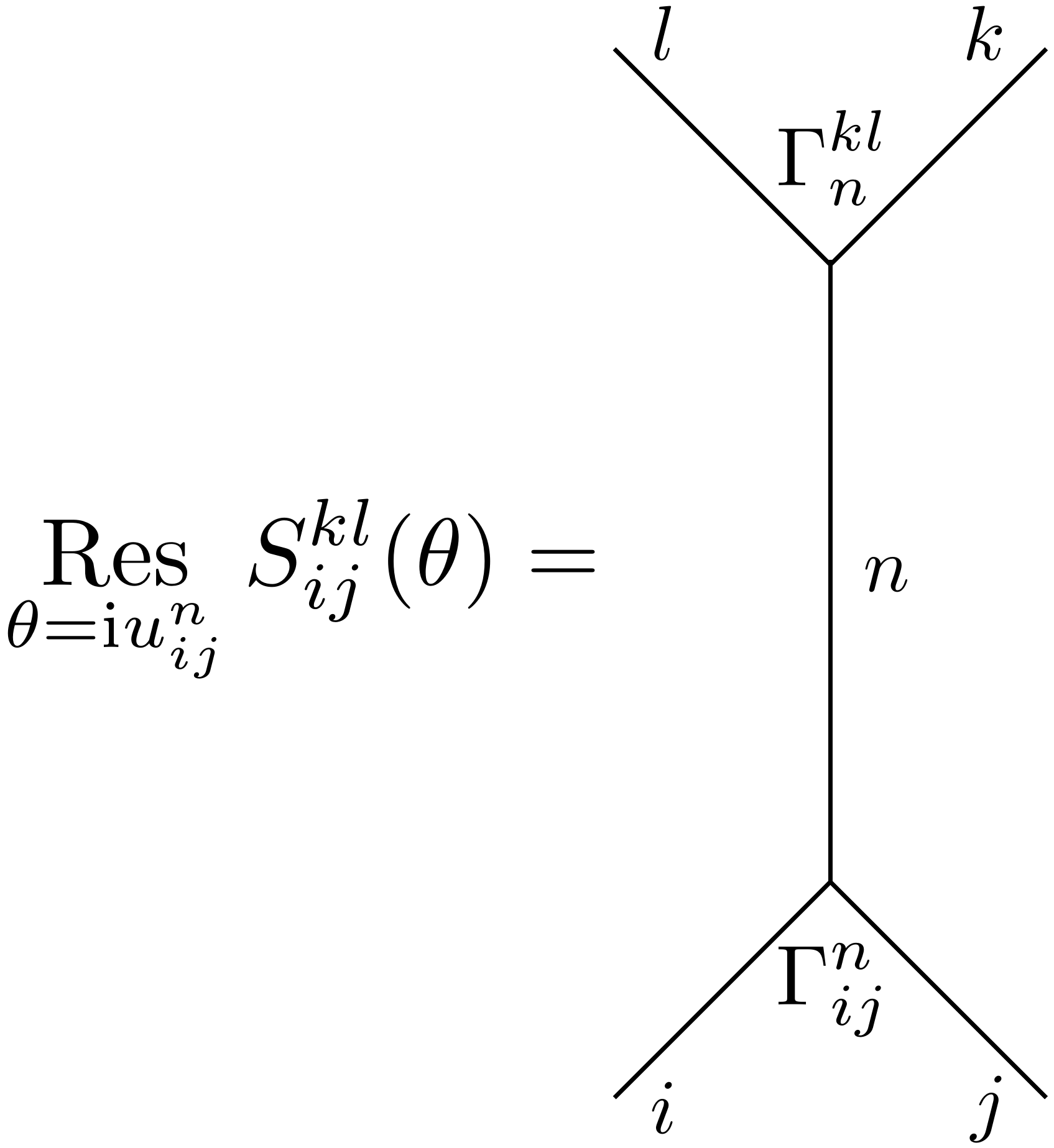}
\par\end{centering}
\caption{Scattering process associated to a bound state.}
\label{fig:pole}
\end{figure}

\begin{figure}
\begin{centering}
\includegraphics[width=0.5\textwidth]{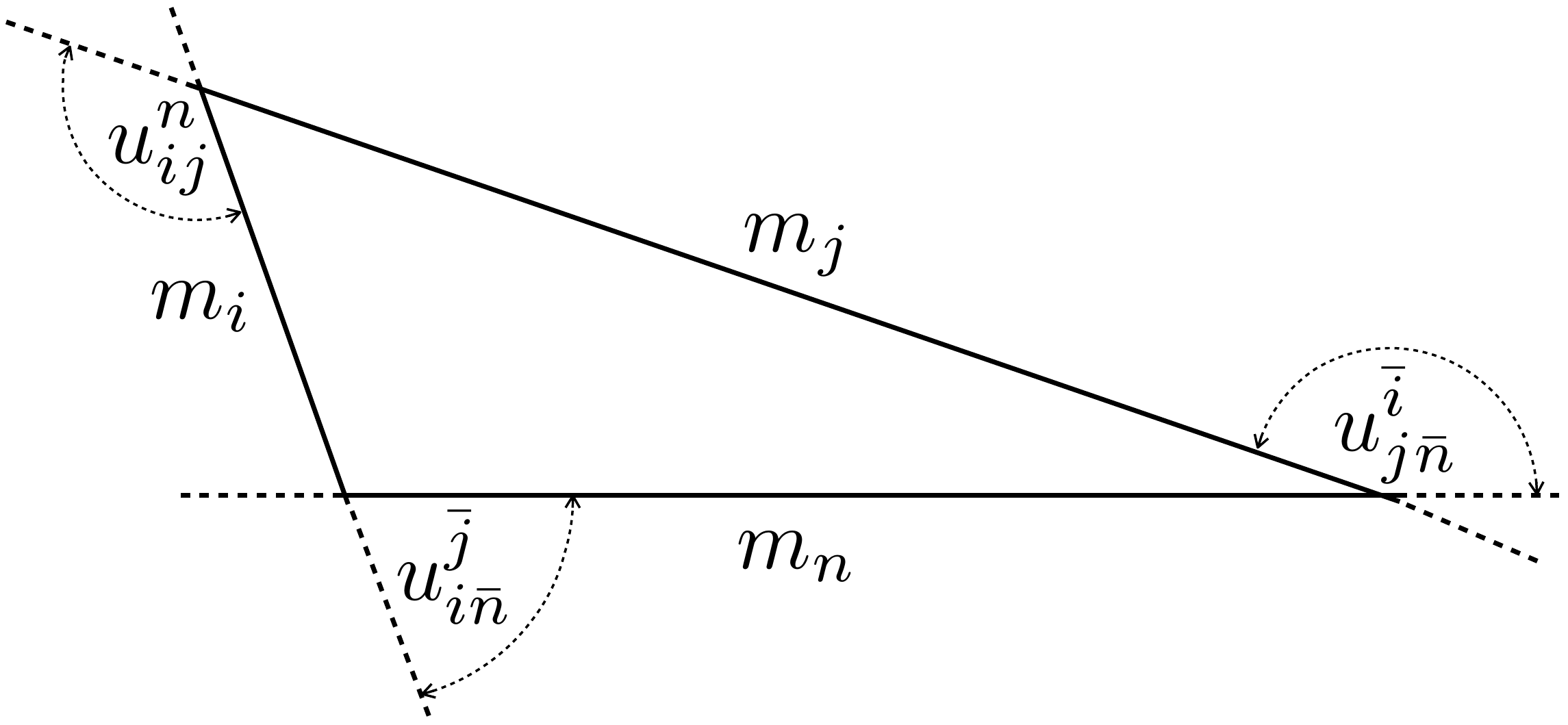}
\par\end{centering}
\caption{Carnot theorem interpretation of relation (\ref{bsmasses}).}
\label{carnot}
\end{figure}

Close to a simple pole $\theta=\mathrm{i}u_{ij}^{n}$, corresponding to a bound state $n$ formed by two particles $i$ and $j$, a generic relativistic S-matrix can be written as (see figure \ref{fig:pole})
\begin{equation}
S_{ij}^{kl}(\theta)\simeq\frac{\Gamma_{ij}^nR_n\Gamma^{kl}_n}{\theta-\mathrm{i}u_{ij}^{n}}\,,
\label{res}
\end{equation}
where $R_n$ is the residue and $\Gamma_{ij}^n, \Gamma^{kl}_n$ are projectors of single particle ($i,j,k$ and $l$) spaces onto the space of the bound state $n$.

The mass of the bound state is given by
\begin{equation}
s=m_{n}^{2}=m_{i}^{2}+m_{j}^{2}+2m_{i}m_{j}\cos u_{ij}^{n}\,.
\label{bsmasses}
\end{equation}
It is interesting to notice that this relation has the geometrical meaning of the Carnot theorem for the triangle of the masses, as illustrated in figure \ref{carnot}.

\begin{figure}
\begin{centering}
\hspace{1cm}
\includegraphics[width=0.6\textwidth]{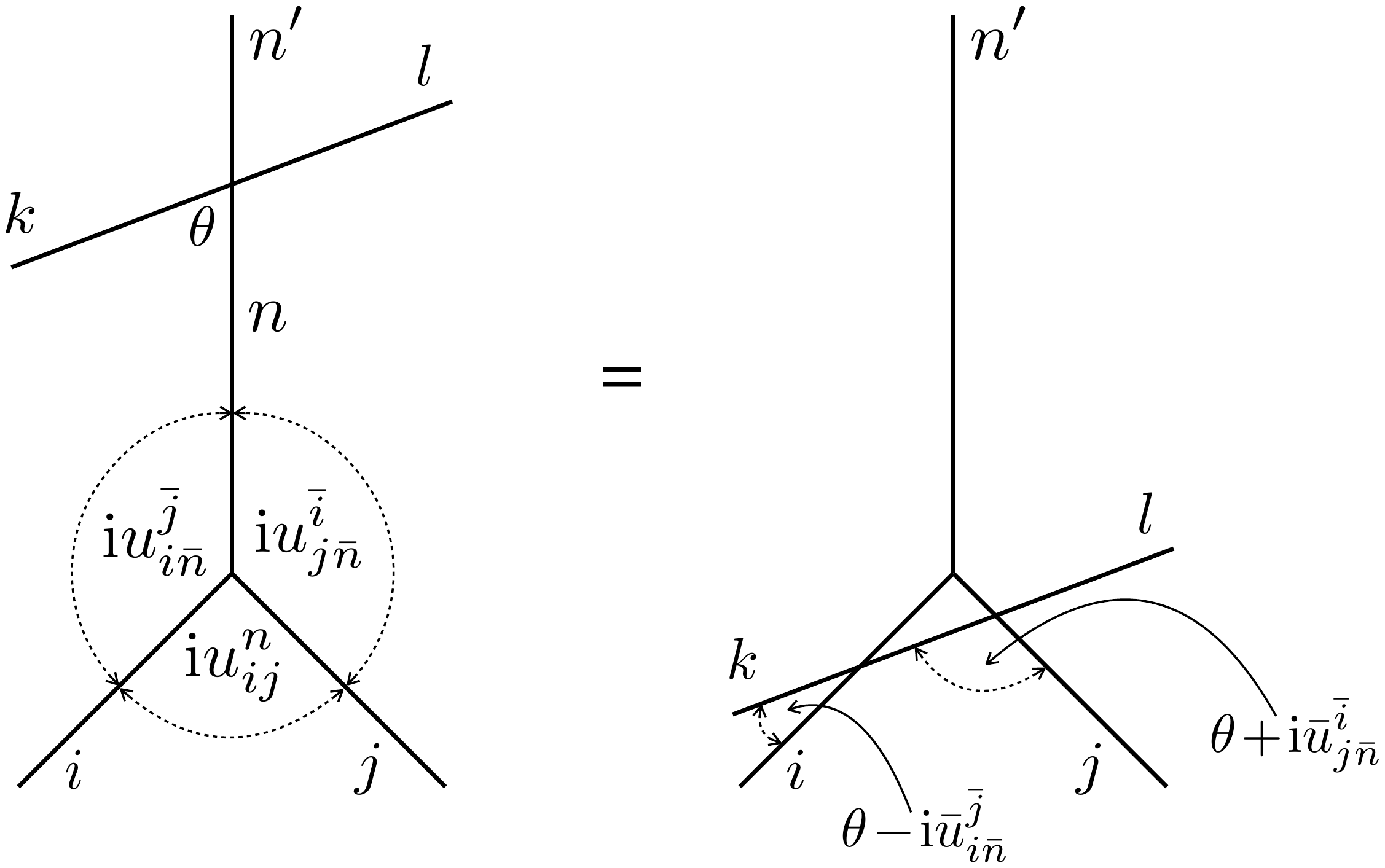}
\par\end{centering}
\caption{Bootstrap equation.}
\label{fig:bootstrap}
\end{figure}

The main idea of the \emph{bootstrap approach} is that the bound states can be considered on the same footing as the asymptotic states describing fundamental particles, even though the bound states can have bigger
masses.
Indeed, the ZF element describing bound states can be formally defined as
\begin{equation}
B_n(\theta)=\lim_{\epsilon\rightarrow0}A_i(\theta-\mathrm{i}\bar u_{i\bar n}^{\bar j}-\epsilon)A_j(\theta+\mathrm{i}\bar u_{j\bar n}^{\bar i}+\epsilon)\,,
\label{bsZF}
\end{equation}
where $\bar u=\pi-u$ and the angles $u_{i\bar n}^{\bar j}$, $u_{j\bar n}^{\bar i}$ are defined according to the l.h.s. of figure \ref{fig:bootstrap}.

Therefore, the S-matrix for the scattering between any particle $k$ and a bound state $n$, formed by the fusion of the particles $i$ and $j$, can be derived by using the new bound state ZF elements (\ref{bsZF}): in a simple diagonal case it is given by the following
product of fundamental diagonal S-matrices:
\begin{equation}
S_{kn}(\theta)=S_{ki}(\theta-\mathrm{i}\bar u_{i\bar n}^{\bar j})S_{kj}(\theta+\mathrm{i}\bar u_{j\bar n}^{\bar i})\,.
\label{bootstrap}
\end{equation}
In a non-diagonal case, the S-matrix is projected onto the bound states channel by the vertex functions defined by (\ref{res}) (see figure \ref{fig:bootstrap}):
\begin{equation}
\Gamma_{ij}^nS_{kn}^{ln'}(\theta)=S_{ki}^{k'i'}(\theta-\mathrm{i}\bar u_{i\bar n}^{\bar j})S_{k'j}^{lj'}(\theta+\mathrm{i}\bar u_{j\bar n}^{\bar i})\Gamma_{i'j'}^{n'}\,,
\label{Sbp}
\end{equation}
where the repeated indices are summed over $1,\dots,N$, with $N$ being the dimension of the symmetry algebra.
In this way we can take into account the possibility to have non-diagonal scattering between fundamental particles and bound states. This is the case of the $SU(3)$ chiral Gross-Neveu model, for instance, that will be discussed in section \ref{CGN}. However, usually bound states and fundamental particles have different masses and then they scatter diagonally. This means that $k=l$ and $n=n'$ in (\ref{Sbp}), which reduces to 
\begin{equation}
\Gamma_{ij}^nS_{kn}(\theta)=S_{ki}^{k'i'}(\theta-\mathrm{i}\bar u_{i\bar n}^{\bar j})S_{k'j}^{kj'}(\theta+\mathrm{i}\bar u_{j\bar n}^{\bar i})\Gamma_{i'j'}^{n}\,.
\label{Sbpdiag}
\end{equation}
Furthermore, the bound state-bound state S-matrix can be calculated by 
\begin{equation}
\Gamma_{ij}^{n}S_{mn}^{m'n'}(\theta)=S_{mi}^{lj'}(\theta-\mathrm{i}\bar u_{i\bar n}^{\bar j})S_{lj}^{m'j'}(\theta+\mathrm{i}\bar u_{j\bar n}^{\bar i})\Gamma_{n'}^{i'j'}\,,
\label{Sbbgen}
\end{equation}
namely by replacing the incoming (outgoing) particle $k$ ($l$) in figure \ref{fig:bootstrap} by the incoming (outgoing) bound state $m$ ($m'$). In this way it is possible to compute all the S-matrices for all the bound-states of the theory. We will see in sections \ref{SGpoles} and \ref{CGNpoles} some concrete use of these equations to derive the corresponding bound states S-matrices.

In terms of the ZF algebra elements, we can rewrite (\ref{bsZF}) in a more formal way as
\begin{equation}
A_{i}(\theta_{1})A_{j}(\theta_{2})=\left.\sum_{n}N_{ij}^{n}B_{n}\left(\frac{\theta_{1}+\theta_{2}}{2}\right)\right|_{\theta_{1}-\theta_{2}=\mathrm{i}\bar u_{ij}^{n}}\,,
\label{fusion}
\end{equation}
where $N_{ij}^{n}$ is 1 if $B_{n}$ is a bound state of $A_{i}$ and $A_{j}$ and 0 otherwise.
The \emph{fusion rules} (\ref{fusion}) must be consistent with the symmetries: $N_{ij}^{n}\neq0$ only if charges $C_i$ satisfy $C_{n}=C_{i}+C_{j}$.
Then the bootstrap entails constraints on the charges.
For example, given some charges eigenvalues with spin $s$ $\omega_{n}^{s}(\theta)=\gamma_{n}^{s}e^{s\theta}$, then these have to satisfy the following \emph{consistency bootstrap equations}
\begin{equation}
\gamma_{n}^{s}=\gamma_{i}^{s}e^{\mathrm{i}\bar u_{i\bar n}^{\bar j}}+\gamma_{j}^{s}e^{-\mathrm{i}\bar u_{j\bar n}^{\bar i}}\,.
\end{equation}

\section{Hopf algebra interpretation}
\label{Hopf}

The Hopf algebras (see part of \cite{Florian} and \cite{Spill:2007bia}
as introductory reviews on this subject) can be an useful tool for writing in a full algebraic way the symmetries of an S-matrix and to determine completely the S-matrix itself.
The basic idea is to add to generic algebras some structures allowing the rigorous definition of operations over tensor products of 
representations, necessary to define multi-particle
states with additive quantum numbers.

Let us consider, as an example, the universal enveloping of a Lie algebra. It is the tensor algebra $T(g)$ of a Lie algebra $g$: $\oplus_{n=0}^{\infty}g^{\otimes n}$. It has a multiplication corresponding to the tensor product
\begin{equation}
 (a_1\otimes\dots\otimes a_n)(b_1\otimes\dots\otimes b_m)=a_1\otimes\dots\otimes a_n\otimes b_1\otimes\dots\otimes b_m\,.
\end{equation}
Then the quotient algebra $U(g)=T(g)/\bf{I}$, where $\bf{I}$ is the ideal generated by elements of the form $AB-BA-[A,B]$, with $A,B\in g$, is a Hopf algebra if a coproduct $\Delta$, a counit $\epsilon$ and an antipode $\Sigma$ are defined 
(see \cite{Florian}). In particular, in this case they are explicitly given, $\forall J\in g$, by
\begin{eqnarray}
 &&\Delta(J)=J\otimes\mathbb{I}+\mathbb{I}\otimes J\,;\ \ \Delta(\mathbb{I})=\mathbb{I}\otimes\mathbb{I}\,;
 \label{coproduct}\\
 &&\epsilon(J)=0\,;\ \ \epsilon(\mathbb{I})=\mathbb{I}\,;\\
 && \Sigma(J)=-J\,;\ \ \Sigma(\mathbb{I})=\mathbb{I}\,.
\end{eqnarray}

So, for example, if applied to the spin operator $S_z$ in a space of two-particle
states classified by the spin eigenvalues $s_1$ and $s_2$, the coproduct gives
\begin{equation}
 \Delta S_z|s_1s_2\rangle=(S_z\otimes\mathbb{I}+\mathbb{I}\otimes S_z)|s_1s_2\rangle=(s_1+s_2)|s_1s_2\rangle\,,
\end{equation}
that is exactly what one expects from the action of a Lie algebra generator on a tensor product state.
In order to generalize the action of Lie algebras on higher tensor products, higher coproducts can be defined as follows
\begin{eqnarray}
\Delta^{(2)}(J)&=&(\mathbb{I}\otimes\Delta)\Delta(J)=(\Delta\otimes\mathbb{I})\Delta(J)=\mathbb{I}\otimes\mathbb{I}\otimes J+\mathbb{I}\otimes J\otimes\mathbb{I}+J\otimes\mathbb{I}
\otimes\mathbb{I}\,,\\
\Delta^{(n)}&=&(\mathbb{I}\otimes\mathbb{I}\otimes\dots\otimes\Delta)\Delta^{(n-1)}\,,
\end{eqnarray}
still giving the desired action of the algebra as a sum of the actions on the single states involved in the tensor product state.

As we have already seen in section \ref{gensol}, when we act with a symmetry generator $J$ on a two-particle state that belongs to a tensor product of two representations, we compute:
\begin{equation}
(J\otimes\mathbb{I}+\mathbb{I}\otimes J)|p_1,p_2\rangle\,.
\end{equation}
Thus, the condition (\ref{syminv}) for the compatibility of the S-matrix with a given symmetry algebra can be rewritten as
\begin{equation}
\left[\Delta(J),S\right]=0\,.
\end{equation}
Moreover, if we equip the symmetry algebra with an antipode $\Sigma$, the antiparticle representation can be derived by
\begin{equation}
\pi\left[\Sigma(J)\right]=\mathcal{C}^{-1}\bar\pi(J)^{t} \mathcal{C}\,,
\label{antipode}
\end{equation}
where $\mathcal{C}$ is the charge-conjugation matrix, $\pi$ denotes the matrix representation and the superscript $t$ means transposition.

Now, let us consider a \emph{quasi cocommutative} Hopf algebra $\mathcal{A}$ (see \cite{Florian} for the particular case of Yangians). By definition, this is equipped with an invertible element $\mathcal{R}$ belonging to $\mathcal{A}\otimes\mathcal{A}$
such that 
\begin{equation}
\Delta^{op}(a)=\mathcal{R}\Delta(a)\mathcal{R}^{-1}\,;\ \ \forall a\in \mathcal{A}\,,
\label{Deltaop}
\end{equation}
where $\Delta^{op}=P\Delta$, $P$ is the permutation operator and $\mathcal{R}$ can be written as the sum $\mathcal{R}=\sum_{i,j}r_{i}\otimes r_{j}$, with $r_{i}\in\mathcal{A}$.
Let us recall the properties satisfied by $\mathcal{R}$: in particular, if we define
\begin{equation}
 \mathcal{R}_{12}=\mathcal{R}\otimes\mathbb{I}\,;\ \mathcal{R}_{23}=\mathbb{I}\otimes \mathcal{R}\,;\ \mathcal{R}_{13}=\sum_{ij}r_i\otimes\mathbb{I}\otimes r_j\,,
 \label{defR}
 \end{equation}
a quasi commutative Hopf algebra is called \emph{quasi triangular} if
\begin{eqnarray}
 &&(\Delta\otimes\mathbb{I})\mathcal{R}=\mathcal{R}_{13}\mathcal{R}_{23}\,,
 \label{qtHa1}\\
 &&(\mathbb{I}\otimes\Delta)\mathcal{R}=\mathcal{R}_{13}\mathcal{R}_{12}\,,
 \label{qtHa2}
\end{eqnarray}
and $\mathcal{R}$ is called the \emph{universal R-matrix}.

It can be shown \cite{Drinfeld:1986in} that the universal R-matrix of a quasi triangular Hopf algebra satisfies
\begin{eqnarray}
 &&\mathcal{R}_{12}\mathcal{R}_{13}\mathcal{R}_{23}=\mathcal{R}_{23}\mathcal{R}_{13}\mathcal{R}_{12}\,,
 \label{qtHath1}\\
 &&(\Sigma\otimes\mathbb{I})\mathcal{R}=(\mathbb{I}\otimes \Sigma^{-1})\mathcal{R}=\mathcal{R}^{-1}\,.
 \label{qtHath2}
\end{eqnarray}
Relation (\ref{qtHath1}) is obtained by comparing the expression of $(\mathbb{I}\otimes\Delta^{op})\mathcal{R}$ written as
\begin{equation}
(\mathbb{I}\otimes P\Delta)\mathcal{R}=(\mathbb{I}\otimes P)(\mathbb{I}\otimes\Delta)\mathcal{R}=(\mathbb{I}\otimes P)\mathcal{R}_{13}\mathcal{R}_{12}=\mathcal{R}_{12}\mathcal{R}_{13}\,,
\label{1xDeltaR1}
\end{equation}
where we used (\ref{qtHa2}), and 
\begin{equation}
(\mathbb{I}\otimes\Delta^{op})\mathcal{R}=(\mathbb{I}\otimes\mathcal{R}\Delta\mathcal{R}^{-1})\mathcal{R}=(\mathbb{I}\otimes\mathcal{R})(\mathbb{I}\otimes\Delta)\mathcal{R}(\mathbb{I}\otimes\mathcal{R}^{-1})=\mathcal{R}_{23}\mathcal{R}_{13}\mathcal{R}_{12}\mathcal{R}_{23}^{-1}\,,
\label{1xDeltaR2}
\end{equation}
where definitions (\ref{Deltaop}) and (\ref{defR}) have been used. Thus, the comparison of (\ref{1xDeltaR2}) with (\ref{1xDeltaR1}) gives (\ref{qtHath1}). For a demonstration of (\ref{qtHath2}), the interested reader can look at section 2.2.1 of \cite{Spill:2012qe}, for instance. 

A spectral parameter $\theta$ can be introduced by an automorphism $D_{\theta}$ of the Hopf algebra $\mathcal{A}$, such that $D_{\theta}D_{\theta'}=D_{\theta+\theta'}$, $D_{0}=1$ and
\begin{equation}
(\mathbb{I}\otimes D_{\theta})\mathcal{R}=(D_{-\theta}\otimes\mathbb{I})\mathcal{R}=\mathcal{R}(\theta)\,.
\end{equation}
Then (\ref{qtHath1}) becomes
\begin{equation}
\mathcal{R}_{12}(\theta)\mathcal{R}_{13}(\theta+\theta')\mathcal{R}_{23}(\theta')=\mathcal{R}_{23}(\theta')\mathcal{R}_{13}(\theta+\theta')\mathcal{R}_{12}(\theta)\,,
\end{equation}
and its matrix representation, with the identification  $S=P \mathcal{R}$, gives the YBE (\ref{YBE}). 

It can be also shown that properties (\ref{qtHath2}) and (\ref{qtHa1})-(\ref{qtHa2}) are respectively equivalent to
the \emph{crossing symmetry} \cite{Delius:1995he} and the \emph{bootstrap equations} (\ref{bootstrap}) for the S-matrix \cite{MacKay}.
Therefore, this algebraic formulation, alternative to the one mentioned in Section \ref{ZF}, can be useful to introduce the concept of crossing symmetry in non-relativistic theories, as done in \cite{Janik:2006dc} for 
the $AdS_5/CFT_4$ case, for instance.

\section{Form factors}
\label{formfactors}

The knowledge of the two-particle S-matrix in an integrable theory is a fundamental step towards the determination of its correlation functions, that are necessary to calculate the physical quantities of the model.

Indeed, an essential ingredient for the full solution of a (1+1)-dimensional integrable theory is the determination of its generalized form factors\footnote{Though the form factor program succeeded in calculating exactly the correlation functions in few cases, such as the Ising model \cite{Wu:1975mw} and the principal chiral model at large $N$ \cite{Orland:2014mya}.}, that are the matrix elements of local operators evaluated 
between \emph{out} and \emph{in} asymptotic states:
\begin{equation}
_{b_{1}\dots b_{m}}^{~~~out}\hspace{-0.1cm}\langle \theta_{1}'\dots \theta_{m}'|\mathcal{O}(x)|\theta_{m+1}\dots \theta_{n}\rangle^{in}_{a_{m+1}\dots a_{n}}\,.
\label{matrel}
\end{equation}
We will see how these are deeply related to the S-matrix and the bootstrap program discussed in the previous sections.

The correlation functions can be related to a special class of generalized form factors by inserting\footnote{In what follows we will collect the color labels $a_{1},\dots,a_{n}$ in the notation $\underline{a}$.}
\begin{equation}
1=\sum_{n=0}^{\infty}\int\frac{d\theta_{1}\dots d\theta_{n}}{n!(2\pi)^{n}}
|\theta_{1},\dots,\theta_{n}\rangle^{in}_{\underline{a}}\,_{~\underline{a}}^{in}\langle\theta_{1},\dots,\theta_{n}|
\end{equation}
into a two-point function
\begin{equation}
\langle\mathcal{O}(x)\mathcal{O}(0)\rangle=\sum_{n=0}^{\infty}\int\frac{d\theta_{1}\dots d\theta_{n}}{n!(2\pi)^{n}}\langle0|\mathcal{O}(x)|\theta_{1},\dots,\theta_{n}\rangle^{in}_{\underline{a}}\,
_{\ \underline{a}}^{in}\langle\theta_{1},\dots,\theta_{n}|\mathcal{O}(0)|0\rangle\,.
\end{equation}
We see indeed that this involves the actual \emph{form factor}
\begin{equation}
F^{\mathcal{O}}_{\underline{a}}(\theta_{1},\dots,\theta_{n})=\langle 0|\mathcal{O}(0)|\theta_{1},\dots,\theta_{n}\rangle^{in}_{\underline{a}}\,,
\end{equation}
that is indeed defined as the matrix element of a local operator placed at the origin, between an $n$-particle state and the vacuum.

\begin{figure}
\begin{centering}
\includegraphics[width=0.75\textwidth]{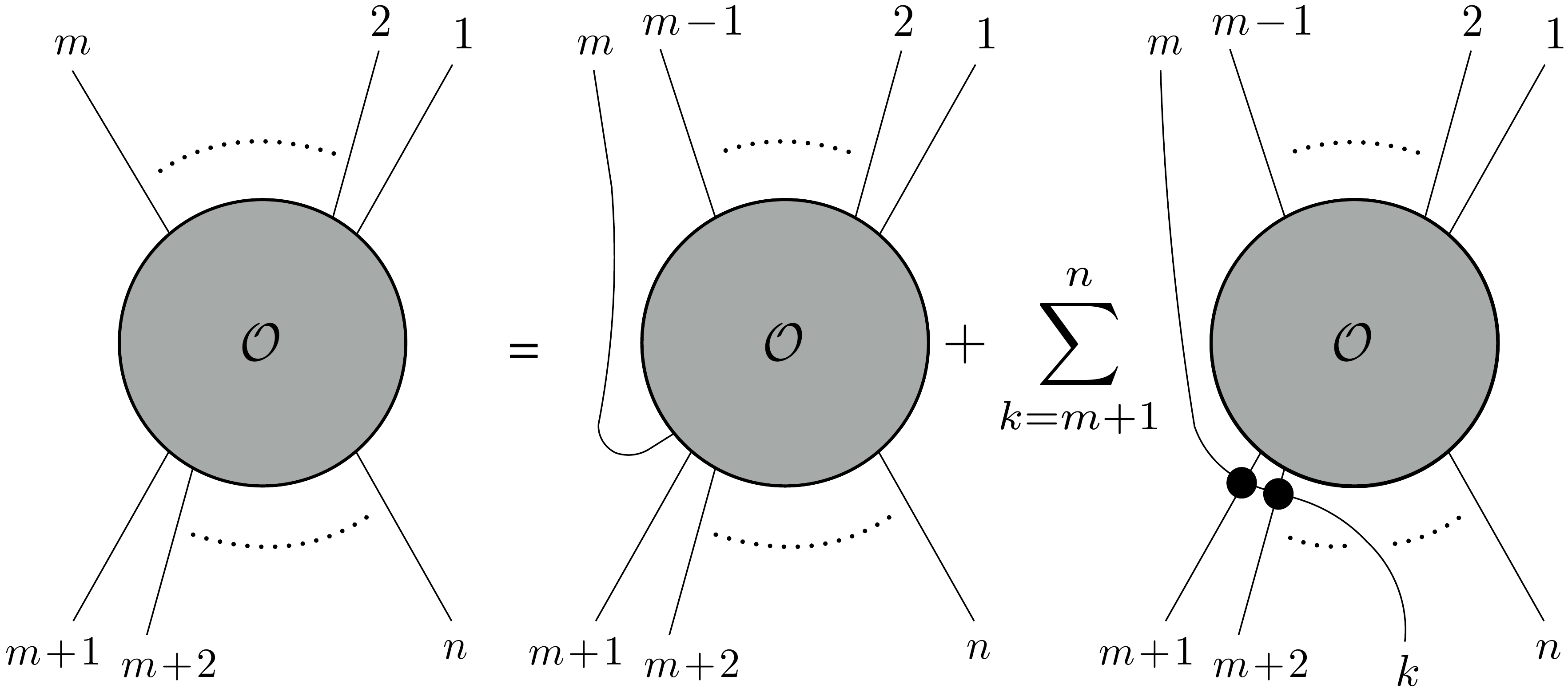}
\par\end{centering}
\caption{Crossing relation for the form factors.}
\label{fig:crossform}
\end{figure}

As for the S-matrix, let us discuss the properties satisfied by the form factors $F^{\mathcal{O}}_{\underline{a}}(\underline{\theta})$. From the constraints given by these properties we will get fundamental hints to find their general solutions. 

First, in the case of local scalar operators $\mathcal{O}(x)$, \emph{relativistic invariance} implies that the form factors are functions of the rapidities differences $\theta_{ij}=\theta_{i}-\theta_{j}$:
\begin{equation}
F_{\underline{a}}^{\mathcal{O}}(\theta_{1},\dots,\theta_{n})=F_{\underline{a}}^{\mathcal{O}}(\theta_{12},\theta_{13},\dots,\theta_{ij},\dots,\theta_{n-1n})\ ;\ \ i<j\,.
\end{equation}
For operators of generic spin $s$, we have instead
\begin{equation}
F_{\underline{a}}^{\mathcal{O}}(\theta_{1}+\Lambda,\dots,\theta_{n}+\Lambda)=e^{s\Lambda}F_{\underline{a}}^{\mathcal{O}}(\theta_{1},\dots,\theta_{n})\,.
\end{equation}
In what follows we will focus on the case of scalar operators.

It is possible to show that \emph{CPT invariance} implies, under replacement of $in$ by $out$ states, the following simple relation 
\begin{equation}
\langle 0|\mathcal{O}(0)|\theta_{1},\dots,\theta_{n}\rangle^{out}_{\underline{a}}=F_{\underline{a}}^{\mathcal{O}}(-\theta_{ij})\ ;\ \ 1\le i<j\le n\,.
\label{CPT}
\end{equation}
The general property satisfied when a particle is moved from the $out$ to the $in$ state, instead, takes the name of \emph{crossing}. It is depicted in figure \ref{fig:crossform} and is formalized by the following relation\footnote{All the formulas of this section, for simplicity, are written for a \emph{diagonal} case with neutral particles.} (see \cite{Pozsgay} for instance):
\begin{eqnarray}
\hspace{-1cm}&&^{~~~out}_{i_{1}\dots i_{m}}\hspace{-0.05cm}\langle \theta'_{1},\dots,\theta'_{m}|\mathcal{O}(0)|\theta_{m+1},\dots,\theta_{n}\rangle^{in}_{j_{m+1}\dots j_{n}}\equiv 
F_{i_{1}\dots i_{m};j_{m+1}\dots j_{n}}^{\mathcal{O}}(\theta'_{1},\dots,\theta'_m|\theta_{m+1},\dots,\theta_{n})\nonumber\\
\hspace{-1cm}&&=F_{i_{1}\dots i_{m-1};i_{m}j_{m+1}\dots j_{n}}^{\mathcal{O}}(\theta'_{1},\dots,\theta'_{m-1}|\theta'_m+\mathrm{i}\pi,\theta_{m+1},\dots,\theta_{n})
+\sum_{k=m+1}^n\delta_{i_mj_{k}}\delta(\theta'_m-\theta_{k})
\label{crossform}\\
\hspace{-1cm}&&\times \prod_{l=1}^{k-1}S_{j_{l}j_{k}}(\theta_{l}-\theta_{k})F_{i_{1}\dots i_{m-1};j_{m+1}\dots j_{k-1}j_{k+1}\dots j_n}
(\theta'_{1},\dots,\theta'_{m-1}|\theta_{m+1},\dots,\theta_{k-1},\theta_{k+1},\dots,\theta_n)\,.\nonumber
\end{eqnarray}

\begin{figure}
\begin{centering}
\includegraphics[width=0.45\textwidth]{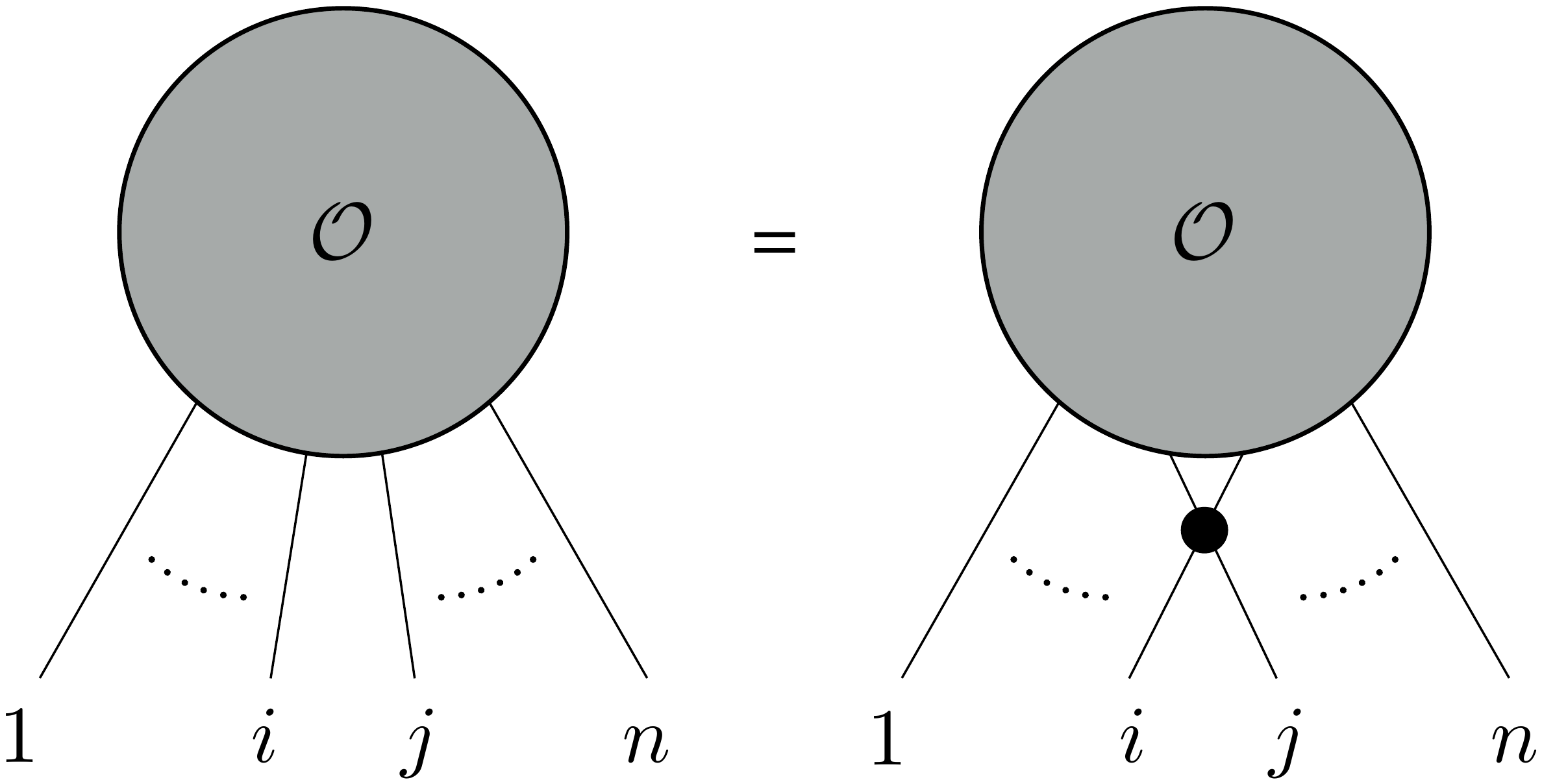}
\par\end{centering}
\caption{Watson equation for permutation of two particles.}
\label{fig:watson1}
\end{figure}

\begin{figure}
\begin{centering}
\includegraphics[width=0.65\textwidth]{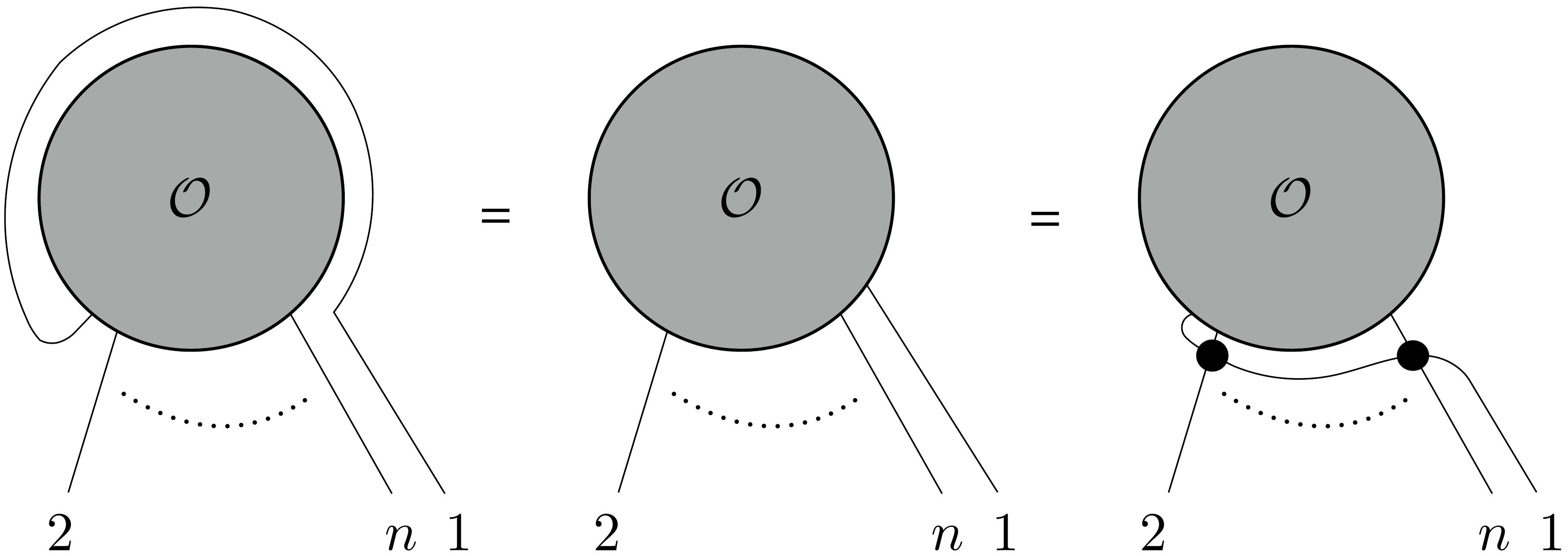}
\par\end{centering}
\caption{Watson equation for periodicity under shifts of $2\pi \mathrm{i}$.}
\label{fig:watson2}
\end{figure}

For example, in the two-particle case, this property reads
\begin{equation}
^{out}_{~a_{1}}\langle\theta_{1}|\mathcal{O}(0)|\theta_{2}\rangle^{in}_{a_{2}}=F_{a_{1}a_{2}}^{\mathcal{O}}(\theta_{12}+\mathrm{i}\pi)+\delta_{a_1a_2}\delta(\theta_{12})\langle\mathcal{O}\rangle\,.
\label{norm}
\end{equation}
Here we just stated formulas (\ref{crossform}) and (\ref{CPT}) without proof, however it is possible to derive them on the basis of the the Lehmann-Symanzik-Zimmermann (LSZ) reduction formalism \cite{Lehmann:1954rq} and the maximal analyticity assumption, $i.e.$ possible singularities of the form factors
can occur only due to physical processes like the formation of bound states, similarly to the analyticity property assumed for the S-matrix. The related derivations 
can be found in Appendix A of \cite{Babujian:1998uw}, for instance.

The symmetry properties satisfied under permutations of $\theta_{i},\theta_{j}$ and shifts by $2\pi \mathrm{i}$, represented in figures \ref{fig:watson1} and \ref{fig:watson2} respectively, are called
\emph{Watson equations} after \cite{Watson}, and in a diagonal case they read
\begin{eqnarray}
&&\hspace{-2cm}F_{a_{1}\dots a_{i}a_{j}\dots a_{n}}^{\mathcal{O}}(\theta_{1},\dots,\theta_{i},\theta_{j},\dots,\theta_{n})\nonumber\\
&&=F_{a_{1}\dots a_{j}a_{i}\dots a_{n}}^{\mathcal{O}}(\theta_{1},\dots,\theta_{j},\theta_{i},\dots,\theta_{n})S_{a_{i}a_{j}}(\theta_{ij})\ ;\ \ j=i+1\,,\label{watson1}\\
&&\hspace{-2cm}F_{\underline{a}}^{\mathcal{O}}(\theta_{1}+2\pi \mathrm{i},\dots,\theta_{n})=F_{a_2\dots a_na_1}^{\mathcal{O}}(\theta_2,\dots,\theta_n,\theta_1)
=\prod_{i=2}^{n}S_{a_{i}a_{1}}(\theta_{i}-\theta_{1})F_{\underline{a}}^{\mathcal{O}}(\theta_{1},\dots,\theta_{n})\,.
\label{watson2}
\end{eqnarray}
They can be derived, in the case $n=2$ for example, by using the definition of the S-matrix, factorization and CPT invariance:
\begin{eqnarray}
F_{a_1a_2}^{\mathcal{O}}(\theta_{12})&=&\langle0|\mathcal{O}(0)|\theta_{1}\theta_{2}\rangle^{in}_{a_1a_2}=\langle0|\mathcal{O}(0)|\theta_{1}\theta_{2}\rangle^{out}_{a_1a_2}S_{a_1a_2}(\theta_{12})\nonumber\\
&=&F_{a_2a_1}^{\mathcal{O}}(-\theta_{12})S_{a_1a_2}(\theta_{12})\,,\\
F_{a_2a_1}^{\mathcal{O}}(\mathrm{i}\pi-\theta_{12})&=&_{~~a_1}^{out}\langle\theta_{1}|\mathcal{O}(0)|\theta_{2}\rangle^{in}_{a_2}=~_{~~a_1}^{~in}\langle \theta_{1}|\mathcal{O}(0)|\theta_{2}\rangle^{out}_{a_2}\nonumber\\
&=&F^{\mathcal{O}}_{a_1a_2}(\mathrm{i}\pi+\theta_{12})\,,
\end{eqnarray}
where the next-to-last identity is due to the triviality of the one-particle S-matrix.

As in the case of the S-matrix, we look for general solutions of the Watson equations and the other conditions listed above in the form of a minimal solution $F_{\underline{a}}^{min}(\theta_{ij})$, without poles and zeros in the physical strip $0\le \mbox{Im}(\theta_{ij})\le\pi$, multiplied by a factor $K_{\underline{a}}(\theta_{ij})$ containing all the information about the poles (zeros) structure. For scalar operators, this reads
\begin{equation}
F^{\mathcal{O}}_{\underline{a}}(\theta_{1},\dots,\theta_{n})=K^{\mathcal{O}}_{\underline{a}}(\theta_{1},\dots,\theta_{n})\prod_{i<j}F^{min}_{a_{i}a_{j}}(\theta_{ij})\,.
\end{equation}
In the case of $n=2$, we are saying that the most general solution of the Watson equations \cite{Karowski:1978vz}
\begin{equation}
F_{a_{1}a_{2}}^{\mathcal{O}}(\theta_{12})=F_{a_{2}a_{1}}^{\mathcal{O}}(-\theta_{12})S_{a_{1}a_{2}}(\theta_{12})\,,\ \ F_{a_{1}a_{2}}^{\mathcal{O}}(\mathrm{i}\pi-\theta_{12})=F_{a_{2}a_{1}}^{\mathcal{O}}(\mathrm{i}\pi+\theta_{12})\,,
\end{equation}
is given by $F_{a_{1}a_{2}}^{\mathcal{O}}(\theta)=K_{a_{1}a_{2}}^{\mathcal{O}}(\theta)F_{a_{1}a_{2}}^{min}(\theta)$, with $K_{a_{1}a_{2}}^{\mathcal{O}}(\theta)$ satisfying
\begin{equation}
K_{a_{1}a_{2}}^{\mathcal{O}}(\theta)=K_{a_{2}a_{1}}^{\mathcal{O}}(-\theta)=K_{a_{1}a_{2}}^{\mathcal{O}}(2\pi \mathrm{i}+\theta)\,.
\end{equation}
If $\pm i\alpha_{1},\dots,\pm \mathrm{i}\alpha_{L}$ are poles of $F_{a_{1}a_{2}}^{\mathcal{O}}(\theta)$ in the physical strip, then
\begin{equation}
K^{\mathcal{O}}_{a_{1}a_{2}}(\theta)=\mathcal{N}^{\mathcal{O}}(\theta)\prod_{k=1}^{L}\frac{1}{\sinh\frac{\theta-\mathrm{i}\alpha_{k}}{2}\sinh\frac{\theta+\mathrm{i}\alpha_{k}}{2}}\,.
\label{K}
\end{equation}
For scalar operators, as we will see in the examples of section \ref{examples}, the normalization factor $\mathcal{N}^{\mathcal{O}}(\theta)$ is a constant and the poles of $K^{\mathcal{O}}(\theta)$ contain all the information about the operator $\mathcal{O}$. If in addition $\langle\mathcal{O}\rangle=0$, $\mathcal{N}^{\mathcal{O}}$ can be fixed using relation (\ref{norm}):
\begin{equation}
_{a}\langle\theta|\mathcal{O}|\theta\rangle_{a}=F_{aa}^{\mathcal{O}}(\mathrm{i}\pi)\,.
\end{equation}
On the other hand, Cauchy theorem implies that, given a contour $C$ enclosing the strip $0\le\mbox{Im}(\theta)\le2\pi$, $F_{a_{1}a_{2}}^{min}(\theta)$ satisfies
\begin{eqnarray}
&&\frac{d}{d\theta}\log F_{a_{1}a_{2}}^{min}(\theta)=\frac{1}{8\pi \mathrm{i}}\int_{C}\frac{dz}{\sinh^{2}\frac{z-\theta}{2}}\log F_{a_{1}a_{2}}^{min}(z)
\label{cauchy}\\
&&=\frac{1}{8\pi \mathrm{i}}\int_{-\infty}^{+\infty}\frac{dz}{\sinh^{2}\frac{z-\theta}{2}}\log \frac{F_{a_{1}a_{2}}^{min}(z)}{F_{a_{1}a_{2}}^{min}(z+2\pi \mathrm{i})}=\frac{1}{8\pi \mathrm{i}}\int_{-\infty}^{+\infty}\frac{dz}{\sinh^{2}\frac{z-\theta}{2}}\log S_{a_{1}a_{2}}(z)\,,\nonumber
\end{eqnarray}
where we used the property (\ref{watson2}) in the last equality. Then we can calculate the minimal solution $F_{a_{1}a_{2}}^{min}(\theta)$ entirely from the S-matrix element $S_{a_{1}a_{2}}(\theta)$.

Regarding the factor $K^{\mathcal{O}}(\theta_{12})$, it has to satisfy the Watson equations with $S(\theta_{12})=1$, then it is symmetric in $\theta_{12}$ and periodic with period $2\pi \mathrm{i}$, {\it i.e.} it is function of $\cosh\theta_{12}$.

\begin{figure}
\begin{centering}
\includegraphics[width=0.8\textwidth]{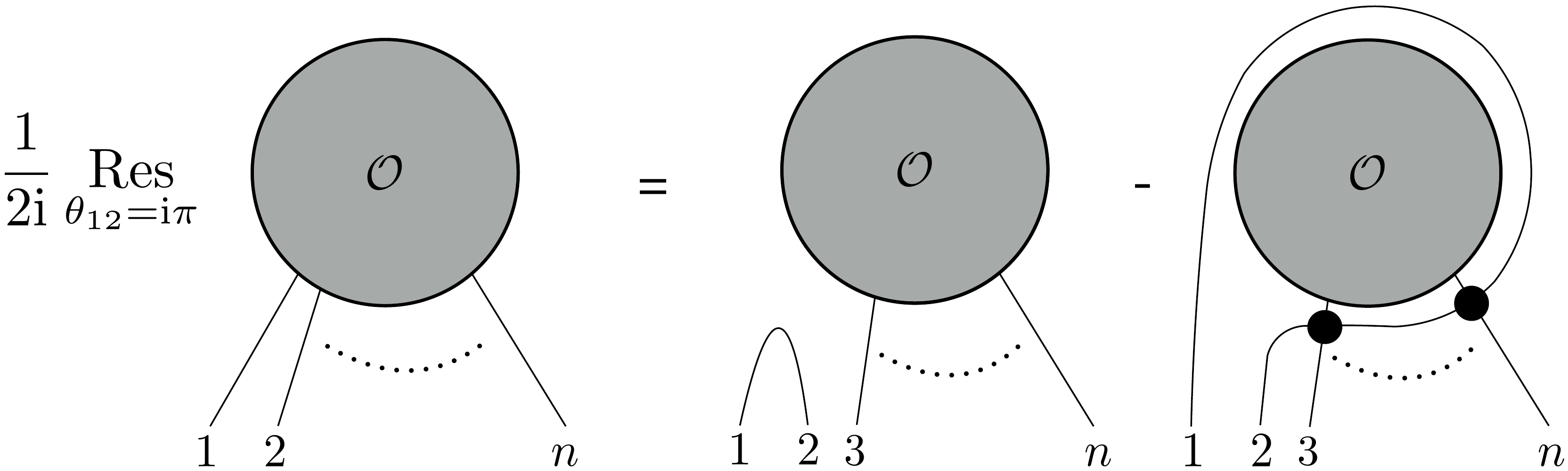}
\par\end{centering}
\caption{Recursive relation from the residue at $\theta_{12}=\mathrm{i}\pi$.}
\label{fig:res1}
\end{figure}

In general, $n$-particle functions $K_{\underline{a}}^{\mathcal{O}}(\theta_{1},\dots,\theta_{n})$ have poles when a cluster of $k$ particles have the kinematic configuration of a one-particle state. In particular, this happens when the set of $in$ particles contains a particle-antiparticle pair with opposite momenta, {\it e.g.} $\theta_{12}=\mathrm{i}\pi$ (see figure \ref{fig:res1}):
\begin{equation}
\mathop{\mathrm{Res}}_{\theta_{12}=\mathrm{i}\pi}F_{\underline{a}}^{\mathcal{O}}(\theta_{1},\dots,\theta_{i},\theta_{j},\dots,\theta_{n})=2\mathrm{i}F_{a_{3}\dots a_{n}}^{\mathcal{O}}(\theta_{3},\dots,\theta_{n})\left[1-\prod_{i=3}^{n}S_{a_{2}a_{i}}(\theta_{2}-\theta_{i})\right]\,,
\label{res1}
\end{equation}
that gives a recursive relation between $n$- and $n-2$-particle form factors.
Property (\ref{res1}) follows from realizing that in (\ref{crossform}) the particle $m$ can be also moved to the end of the $in$ particles set:
\begin{eqnarray}
\hspace{-1cm}&&F_{i_{1}\dots i_{m};j_{m+1}\dots j_{n}}^{\mathcal{O}}(\theta'_{1},\dots,\theta'_m|\theta_{m+1},\dots,\theta_{n})\label{crossform2}\\
\hspace{-1cm}&&=F_{i_{1}\dots i_{m-1};j_{m+1}\dots j_{n}i_{m}}^{\mathcal{O}}(\theta'_{1},\dots,\theta'_{m-1}|\theta_{m+1},\dots,\theta_{n},\theta'_{m}-\mathrm{i}\pi)+\sum_{k=m+1}^n\delta_{i_mj_{k}}\delta(\theta'_m-\theta_{k})\nonumber\\
\hspace{-1cm}&&\times\prod_{l=k+1}^{n}S_{j_{l}j_{k}}(\theta_{l}-\theta_{k})F_{i_{1}\dots i_{m-1};j_{m+1}\dots j_{k-1}j_{k+1}\dots j_n}^{\mathcal{O}}
(\theta'_{1},\dots,\theta'_{m-1}|\theta_{m+1},\dots,\theta_{k-1},\theta_{k+1},\dots,\theta_n)\,.\nonumber
\end{eqnarray}
Thus, comparing the analytic parts of the crossing relations (\ref{crossform}) and (\ref{crossform2}) we can obtain the first periodicity relation in (\ref{watson2}), and if we evaluate that at $\theta_{1}\sim\theta_{2}$ we get
\begin{eqnarray}
F_{a_{1}a_{2}\dots a_{n}}^{\mathcal{O}}(\theta_{1}+\mathrm{i}\pi,\theta_{2},\dots,\theta_{n})\sim\frac{f(\theta_{2},\dots,\theta_{n})}{\theta_{1}-\theta_{2}-\mathrm{i}\epsilon}\,,
\label{lim1}\\
F_{a_{2}\dots a_{n}a_{1}}^{\mathcal{O}}(\theta_{2},\dots,\theta_{n},\theta_{1}-\mathrm{i}\pi)\sim\frac{f(\theta_{2},\dots,\theta_{n})}{\theta_{1}-\theta_{2}+\mathrm{i}\epsilon}\,,
\label{lim2}
\end{eqnarray}
for some function $f$ and small $\epsilon$. Hence, plugging (\ref{lim1}) and (\ref{lim2}) into (\ref{crossform}) and (\ref{crossform2})  respectively, in the case $m=1$ and evaluated at $\theta_1\sim\theta_2$, and comparing the $\delta$-function parts, one obtains
\begin{equation}
f(\theta_{2},\dots,\theta_{n})=2\mathrm{i}F_{a_{3}\dots a_{n}}^{\mathcal{O}}(\theta_{3},\dots,\theta_{n})\left[1-\prod_{i=3}^{n}S_{a_{2}a_{i}}(\theta_{2}-\theta_{i})\right]\,,
\end{equation}
and then (\ref{res1}).

\begin{figure}
\begin{centering}
\includegraphics[width=9cm]{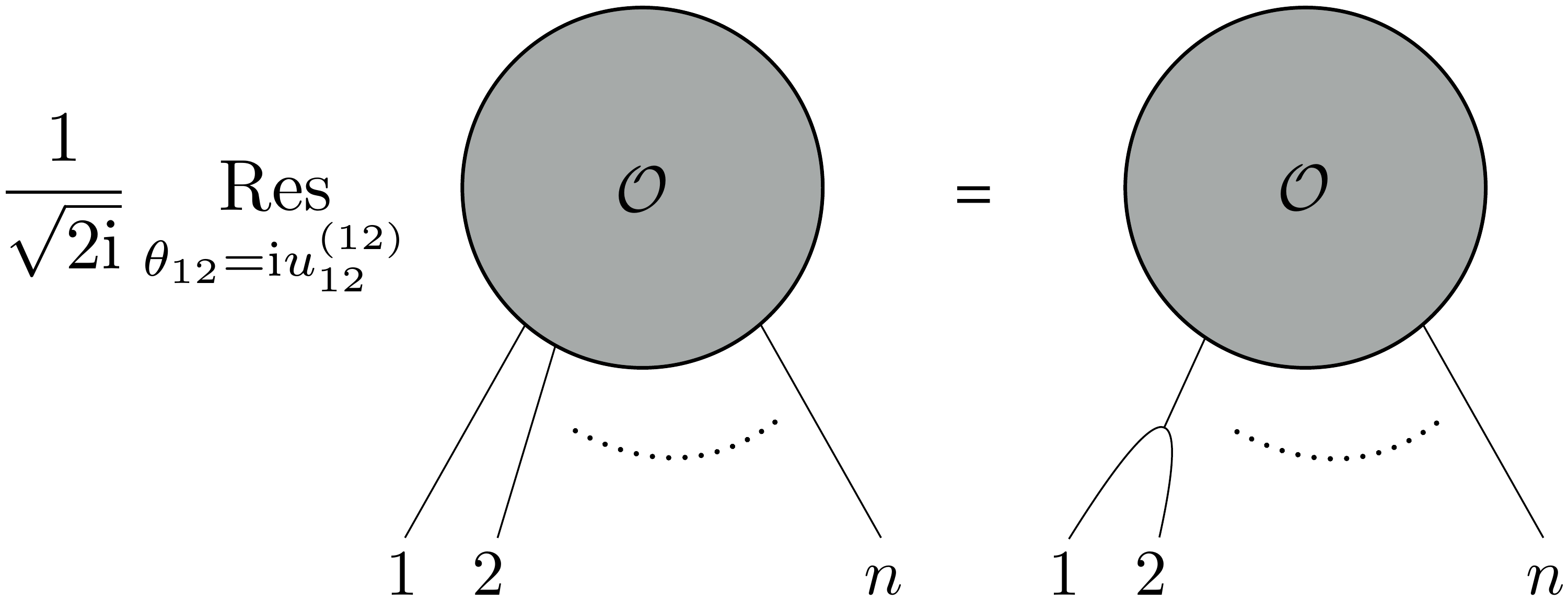}
\par\end{centering}
\caption{Recursive relation from the residue at $\theta_{12}=\mathrm{i}u_{12}^{(12)}$.}
\label{fig:res2}
\end{figure}

A further recursive relation, depicted in figure \ref{fig:res2}, connects $n$- and $n-1$-particle form factors if there is a bound state pole at $\theta_{12}=\mathrm{i}u_{12}^{(12)}$, for example\footnote{By $\mathrm{i}u_{12}^{(12)}$, we denote the position of the pole corresponding to the bound state $(12)$ made of particles 1 and 2.}: 
\begin{equation}
\mathop{\mathrm{Res}}_{\theta_{12}=\mathrm{i}u_{12}^{(12)}}F_{\underline{a}}^{\mathcal{O}}(\theta_{1},\dots,\theta_{n})=\sqrt{2\mathrm{i}R_{(12)}}\Gamma^{(12)}_{12}F_{a_{(12)}a_{3}\dots a_{n}}^{\mathcal{O}}(\theta_{(12)},\theta_{3}\dots,\theta_{n})\,,
\label{FFbs}
\end{equation}
where $\theta_{(12)}=(\theta_{1}+\theta_{2})/2$, $R_{(12)}$ is the residue of the S-matrix at $\theta=\mathrm{i}u_{12}^{(12)}$ and $\Gamma^{(12)}_{12}$ projects the spaces of particles $1$ and $2$ onto the space of the bound state
$(12)$, as defined in (\ref{res}). 

A derivation of (\ref{FFbs}), making use of two-point correlators, the Watson equation (\ref{watson1}) and the residue of the S-matrix (\ref{res}), can be found, as all the others discussed 
in this section, in Appendix A of \cite{Babujian:1998uw}.
A few examples of solutions in very simple cases are given in sections \ref{SGform} and \ref{CGNform}. The interested reader can look at \cite{Mussardobook, Karowski:1978vz, Babujian:2006km, Babujian:1998uw, Smirnovbook, Babujian:2009zg}
for further details.

\section{Examples}
\label{examples}

As promised, in this section we specialize the properties and results  of S-matrices and form factors, discussed above for generic (1+1)-dimensional integrable theories, to two relevant examples of quantum integrable 
relativistic models: sine-Gordon and chiral Gross-Neveu.
At the end of the section, we will also summarize recent developments about the S-matrices of $AdS/CFT$ correspondences.

\subsection{Sine-Gordon}

The quantum sine-Gordon model (see \cite{Alessandro} for the discussion of the classical theory) is a (1+1)-dimensional integrable\footnote{Its quantum integrability has been shown in \cite{Faddeev:1973aj, Kulish:1976xi}.} theory of a bosonic scalar field $\phi$, described by the following Lagrangian density:
\begin{equation}
\mathcal{L}_{sG}=\frac{1}{2}(\partial_{\mu}\phi)^{2}+\frac{m^{2}}{\beta^{2}}(\cos\beta\phi-1)\,,
\label{LsG}
\end{equation}
where $\mu=0,1$ and $\beta$ is a coupling constant. In what follows we will use a parameter $\xi$ given by
\begin{equation}
\xi=\frac{\beta^{2}}{8}\frac{1}{1-\frac{\beta^{2}}{8\pi}}\,.
\end{equation}
In particular, the coupling constant defines two distinct regions for $\beta^{2}<4\pi$ $(\xi<\pi)$ and $\beta^{2}>4\pi$ $(\xi>\pi)$, which are called respectively \emph{attractive} and \emph{repulsive} regimes. 
These names are due to the presence of bound state solutions in the attractive case and their absence in the repulsive one.
As we will see at the end of this section, the elementary excitation of the bosonic field $\phi$ corresponds to the bound state of a soliton and an antisoliton, which classically are solutions of the equation of motion associated to the 
Lagrangian (\ref{LsG}), reviewed in \cite{Alessandro}.
The quantized solitons are the fundamental excitations interacting through the S-matrix that we are going to study in the next section. As shown in \cite{Coleman:1974bu, Mandelstam:1975hb}, they can be put in correspondence 
with the self-interacting Dirac fermions described by a (1+1)-dimensional theory, called the massive Thirring model (MTM), defined by the following Lagrangian density
\begin{equation}
\mathcal{L}_{MTM}=\bar\psi(\mathrm{i}\gamma_{\mu}\partial^{\mu}-m)\psi-\frac{g}{2}\bar\psi\gamma^{\mu}\psi\bar\psi\gamma_{\mu}\psi\,,
\end{equation}
where $\gamma^{\mu}$ are the two-dimensional Dirac matrices and $g$ is a coupling constant related to the sine-Gordon $\beta$ as \cite{Coleman:1974bu}
\begin{equation}
\frac{4\pi}{\beta^{2}}=1+\frac{g}{\pi}\,.
\end{equation} 
In particular, at $\beta^2=4\pi$ ($\xi=\pi$) the theory describes a free fermion.

The sine-Gordon model possesses an $O(2)$ symmetry, and we will use this to constrain the matrix form of the S-matrix. In general, the $O(N)$ symmetry tells us that the spectrum of the fundamental excitations consists of a multiplet of $N$ particles of equal mass, denoted by $A_{i}\,, i=1,\dots,N$.
Moreover, the commutation relations corresponding to (\ref{ZFcr}) are constrained to be \cite{ZZ}
\begin{eqnarray}
&&A_{i}(\theta_{1})A_{j}(\theta_{2})=\delta_{ij}S_{1}(\theta_{1}-\theta_{2})\sum_{k=1}^{N}A_{k}(\theta_{2})A_{k}(\theta_{1})\\
&&+S_{2}(\theta_{1}-\theta_{2})A_{j}(\theta_{2})A_{i}(\theta_{1})+S_{3}(\theta_{1}-\theta_{2})A_{i}(\theta_{2})A_{j}(\theta_{1})\,.
\label{ONcr}
\end{eqnarray}

\subsubsection{Solution for the exact S-matrix}
\label{SGSmatrix}

The $O(2)$ symmetry group is the group of orthogonal matrices in two dimensions, and its Lie algebra is generated by
\begin{equation}
J=\left(
\begin{array}{cc}
0 & -1  \\
 1 &  0 
\end{array}
\right)\,.
\end{equation}
Now, we can equip this algebra with the operations and properties of a Hopf algebra and in particular we can impose the invariance of the S-matrix under $O(2)$ by using the coproduct (\ref{coproduct})
\begin{equation}
\left[\Delta(J),S\right]=0\,.
\label{O2inv}
\end{equation}
Solving the system of equations given by (\ref{O2inv}) and requiring parity and time reversal
invariances, one gets the following matrix structure:
\begin{equation}
S_{sG}=\left(
\begin{array}{cccc}
 S_1+S_2+S_3 &   &  & S_1\\
  & S_2  & S_3 & \\
  & S_3  & S_2 & \\
S_1  &   &  & S_1+S_2+S_3 
\end{array}
\right)\,.
\label{sGmatrix123}
\end{equation}
Written in terms of the ZF elements $A_1(\theta), A_2(\theta)$, this is equivalent to (\ref{ONcr}) for $N=2$.
Following \cite{ZZ}, we define the soliton and antisoliton ZF elements as
\begin{eqnarray}
&\mbox{soliton}&s(\theta)=A_{1}(\theta)+\mathrm{i}A_{2}(\theta)\,,\\
&\mbox{antisoliton}&\bar s(\theta)=A_{1}(\theta)-\mathrm{i}A_{2}(\theta)\,.
\label{sGex}
\end{eqnarray}
In this new basis, the ZF commutation relations become
\begin{eqnarray}
s(\theta_{1})\bar s(\theta_{2})&=&S_{T}(\theta_{1}-\theta_{2})\bar s(\theta_{2})s(\theta_{1})+S_{R}(\theta_{1}-\theta_{2})s(\theta_{2})\bar s(\theta_{1})\,,\\
s(\theta_{1})s(\theta_{2})&=&S(\theta_{1}-\theta_{2})s(\theta_{2})s(\theta_{1})\,,\\
\bar s(\theta_{1})\bar s(\theta_{2})&=&S(\theta_{1}-\theta_{2})\bar s(\theta_{2})\bar s(\theta_{1})\,,
\end{eqnarray}
where $S_{T}$ and $S_{R}$ denote the transmission and reflection amplitudes respectively, and in terms of $S_1,S_2$ and $S_3$ they read
\begin{eqnarray}
 S(\theta)&=& S_3(\theta)+S_2(\theta)\,,\\
 S_T(\theta)&=& S_1(\theta)+S_2(\theta)\,,\\
 S_R(\theta)&=& S_1(\theta)+S_3(\theta)\,.
\end{eqnarray}
Then the S-matrix  takes the form
\begin{equation}
S_{sG}=\left(
\begin{array}{cccc}
 S &   &  & \\
  & S_T  & S_R & \\
  & S_R  & S_T & \\
  &   &  & S 
\end{array}
\right)\,.
\label{sGmatrix}
\end{equation}
Imposing \emph{crossing symmetry} on this S-matrix and using the charge conjugation matrix $\mathcal{C}_{ij}=\delta_{\bar ij}$, one obtains
\begin{equation}
S(\theta)=S_{T}(\mathrm{i}\pi-\theta)\,,\ \ S_{R}(\theta)=S_{R}(\mathrm{i}\pi-\theta)\,,
\label{sGcross}
\end{equation}
while \emph{unitarity} entails
\begin{eqnarray}
&&S(\theta)S(-\theta)=1\,,
\label{sGuni1}\\
&&S_{T}(\theta)S_{T}(-\theta)+S_{R}(\theta)S_{R}(-\theta)=1\,,
\label{sGuni2}\\
&&S_{T}(\theta)S_{R}(-\theta)+S_{R}(\theta)S_{T}(-\theta)=0\,.
\label{sGuni3}
\end{eqnarray}

The YBE (\ref{YBEd}) fixes the ratio $S_{T}/S_{R}$, as mentioned in section \ref{gensol}.
In details, imposing the condition (\ref{YBEd}), one obtains
\begin{eqnarray}
&&S_{R}(\theta_{12})S_{R}(\theta_{13})S_{T}(\theta_{23})+S_{T}(\theta_{12})S(\theta_{13})S_{R}(\theta_{23})-S(\theta_{12})S_{T}(\theta_{13})S_{R}(\theta_{23})=0\,,~~~~\\
&&S_{R}(\theta_{12})S(\theta_{13})S_{R}(\theta_{23})+S_{T}(\theta_{12})S_{R}(\theta_{13})S_{T}(\theta_{23})-S(\theta_{12})S_{R}(\theta_{13})S(\theta_{23})=0\,.
\label{YBEsG}
\end{eqnarray}
These constraints were solved, in terms of the ratios  $S_{2}/S_{3}$ and $S_{1}/S_{3}$ of the elements appearing in (\ref{sGmatrix123}), in the Appendix A of \cite{ZZ}. In particular, those ratios were respectively given as solutions of differential equations obtained by differentiating the YBE (\ref{YBEsG}), with boundary conditions satisfying crossing (the second of (\ref{sGcross})) and unitarity (\ref{sGuni2})-(\ref{sGuni3}). 
Fulfilling all these constraints actually leaves a free parameter, which can be fixed to be proportional to $\xi$ by comparison with semi-classical results \cite{ZZ}.
Finally, $S_{R}$ and $S_{T}$ result to depend on $S$ in the following way
\begin{equation}
S_{T}(\theta)=\frac{\sinh\frac{\pi\theta}{\xi}}{\sinh\frac{\pi(\mathrm{i}\pi-\theta)}{\xi}}S(\theta)\,,\ \ S_{R}(\theta)=\frac{\mathrm{i}\sin\frac{\pi^{2}}{\xi}}{\sinh\frac{\pi(\mathrm{i}\pi-\theta)}{\xi}}S(\theta)\,.
\label{sGsol}
\end{equation}
Hence the crossing relation for $S(\theta)$ can be written as
\begin{equation}
S(\theta)=\frac{\sinh\frac{\pi(\mathrm{i}\pi-\theta)}{\xi}}{\sinh\frac{\pi\theta}{\xi}}S(\mathrm{i}\pi-\theta)\,.
\label{S0cross}
\end{equation}
Now, the first step to find a minimal solution of (\ref{sGuni1}) and (\ref{S0cross}) for $S(\theta)$ is to write (\ref{S0cross}) in terms of $\Gamma$ functions by using the property $\sinh\pi x=\pi\left[\Gamma(1+\mathrm{i}x)\Gamma(-\mathrm{i}x)\right]^{-1}$:
\begin{equation}
S(\theta)=\frac{\Gamma\left(1+\mathrm{i}\frac{\theta}{\xi}\right)\Gamma\left(-\mathrm{i}\frac{\theta}{\xi}\right)}{\Gamma\left(1-\frac{\pi}{\xi}-\mathrm{i}\frac{\theta}{\xi}\right)\Gamma\left(\frac{\pi}{\xi}+\mathrm{i}\frac{\theta}{\xi}\right)}S(\mathrm{i}\pi-\theta)\,.
\label{crossgamma}
\end{equation}
Then, taking an ansatz for $S(\theta)$ satisfying (\ref{crossgamma}) 
\begin{equation}
S(\theta)=\frac{\Gamma\left(1+\mathrm{i}\frac{\theta}{\xi}\right)}{\Gamma\left(\frac{\pi}{\xi}+\mathrm{i}\frac{\theta}{\xi}\right)}\,,
\end{equation}
we multiply it by a factor $f(\theta)$ such that the corrected $S(\theta)$ now satisfies unitarity (\ref{sGuni1})
\begin{equation}
f(\theta)f(-\theta)\frac{\Gamma\left(1+\mathrm{i}\frac{\theta}{\xi}\right)}{\Gamma\left(\frac{\pi}{\xi}+\mathrm{i}\frac{\theta}{\xi}\right)}\frac{\Gamma\left(1-\mathrm{i}\frac{\theta}{\xi}\right)}{\Gamma\left(\frac{\pi}{\xi}-\mathrm{i}\frac{\theta}{\xi}\right)}=1
\ \Rightarrow\ f(\theta)=\frac{\Gamma\left(\frac{\pi}{\xi}-\mathrm{i}\frac{\theta}{\xi}\right)}{\Gamma\left(1-\mathrm{i}\frac{\theta}{\xi}\right)}g(\theta)\,,
\end{equation}
with $g(\theta)$ such that $S(\theta)$ satisfies crossing again
 \begin{equation}
\frac{\Gamma\left(\frac{\pi}{\xi}-\mathrm{i}\frac{\theta}{\xi}\right)}{\Gamma\left(1-\mathrm{i}\frac{\theta}{\xi}\right)}g(\theta)=
\frac{\Gamma\left(\frac{2\pi}{\xi}+\mathrm{i}\frac{\theta}{\xi}\right)}{\Gamma\left(1+\frac{\pi}{\xi}+\mathrm{i}\frac{\theta}{\xi}\right)}g(\mathrm{i}\pi-\theta)\ \Rightarrow\ 
g(\theta)=\frac{\Gamma\left(\frac{2\pi}{\xi}+\mathrm{i}\frac{\theta}{\xi}\right)}{\Gamma\left(1+\frac{\pi}{\xi}+\mathrm{i}\frac{\theta}{\xi}\right)}h(\theta)\,,
\end{equation}
and so on. At the end of this recursive procedure, one gets the infinite product
\begin{eqnarray}
\label{sGS0}
S(\theta)&=&-\prod_{k=0}^{\infty}\frac{\Gamma\left(1+(2k+1)\frac{\pi}{\xi}-\mathrm{i}\frac{\theta}{\xi}\right)\Gamma\left(1+2k\frac{\pi}{\xi}+\mathrm{i}\frac{\theta}{\xi}\right)}{\Gamma\left(1+(2k+1)\frac{\pi}{\xi}+\mathrm{i}\frac{\theta}{\xi}\right)\Gamma\left(1+2k\frac{\pi}{\xi}-\mathrm{i}\frac{\theta}{\xi}\right)}\nonumber\\
&\times&\frac{\Gamma\left((2k+1)\frac{\pi}{\xi}-\mathrm{i}\frac{\theta}{\xi}\right)\Gamma\left((2k+2)\frac{\pi}{\xi}+\mathrm{i}\frac{\theta}{\xi}\right)}{\Gamma\left((2k+1)\frac{\pi}{\xi}+\mathrm{i}\frac{\theta}{\xi}\right)\Gamma\left((2k+2)\frac{\pi}{\xi}-\mathrm{i}\frac{\theta}{\xi}\right)}\,,
\label{infprod}
\end{eqnarray}
where we put an overall minus sign since the sine-Gordon S-matrix, from the discussion in \cite{Klassen:1992eq}, has to satisfy $S_{aa}^{aa}(0)=-1$.
The result (\ref{sGS0}) can be also derived using the technique explained in \cite{Volin:2009uv, Vieira:2010kb}: introducing the shift operator $D\equiv e^{\frac{\mathrm{i\pi}}{2}\partial_{\theta}}$, such that $Df(\theta)=f(\theta+\mathrm{i}\pi/2)$ and $f^D=e^{D\log f}$, we can write the crossing relation (\ref{crossgamma}) as 
\begin{equation}
S(\theta)^{D+D^{-1}}= \frac{\left[\Gamma\left(1+\mathrm{i}\frac{\theta}{\xi}\right)\Gamma\left(-\mathrm{i}\frac{\theta}{\xi}\right)\right]^D}{\left[\Gamma\left(1-\mathrm{i}\frac{\theta}{\xi}\right)\Gamma\left(\mathrm{i}\frac{\theta}{\xi}\right)\right]^{D^{-1}}}\,,
\end{equation}
that is formally solved by
\begin{equation}
S(\theta)= \frac{\left[\Gamma\left(1+\mathrm{i}\frac{\theta}{\xi}\right)\Gamma\left(-\mathrm{i}\frac{\theta}{\xi}\right)\right]^{\frac{D}{D+D^{-1}}}}{\left[\Gamma\left(1-\mathrm{i}\frac{\theta}{\xi}\right)\Gamma\left(\mathrm{i}\frac{\theta}{\xi}\right)\right]^{\frac{D^{-1}}{D+D^{-1}}}}\,.
\label{S0sGD}
\end{equation}
The exponents $D^{\pm1}/(D+D^{-1})$ can be expanded at small or at large $D$: the choice should be consistent with the minimality condition, {\it i.e.} the absence of zeros and poles in the physical strip, of the resulting $S(\theta)$. In particular, the factors in the r.h.s. of (\ref{S0sGD}) can be written as
\begin{eqnarray}
 &&\Gamma\left(1+\mathrm{i}\frac{\theta}{\xi}\right)^{\frac{D}{D+D^{-1}}}=\exp\left[-\sum_{n=1}^{\infty}(-1)^{n}D^{-2n+2}\log\Gamma\left(1+\mathrm{i}\frac{\theta}{\xi}\right)\right]\,,\\
 &&\Gamma\left(-\mathrm{i}\frac{\theta}{\xi}\right)^{\frac{D}{D+D^{-1}}}=\exp\left[-\sum_{n=1}^{\infty}(-1)^{n}D^{2n}\log\Gamma\left(-\mathrm{i}\frac{\theta}{\xi}\right)\right]\,,\\
 &&\Gamma\left(1-\mathrm{i}\frac{\theta}{\xi}\right)^{\frac{D^{-1}}{D+D^{-1}}}=\exp\left[-\sum_{n=1}^{\infty}(-1)^{n}D^{2n-2}\log\Gamma\left(1-\mathrm{i}\frac{\theta}{\xi}\right)\right]\,,\\
 &&\Gamma\left(\mathrm{i}\frac{\theta}{\xi}\right)^{\frac{D^{-1}}{D+D^{-1}}}=\exp\left[-\sum_{n=1}^{\infty}(-1)^{n}D^{-2n}\log\Gamma\left(-\mathrm{i}\frac{\theta}{\xi}\right)\right]\,.
\end{eqnarray}
Thus, we get the product (\ref{infprod}). Regularizing the sums in the exponents introduces an overall constant, that is set to $-1$ by the aforementioned condition $S_{aa}^{aa}(0)=-1$.
Moreover, using the following integral representation of $\log\Gamma$ 
\begin{equation}
\log\Gamma(x)=\int_{0}^{\infty}\frac{dt}{t}\left[(x-1)e^{-t}+\frac{e^{-tx}-e^{-t}}{1-e^{-t}}\right]\,,
\end{equation}
(\ref{infprod}) can be recast in the following compact integral form
\begin{equation}
S(\theta)=-\exp\left[-\mathrm{i}\int_{0}^{\infty}\frac{dt}{t}\frac{\sinh\frac{t(\pi-\xi)}{2}}{\sinh\frac{\xi t}{2}\cosh\frac{\pi t}{2}}\sin\theta t\right]\,.
\label{S0sG}
\end{equation}
At the specific value $\xi=\pi/N$, the soliton-soliton amplitude was already determined in \cite{Korepin:1975zu, Korepin:1975vd}, later confirmed by the exact derivation, on the basis of crossing and unitarity, of \cite{Zamolodchikov:1976uc}.
In the limit of $\xi\rightarrow0$, expressions (\ref{sGsol}) and (\ref{sGS0}) agree with the semi-classical results of \cite{Korepin:1975zu, Jackiw:1975im}.

Another way to determine (\ref{S0sG}), that can be found in \cite{Karowski:1977th, Karowski:1977tv}, uses a trick similar to that used for the derivation of the two-particle minimal form factor (\ref{cauchy}):
\begin{equation}
\log S_{T}(\theta)=\int_{C}\frac{dz}{2\pi \mathrm{i}}\frac{\log S_{T}(z)}{\sinh(z-\theta)}=\int_{-\infty}^{\infty}\frac{dz}{2\pi \mathrm{i}}\frac{\log S_{T}(z)S_{T}(\mathrm{i}\pi+z)}{\sinh(z-\theta)}\,,
\label{S0cauchy}
\end{equation}
where $C$ is a contour encircling the strip $0<\mbox{Im}(\theta)<\pi$. 
Unitarity and crossing relations imply
\begin{equation}
S_{T}(z)S_{T}(\mathrm{i}\pi+z)=\frac{S_{T}(z)S_{R}(\mathrm{i}\pi-z)}{S_{R}(z)S_{T}(\mathrm{i}\pi-z)}\,.
\label{STST}
\end{equation}
The ratio $S_{T}/S_{R}$ can be obtained by solving (\ref{YBEsG}), that gives 
\begin{equation}
\frac{S_{T}(\theta)}{S_{R}(\theta)}=\frac{\sinh\lambda\theta}{\sinh\lambda \mathrm{i}\pi}\,.
\label{ST/SR}
\end{equation}
Again, $\lambda$ is a free parameter that can be fixed to $\lambda=\xi/\pi$ by comparison with the known semi-classical expansion of the bound states masses \cite{ZZ}, that will be discussed in the next section.
Then, plugging (\ref{ST/SR}) into (\ref{STST}) and using (\ref{S0cauchy}), one easily gets (\ref{S0sG}).

\subsubsection{Pole structure and bound states}
\label{SGpoles}

It can be easily seen in (\ref{sGS0}) that $S(\theta)$ has a set of poles in $\theta=\mathrm{i}n\xi$, for $n=1,2,\dots$.
On the other hand, $S_T(\theta)$ and $S_R(\theta)$ are singular respectively in $\theta=\mathrm{i}(\pi-n\xi)$ and $\theta=\mathrm{i}(\pi-n\xi), \theta=\mathrm{i}n\xi$, with $n=1,2,\dots$.
These poles belong to the physical strip $0\le\mbox{Im}(\theta)\le\pi$ only if $\xi<\pi$: as anticipated above, this is indeed the so-called \emph{attractive} regime. This implies also that $S$ has poles in the $s$-channel, 
while $S_T$ and $S_R$ in the $t$-channel. In the so-called \emph{repulsive} regime $\xi>\pi$, instead, the poles move out of the physical strip, and therefore do not correspond to particle excitations.

If we consider the following combinations of  S-matrix elements with defined charge-con\-ju\-ga\-tion parity
\begin{equation}
S_{\pm}(\theta)=S_T(\theta)\pm S_R(\theta)\,,
\label{singeig}
\end{equation}
then $S_+$ has poles in $\theta=\mathrm{i}(\pi-n\xi)$ for even $n$, $S_-$ for odd $n$.
These bound states are called \emph{breathers}, with mass spectrum given by (\ref{bsmasses}) with $u_{ij}^n=\pi-n\xi$:
\begin{equation}
m_n=2M\sin\frac{n\xi}{2}\ ;\ \ n=1,\dots,N\ ;\ \ N<\left[\frac{\pi}{\xi}\right]\,,
\end{equation}
where $[x]$ denotes the integer part of $x$.

The S-matrices for the bound states can be derived by defining the following ZF operators $B_n$
\begin{eqnarray}
&&B_n\left(\frac{\theta_1+\theta_2}{2}\right)=\left.\left[s(\theta_2)\bar s(\theta_1)+\bar s(\theta_2)s(\theta_1)\right]\right|_{\theta_1-\theta_2=\mathrm{i}(\pi-n\xi)}\,;\ \ \mbox{for $n$ even}\,, \\
&&B_n\left(\frac{\theta_1+\theta_2}{2}\right)=\left.\left[s(\theta_2)\bar s(\theta_1)-\bar s(\theta_2)s(\theta_1)\right]\right|_{\theta_1-\theta_2=\mathrm{i}(\pi-n\xi)}\,;\ \ \mbox{for $n$ odd}\,,
\end{eqnarray}
that create the $n$th breathers.
Then the bootstrap equations (\ref{Sbp}) can be written as commutation relations of bound state and soliton, or antisoliton,  ZF generators
\begin{eqnarray}
&&s(\theta_1)B_n(\theta_2)=S^{(n)}(\theta_{12})B_n(\theta_2)s(\theta_1)\,,
\label{ZFB1}\\
&&\bar s(\theta_1)B_n(\theta_2)=S^{(n)}(\theta_{12})B_n(\theta_2)\bar s(\theta_1)\,,
\end{eqnarray}
while the S-matrices for scattering between bound states are calculated by
\begin{equation}
B_n(\theta_1)B_m(\theta_2)=S^{(nm)}(\theta_{12})B_m(\theta_2)B_n(\theta_1)\,.
\end{equation}
Alternatively, the breather-particle S-matrix can be calculated using (\ref{Sbpdiag}). The projector $\Gamma_{ij}^n$ is the eigenvector of the S-matrix corresponding to its singular eigenvalue 
 \cite{Karowski:1977fu}. Indeed, the S-matrix is diagonalized as follows
\begin{equation}
S_{ij}^{kl}(\theta)=\sum_{e}\Gamma_{e}^{kl}S_e(\theta)\Gamma_{ij}^{e}\,,
\end{equation}
where $S_{e}$, with $e=1,\dots,4$, are the eigenvalues and $\Gamma_{e}^{ij}$ the corresponding eigenvectors. One of the eigenvalues turns out to be the singular combination $S_{-}(\theta)$ as defined in (\ref{singeig}), while $\Gamma_{-}^{ij}=(0,-1/\sqrt{2},1/\sqrt{2},0)$.
Then one gets the following amplitude for the lowest bound state $B_1$:
\begin{eqnarray}
S^{(1)}\left(\theta_1-\theta_2\right)&=&\left.\frac{1}{2}\left[S_T(\theta_{13})S(\theta_{14})-S_R(\theta_{13})S_R(\theta_{14})+S(\theta_{13})S_T(\theta_{14})\right]\right|_{\theta_3-\theta_4=\mathrm{i}(\pi-\xi)}\nonumber\\
 &=&\frac{\sinh\theta_{12}+\mathrm{i}\sin\frac{\pi+\xi}{2}}{\sinh\theta_{12}-\mathrm{i}\sin\frac{\pi+\xi}{2}}\,.
\label{tbp}
 \end{eqnarray}
In fact, this is the only amplitude needed to describe the single breather-particle scattering, since only $S_{-}$ has a pole at $\theta=\mathrm{i}(\pi-n\xi)$, for $n=1$.

On the other hand, using (\ref{Sbbgen}), one can get the following breather-breather amplitude 
\begin{equation}
 S^{(11)}(\theta)=\frac{\sinh\theta+\mathrm{i}\sin\xi}{\sinh\theta-\mathrm{i}\sin\xi}\,,
 \label{Sbb}
\end{equation}
whose expansion in powers of $\beta^2$ has been successfully compared to the perturbation theory for the Lagrangian (\ref{LsG}), since $B_1$ is actually 
a pseudo-scalar particle corresponding to the fundamental field of sine-Gordon \cite{Arefeva:1974bk}.
\begin{framed}
\paragraph*{Exercises}
\begin{enumerate}
 \item Derive $S^{(1)}(\theta)$ using relation (\ref{ZFB1}) and the identities 
 \begin{equation}
  \hspace{-1cm}S(\theta_{32})S_R(\theta_{31})=S_R(\theta_{32})S_T(\theta_{31})\,,\quad S(\theta_{32})S_T(\theta_{31})-S_R(\theta_{31})S_R(\theta_{31})=S_T(\theta_{32})S(\theta_{31})\,,\nonumber
 \end{equation}
valid for $\theta_{12}=\mathrm{i}(\pi-\xi)$, and verify the explicit expression given in (\ref{tbp}). Finally, using the fusion of two amplitudes $S^{(1)}(\theta)$, check expression (\ref{Sbb}).
 \item Derive (\ref{tbp}) using (\ref{Sbpdiag}): verify that $S_-(\theta)$ in (\ref{singeig}) is the only singular eigenvalue in the case of one breather, $\Gamma_{-}^{ij}=(0,-1/\sqrt{2},1/\sqrt{2},0)$ and 
 \begin{equation}
  \mathop{\mathrm{Res}}_{\theta=\mathrm{i}(\pi-\xi)}(S_{sG})_{ij}^{kl}(\theta)\propto\Gamma_{ij}^-\Gamma_-^{kl}\,.\nonumber                                                                                                                                                                                                              
 \end{equation}
 \end{enumerate}
\end{framed}

\subsubsection{Form Factors}
\label{SGform}

The soliton-soliton form factor of sine-Gordon satisfies the following Watson equations
\begin{equation}
F_{ss}(\theta)=F_{ss}(-\theta)S(\theta)=F_{ss}(2\pi \mathrm{i}-\theta)\,,
\label{2pwatson}
\end{equation}
where $S(\theta)$ is the sine-Gordon soliton-soliton amplitude (\ref{S0sG}).
The minimal solution of (\ref{2pwatson}) can be found in a way analogous to the procedure, discussed in Section \ref{SGSmatrix}, which was used to fix the soliton-soliton amplitude of sine-Gordon as a solution 
of the crossing and unitarity constraints (\ref{S0cross}), (\ref{sGuni1}). The result is  \cite{Weisz:1977ii}
\begin{equation}
F_{ss}^{min}(\theta)=-\mathrm{i}\sinh\frac{\theta}{2}\exp\left[\int_{0}^{\infty}\frac{dt}{t}\frac{\sinh\frac{(\pi-\xi)t}{2\pi}}{\sinh\frac{\xi t}{2\pi}\cosh\frac{t}{2}}\frac{1-\cosh t\left(1-\frac{\theta}{\mathrm{i}\pi}\right)}{2\sinh t}\right]
\,,
\label{solwatson}
\end{equation}
where the factor $-\mathrm{i}\sinh\frac{\theta}{2}$ is due to the overall minus sign in (\ref{S0sG}).
The solution (\ref{solwatson}) can be derived in a simpler way by applying equation (\ref{cauchy}) to the soliton-soliton amplitude (\ref{S0sG}) \cite{Karowski:1978vz}.
In general, with an amplitude given by
\begin{equation}
S(\theta)=\exp\left[\int_{0}^{\infty}dt f(t)\sinh\frac{t\theta}{\mathrm{i}\pi}\right]\,,
\label{Sgen}
\end{equation}
the corresponding minimal solution for the form factor is
\begin{equation}
F^{min}(\theta)=\exp\left[\int_{0}^{\infty}dt f(t)\frac{1-\cosh t\left(1-\frac{\theta}{\mathrm{i}\pi}\right)}{2\sinh t}\right]\,.
\label{fmingen}
\end{equation}
Full expressions of soliton-soliton form factors are given then by the minimal solution (\ref{solwatson}) multiplied by normalization constants and factors giving additional zeros/poles in the physical strip:
both of these objects depend crucially on the operator connecting the soliton-soliton state to the vacuum, as mentioned in section \ref{formfactors}.

For example, the breather-breather form factors are given by
\begin{equation}
F_{bb}^{\mathcal{O}}(\theta_{12})=\mathcal{N}^{\mathcal{O}}K_{bb}^{\mathcal{O}}(\theta_{12})F_{bb}^{min}(\theta_{12})\,,
\label{bbform}
\end{equation}
whose minimal solution can be derived just from the corresponding amplitude (\ref{Sbb}) by using (\ref{Sgen}), (\ref{fmingen}), as in the previous case.   
Indeed (\ref{Sbb}) can be written as (\ref{Sgen}), with 
\begin{equation}
f(t)=2\frac{\cosh\frac{(\pi-\xi)t}{\pi}}{t\cosh\frac{t}{2}}\,.
\end{equation}
If the operator is $O(x)=\phi^2(x)$, then $K_{bb}^{\phi^2}(\theta)$ turns out to be (\ref{K}) with $n=1,\alpha_{1}=\xi$, while $\mathcal{N}^{\phi^2}$ can be fixed by matching the large $\theta$ 
asymptotic behavior of (\ref{bbform}) with the corresponding small-$\beta$ diagrammatic perturbative result \cite{Karowski:1978vz}.

\subsection{Chiral Gross-Neveu}
\label{CGN}

The $SU(N)$ chiral Gross-Neveu (cGN) model \cite{Gross:1974jv} is described by the Lagrangian
(see \cite{Florian} and \cite{Stjin})
\begin{equation}
\mathcal{L}_{cGN}=\mathrm{i}\sum_{i=1}^{N}\bar\psi_{i}\slashed{\partial}\psi_{i}+\frac{g^{2}_{cGN}}{2}\left[\left(\sum_{i=1}^{N}\bar\psi_{i}\psi_{i}\right)^{2}-\left(\sum_{i=1}^{N}\bar\psi_{i}\gamma^{5}\psi_{i}\right)^{2}\right]\,,
\end{equation}
and its particle spectrum consists of $N-1$ multiplets with masses
\begin{equation}
m_{n}=m_{1}\frac{\sin \frac{n\pi}{N}}{\sin\frac{\pi}{N}}\,\quad n=1,\dots,N-1\,.
\label{bsmassSUN}
\end{equation}
The form of the S-matrix for two fundamental particles is constrained by the $SU(N)$ symmetry to be \cite{Berg:1977dp, Kurak:1978su}
\begin{equation}
S_{ij}^{kl}(\theta)=S_{0}^{(N)}(\theta)\left(b_{N}(\theta)\delta_{i}^{l}\delta_{j}^{k}+c_{N}(\theta)\delta_{i}^{k}\delta_{j}^{l}\right)\,,
\end{equation}
with indices $i, j, k, l$ running over $1,\dots,N$.
The overall scalar factor and the ratio between transmission and reflection amplitudes are instead given by
\begin{equation}
S_{0}^{(N)}(\theta)=-\frac{\Gamma\left(1+\frac{\mathrm{i}\theta}{2\pi}\right)\Gamma\left(1-\frac{\mathrm{i}\theta}{2\pi}-\frac{1}{N}\right)}{\Gamma\left(1-\frac{\mathrm{i}\theta}{2\pi}\right)\Gamma\left(1+\frac{\mathrm{i}\theta}{2\pi}-\frac{1}{N}\right)}\,,
\ \ c_{N}(\theta)=-\frac{2\pi\mathrm{i}}{N\theta}b_{N}(\theta)\,,
\label{S0CGN}
\end{equation}
which are determined by unitarity, crossing symmetry and  the YBE (\ref{YBE}), which in particular fixes the proportionality factor between $c_{N}(\theta)$ and $b_{N}(\theta)$.

\subsubsection{Solutions for the $SU(2)$ and $SU(3)$ S-matrices}

In particular, for $N=2$ the particle-particle S-matrix turns out to be the limit $\xi\rightarrow\infty$ (or $\beta^{2}\rightarrow8\pi$) of the sine-Gordon S-matrix (\ref{sGmatrix})-(\ref{sGsol})-(\ref{sGS0}): the commutation conditions with the coproducts (\ref{coproduct}) built on the $SU(2)$ generators (Pauli matrices) restrict the S-matrix to be
\begin{equation}
S_{cGN}^{SU(2)}(\theta)=S_{0}^{(2)}(\theta)\left(
\begin{array}{cccc}
 a_{2}(\theta) &   &  & \\
  & b_{2}(\theta)  & c_{2}(\theta) & \\
  & c_{2}(\theta)  & b_{2}(\theta) & \\
  &   &  &  a_{2}(\theta)
\end{array}
\right)\,,\ \ a_{2}(\theta)=b_{2}(\theta)+c_{2}(\theta)\,.
\label{CGNmatrix2}
\end{equation}
We consider also the $SU(3)$ case (that will be useful for \cite{Fedor}), whose symmetry algebra is generated by the eight Gell-Mann matrices. Imposing the commutation with four of them is enough to fix the following structure of 
the S-matrix:
\begin{equation}
S_{cGN}^{SU(3)}(\theta)=S_{0}^{(3)}(\theta)\left(
\begin{array}{ccccccccc}
 a_{3}(\theta) &   &  & & & & & & \\
  & b_{3}(\theta)  & & c_{3}(\theta) & & & & & \\
  & & b_{3}(\theta) & & & & c_{3}(\theta) & & \\
  & c_{3}(\theta) & & b_{3}(\theta) & & & & & \\
  &   &  &  & a_{3}(\theta) & & & & \\
  &   &  &  & & b_{3}(\theta) & & c_{3}(\theta) &\\
  &  & c_{3}(\theta) & & & & b_{3}(\theta) & &\\
  &  &  & & & c_{3}(\theta) & & b_{3}(\theta) &\\
  & & & & & & & & a_{3}(\theta) 
\end{array}
\right)\,,
\label{CGNmatrix3}
\end{equation}
with $a_{3}(\theta)=b_{3}(\theta)+c_{3}(\theta)$.
The matrix elements and the (minimal) scalar factors are determined by the YBE, unitarity and crossing symmetry, using the charge conjugation matrix
\begin{equation}
\mathcal{C}_{a_{1}\dots a_{N}}=\epsilon_{a_{1}\dots a_{N}}\,;\quad \mathcal{C}^{a_{1}\dots a_{N}}=(-1)^{N-1}\epsilon^{a_{1}\dots a_{N}}\,,
\end{equation}
with $\epsilon_{a_{1}\dots a_{N}}$ and $\epsilon^{a_{1}\dots a_{N}}$  totally antisymmetric tensors.
For the $SU(2)$ case the resulting elements read 
\begin{equation}
a_{2}(\theta)=1\,,\quad b_{2}(\theta)=\frac{\theta}{\theta-\mathrm{i}\pi}\,,\quad c_{2}(\theta)=\frac{-\mathrm{i}\pi}{\theta-\mathrm{i}\pi}\,,
\end{equation}
while for $SU(3)$ one finds 
\begin{equation}
a_{3}(\theta)=1\,,\quad b_{3}(\theta)=\frac{3\theta}{3\theta-2\mathrm{i}\pi}\,,\quad c_{3}(\theta)=\frac{-2\mathrm{i}\pi}{3\theta-2\mathrm{i}\pi}\,.
\end{equation}
The scalar factors are given by (\ref{S0CGN}) with $N=2$ and $N=3$ respectively. 

\subsubsection{Pole structure and bound states}
\label{CGNpoles}

Since the $SU(2)$ cGN S-matrix can be thought of as the $\xi\rightarrow\infty$ limit of the sine-Gordon one, and this limit corresponds to the highest repulsive regime for sine-Gordon, 
it can be easily understood that there are no bound states in this theory. Otherwise, it can be also verified that the S-matrix does not have any pole in the strip 
$0\le\mbox{Im}(\theta)\le\pi$.

On the other hand, the $SU(3)$ S-matrix has a pole in the physical strip at $\theta=\frac{2\mathrm{i}\pi}{3}$, corresponding to a bound state with mass $m_{2}=m_{1}$, then equal to the fundamental particle mass. For generic $N$, bound states have masses given by the expression (\ref{bsmassSUN}), so that $m_{N-1}=m_{1}$.
Indeed, it is possible to show that in the $SU(N)$ cGN the antiparticles are bound states of $N-1$ particles and vice versa \cite{Kurak:1978su}. 

We can now derive the particle-bound state S-matrix using (\ref{Sbp}) and finding $\Gamma_{ij}^n$, given by the  antisymmetric tensor $\epsilon_{nij}$, as the three eigenvectors corresponding to the singular eigenvalue $b(\theta)-c(\theta)$. The result is
\begin{equation}
S_{cGN}^{bp}(\theta)=S_{0}^{bp}(\theta)\left(
\begin{array}{ccccccccc}
 A(\theta) &   &  & & C(\theta) & & & & C(\theta)\\
  & B(\theta)  & & & & & & & \\
  & & B(\theta) & & & & & & \\
  &  & & B(\theta) & & & & & \\
 C(\theta) &   &  &  & A(\theta) & & & & C(\theta)\\
  &   &  &  & & B(\theta) & & &\\
  &  &  & & & & B(\theta) & &\\
  &  &  & & &  & & B(\theta) &\\
 C(\theta) & & & & C(\theta) & & & & A(\theta) 
\end{array}
\right)\,,
\label{CGNmatrixbp}
\end{equation}
where $A(\theta)=B(\theta)+C(\theta)$, the scalar factor is given by the fusion of the fundamental ones
\begin{equation}
 S_{0}^{bp}(\theta)=S_{0}^{(3)}(\theta-\mathrm{i}\pi/3)S_{0}^{(3)}(\theta+\mathrm{i}\pi/3)=\frac{\Gamma\left(\frac{1}{2}-\mathrm{i}\frac{\theta}{2\pi}\right)\Gamma\left(\frac{7}{6}+\mathrm{i}\frac{\theta}{2\pi}\right)}{\Gamma\left(\frac{1}{2}+\mathrm{i}\frac{\theta}{2\pi}\right)\Gamma\left(\frac{7}{6}-\mathrm{i}\frac{\theta}{2\pi}\right)}\,,
\end{equation}
and the remaining matrix elements read
\begin{eqnarray}
&&\hspace{-1cm}B(\theta)=\frac{1}{2}\mathop{\mathrm{Res}}_{\theta_{12}=2\mathrm{i}\pi/3}\left(b(\theta_{12})-c(\theta_{12})\right)\left(2b(\theta_{13})b(\theta_{23})+b(\theta_{13})c(\theta_{23})+c(\theta_{13})b(\theta_{23})\right.\nonumber\\
&&\left.-c(\theta_{13})c(\theta_{23})\right)\,,~~~~~~~~
\label{B}\\
&&\hspace{-1cm}C(\theta)=-\frac{1}{2}\mathop{\mathrm{Res}}_{\theta_{12}=2\mathrm{i}\pi/3}\left(b(\theta_{12})-c(\theta_{12})\right)\left(b(\theta_{13})c(\theta_{23})+c(\theta_{13})b(\theta_{23})-c(\theta_{13})c(\theta_{23})\right)\,.~~~~~~~~
\end{eqnarray}
Finally, the bound state-bound state amplitudes can be derived using (\ref{Sbbgen}): we leave this as an exercise to the interested reader.

\subsubsection{Form Factors}
\label{CGNform}

Here we show the simplest example of form factors in $SU(N)$ cGN models, that is the minimal solution to the two-particle Watson equations (\ref{2pwatson}), corresponding to the amplitudes $a_{2,3}(\theta)=1$ of the S-matrices (\ref{CGNmatrix2}) and (\ref{CGNmatrix3}). Then the amplitudes are actually $S_{0}^{(N)}$ as written in (\ref{S0CGN}) and, using the trick given by equations (\ref{Sgen}) and (\ref{fmingen}), one can easily find
\begin{equation}
F_{11}^{min}(\theta)=c\exp\left[\int_{0}^{\infty}\frac{dt}{t}\frac{e^{\frac{t}{N}}\sinh t\left(1-\frac{1}{N}\right)}{\sinh^{2}t}\left(1-\cosh t\left(1-\frac{\theta}{\mathrm{i}\pi}\right)\right)\right]\,.
\end{equation}
A generic $n$-particle form factor for scalar operators would be
\begin{equation}
F_{\underline{a}}^{\mathcal{O}}(\theta_1,\dots,\theta_{n})=K_{\underline{a}}^{\mathcal{O}}(\theta_1,\dots,\theta_{n})\prod_{1\le i<j\le n}F_{a_{i}a_{j}}^{min}(\theta_{ij})\,,
\end{equation}
where the function $K_{\underline{a}}^{\mathcal{O}}$ contains the pole structure and is partially fixed by the Watson equations (\ref{watson1})-(\ref{watson2}), with the amplitude $S$ replaced by $\tilde S=S/S_{0}$ \cite{Babujian:2006md}:
\begin{eqnarray}
\hspace{-1cm}&&K_{\underline{a}}^{\mathcal{O}}(\theta_{1},\dots,\theta_{i},\theta_{j},\dots,\theta_{n})=K_{a_{1},\dots,a_{j},a_{i},\dots,a_{n}}^{\mathcal{O}}(\theta_{1},\dots,\theta_{j},\theta_{i},\dots,\theta_{n})\tilde S_{a_{i}a_{j}}(\theta_{ij})\,,\ \ j=i+1\,,\nonumber\\
\hspace{-1cm}&&K_{\underline{a}}^{\mathcal{O}}(\theta_{1}+2\pi i,\dots,\theta_{n})=K_{\underline{a}}^{\mathcal{O}}(\theta_{1},\dots,\theta_{n})\prod_{i=2}^{n}\tilde S_{a_{i}a_{1}}(\theta_{i}-\theta_{1})\,.
\end{eqnarray}
The only solutions of these equations we found in literature for $SU(N)$ cGN were obtained using the so-called ``off-shell'' nested Bethe ansatz method (see \cite{Babujian:1998uw, Babujian:2006km, Babujian:2006md, Babujian:2009zg} for example), that is beyond the scope of these lectures.

\subsection{AdS/CFTs}

In $AdS_{5}/CFT_{4}$, the dynamics of string excitations is described by an integrable non-linear $\sigma$-model defined on the super coset $\frac{PSU(2,2|4)}{SO(4,1)\times SO(5)}$ \cite{Bena:2003wd}, while, on the gauge side, 
the fields composing single-trace operators correspond to the excitations of an integrable super spin chain \cite{Minahan:2002ve} (see also the aforementioned reviews \cite{AF} and \cite{Beisert:2010jr}).

Then such excitations interact via a factorized S-matrix, depending on the 't Hooft coupling $\lambda$, whose matrix elements were fixed in \cite{Beisert:2005tm}, up to an overall scalar factor, by imposing the invariance of 
the S-matrix under two copies of the centrally extended $su(2|2)$ superalgebra, that is the symmetry algebra leaving invariant the vacuum.

In order to determine the scalar factor, crossing symmetry has been imposed in the algebraic ways explained in Sections \ref{sec:nonrel} and \ref{Hopf}, in \cite{Arutyunov:2006yd} and \cite{Janik:2006dc} respectively. The equation arising from such condition was satisfied by the conjecture
of \cite{Arutyunov:2004vx} and was finally solved in \cite{Volin:2009uv} (see also \cite{Vieira:2010kb} for a review).

The bound states S-matrices have been determined in \cite{Arutyunov:2009mi} by using the Yangian symmetry $Y(su(2|2))$ \cite{Florian}. Recently, the usual bootstrap procedure was generalized to 
the $AdS_{5}/CFT_{4}$ case \cite{Beisert:2015msa}.

Determining the exact, all-loop S-matrix in $AdS_{5}/CFT_{4}$ has been of essential importance, as in any other integrable theory, to study its exact finite volume spectrum. From the S-matrix of \cite{Beisert:2005tm}, indeed, the asymptotic Bethe equations conjectured in \cite{Beisert:2005fw} could be derived \cite{Beisert:2005tm, Martins:2007hb}.
Then, on the basis of the same S-matrix, it was possible to study and compute the leading order finite-size corrections \cite{Janik:2007wt} and the exact spectrum via the TBA \cite{TBA},
that was recently reduced to a simple set of non-linear Riemann-Hilbert equations in \cite{Gromov:2013pga}, the so-called quantum spectral curve (QSC) equations.

Concerning $AdS_4/CFT_3$, the exact S-matrix was determined on the basis of a symmetry superalgebra still related to $su(2|2)$, while the scalar factors were fixed by
slightly different crossing symmetry relations \cite{Ahn:2008aa}. This S-matrix gives the Bethe equations conjectured in \cite{Gromov:2008qe} and was used to derive L\"uscher-like corrections \cite{Bombardelli:2008qd}
and the corresponding TBA \cite{Bombardelli:2009xz} (see also the review \cite{Klose:2010ki}). Finally, also in this example of integrable $AdS/CFT$ correspondence, it was possible to reduce the spectral problem to the solution of QSC equations 
\cite{Cavaglia:2014exa}.

In the case of $AdS_3/CFT_2$, two string backgrounds were studied, $AdS_3\times S^3\times S^3\times S^1$ and $AdS_3\times S^3\times T^4$. Both of them involve massless string modes, a new feature compared to $AdS_5$ and $AdS_4$. A set of all-loop Bethe equations for the massive modes of $AdS_3\times S^3\times S^3\times S^1$ were conjectured in \cite{Babichenko:2009dk} and later derived from the S-matrix proposed in \cite{Ahn:2012hw}. However, this S-matrix could describe only a sector of the theory. Imposing the commutation with the generators of the full (centrally extended $su(1|1)^2$) symmetry algebra allowed \cite{Borsato:2012ud} to determine the complete S-matrix for massive excitations and the consequent all-loop Bethe equations \cite{Borsato:2012ss} describing
the large volume limit of the $AdS_3\times S^3\times S^3\times S^1$ massive spectrum. Massless modes were included in the integrability framework in \cite{Borsato:2015mma}, while for the 
$AdS_3\times S^3\times T^4$ case the reader can look at the complete S-matrix determined in \cite{Borsato:2014exa}. 
These S-matrices are substantially more involved than the ones appearing in the higher-dimensional holographic pairs, due to the presence of several distinct scalar factors with novel properties. An all-loop proposal exists for  the scalar factors, involving both massive \cite{Borsato:2013hoa} and massless \cite{Borsato:2016kbm} modes, of the $AdS_3\times S^3\times T^4$ S-matrix. Furthermore, finite-size corrections due to massless modes seem to play a new role in the calculation of the large volume spectrum \cite{Abbott:2015pps}. See \cite{Sfondrini:2014via} for a review about these and other developments of integrability in $AdS_3/CFT_2$.

Finally, in $AdS_2/CFT_{1}$ the determination of an exact S-matrix and related Bethe equations is even more difficult due to the presence of more massless modes and less supersymmetry, while crossing symmetry relations are still understood only formally and it is not clear what is the $CFT_1$ involved in this duality \cite{Hoare:2014kma}.   

\section*{Acknowledgements}

It is a pleasure to thank the GATIS network and the organizers of the ``Young Researcher Integrability School'' for giving me the extraordinary opportunity to give these lectures at the Department of Mathematical Sciences
at Durham University, which I thank for hospitality.
I would like to thank the other lecturers and the participants of the school for creating a very stimulating atmosphere, for suggestions and discussions.
I am especially grateful to Zolt\'an Bajnok, Andrea Cavagli\`a, Alessandro Sfondrini and Roberto Tateo for helpful comments on the manuscript and discussions.
I thank also Emanuele Latini, Francesco Ravanini, Patricia Ritter for discussions and the referees of \emph{J. Phys. A: Math. Theor.} for useful comments.
The work of the author has been partially funded by the INFN grants GAST and FTECP, and the research grant UniTo-SanPaolo Nr TO-Call3-2012-0088 ``Modern
Applications of String Theory'' (MAST).


\begin{thebibliography}{99}

\bibitem{Wheeler}
  J.~A.~Wheeler,
  ``On the mathematical description of light nuclei by the method of resonating group structure'',
  Phys.\ Rev.\  {\bf 52} (1937) 1107.

\bibitem{Heisenberg}
W.~Heisenberg,
``Der mathematische Rahmen der Quantentheorie der Wellenfelder'', Zeit. f\"ur Naturforschung {\bf 1} (1946) 608.

\bibitem{Chew}
G.~Chew, ``The S-matrix theory of strong interaction'', W. A. Benjamin Inc., New York (1961).

\bibitem{Mandelstam:1958xc}
  S.~Mandelstam,
  ``Determination of the pion - nucleon scattering amplitude from dispersion relations and unitarity. General theory'',
  Phys.\ Rev.\  {\bf 112} (1958) 1344.

\bibitem{Coleman:1967ad}
  S.~R.~Coleman and J.~Mandula,
  ``All possible symmetries of the S-matrix'',
  Phys.\ Rev.\  {\bf 159} (1967) 1251.
  
\bibitem{Fedor}
F.~Levkovich-Maslyuk, ``Lectures on the Bethe Ansatz'', in: ``An integrability primer for the gauge-gravity correspondence'', ed.: A.~Cagnazzo, R.~Frassek, A.~Sfondrini,
I.~M.~Sz\'ecs\'enyi and S.~J.~van Tongeren, J.\ Phys.\ A {\bf 49} (2016) no.32,  323004
  [arXiv:1606.02950 [hep-th]].

\bibitem{Luscher:1983rk}
  M.~L\"uscher,
  ``On a relation between finite size effects and elastic scattering processes'', in the proceedings of the Cargese Summer Institute: ``Progress in gauge field theory'', edited by G. 't Hooft, A. Singer, R. Stora, Plenum Press, New York (1984);
  M.~L\"uscher,
  ``Volume dependence of the energy spectrum in massive quantum field theories. 1. Stable particle states'',
  Commun.\ Math.\ Phys.\  {\bf 104} (1986) 177;
  T.~R.~Klassen and E.~Melzer,
  ``On the relation between scattering amplitudes and finite size mass corrections in QFT'',
  Nucl.\ Phys.\ B {\bf 362} (1991) 329.

\bibitem{Stjin}
S.~J.~van Tongeren, ``Introduction to the thermodynamic Bethe ansatz'', in ``An integrability primer for the gauge-gravity correspondence'', ed.: A.~Cagnazzo, R.~Frassek, A.~Sfondrini,
I.~M.~Sz\'ecs\'enyi and S.~J.~van Tongeren, J.\ Phys.\ A {\bf 49} (2016) no.32,  323005
  [arXiv:1606.02951 [hep-th]].
  
\bibitem{Mussardobook} 
 G.~Mussardo,
  ``Statistical field theory, an introduction to exactly solved models in statistical physics'', Oxford University Press, New York (2010).
  
\bibitem{Dorey}
  P.~Dorey,
  ``Exact S-matrices'', in the proceedings of the E\"otv\"os Summer School in Physics: ``Conformal field theories and integrable models'', edited by Z.~Horvath, L.~Palla, Springer, Berlin (1997) [hep-th/9810026].
  
\bibitem{ZZ}
  A.~B.~Zamolodchikov and A.~B.~Zamolodchikov,
  ``Factorized S-matrices in two-dimensions as the exact solutions of certain relativistic quantum field models'',
  Annals Phys.\  {\bf 120} (1979) 253.
  
  \bibitem{AF}
  G.~Arutyunov and S.~Frolov,
  ``Foundations of the $AdS_5 \times S^5$ superstring. Part I'',
  J.\ Phys.\ A {\bf 42} (2009) 254003
  [arXiv:0901.4937 [hep-th]].
  
\bibitem{Ahn:2010ka}
  C.~Ahn and R.~I.~Nepomechie,
  ``Review of AdS/CFT integrability, Chapter III.2: Exact worldsheet S-matrix'',
  Lett.\ Math.\ Phys.\  {\bf 99} (2012) 209 
    [arXiv:1012.3991 [hep-th]].
    
\bibitem{Vieira:2010kb}
  P.~Vieira and D.~Volin,
  ``Review of AdS/CFT Integrability, Chapter III.3: The Dressing factor,''
  Lett.\ Math.\ Phys.\  {\bf 99} (2012) 231
  [arXiv:1012.3992 [hep-th]].
  
\bibitem{Staudacher:2004tk}
  M.~Staudacher,
  ``The factorized S-matrix of CFT/AdS'',
  JHEP {\bf 0505} (2005) 054
  [hep-th/0412188].
  
\bibitem{Beisert:2005tm}
  N.~Beisert,
  ``The SU($2|2$) dynamic S-matrix'',
  Adv.\ Theor.\ Math.\ Phys.\  {\bf 12} (2008) 945
  [hep-th/0511082];
  N.~Beisert,
  ``The analytic Bethe Ansatz for a chain with centrally extended su($2|2$) Symmetry'',
  J.\ Stat.\ Mech.\  {\bf 0701} (2007) P01017
  [nlin/0610017 [nlin.SI]].
  
\bibitem{Janik:2006dc}
  R.~A.~Janik,
  ``The AdS$_{5} \times$S$^5$ superstring worldsheet S-matrix and crossing symmetry'',
  Phys.\ Rev.\ D {\bf 73} (2006) 086006
  [hep-th/0603038].

\bibitem{Arutyunov:2004vx}
  G.~Arutyunov, S.~Frolov and M.~Staudacher,
  ``Bethe Ansatz for quantum strings'',
  JHEP {\bf 0410} (2004) 016
  [hep-th/0406256];
  N.~Beisert, B.~Eden and M.~Staudacher,
  ``Transcendentality and crossing'',
  J.\ Stat.\ Mech.\  {\bf 0701} (2007) P01021
  [hep-th/0610251].
  
\bibitem{Arutyunov:2006yd}
  G.~Arutyunov, S.~Frolov and M.~Zamaklar,
  ``The Zamolodchikov-Faddeev algebra for AdS$_{5} \times$S$^{5}$ superstring'',
  JHEP {\bf 0704} (2007) 002
  [hep-th/0612229].
  
\bibitem{Beisert:2010jr}
  N.~Beisert {\it et al.},
  ``Review of AdS/CFT integrability: an overview'',
  Lett.\ Math.\ Phys.\  {\bf 99} (2012) 3
  [arXiv:1012.3982 [hep-th]].
  

\bibitem{Karowski:1978vz}
  M.~Karowski and P.~Weisz,
  ``Exact form factors in (1+1)-dimensional field theoretic models with soliton behavior'',
  Nucl.\ Phys.\ B {\bf 139} (1978) 455.
  
\bibitem{Babujian:2006km}
  H.~M.~Babujian, A.~Foerster and M.~Karowski,
  ``The form factor program: A review and new results: The nested SU(N) off-shell Bethe Ansatz'',
  SIGMA {\bf 2} (2006) 082
  [hep-th/0609130].
  
  \bibitem{Yurov:1990kv}
  V.~P.~Yurov and A.~B.~Zamolodchikov,
  ``Correlation functions of integrable 2-D models of relativistic field theory. Ising model'',
  Int.\ J.\ Mod.\ Phys.\ A {\bf 6} (1991) 3419.
  
\bibitem{Babujian:1998uw}
  H.~M.~Babujian, A.~Fring, M.~Karowski and A.~Zapletal,
  ``Exact form-factors in integrable quantum field theories: The sine-Gordon model'',
  Nucl.\ Phys.\ B {\bf 538} (1999) 535
  [hep-th/9805185].
  
\bibitem{Smirnovbook}
F.~A.~Smirnov,``Form factors in completely integrable models of quantum field theory'', Adv. Series in Math. Phys. {\bf14}, World Scientific, Singapore (1992).

\bibitem{Delfino:2015ria}
  G.~Delfino,
  ``Fields, particles and universality in two dimensions'',
  Annals Phys.\  {\bf 360} (2015) 477
  [arXiv:1502.05538 [cond-mat.stat-mech]].
  
\bibitem{Polyakov}
  A.~M.~Polyakov,
  ``Hidden symmetry of the two-dimensional chiral fields'',
  Phys.\ Lett.\ B {\bf 72} (1977) 224.
 
\bibitem{ShankarWitten}
  R.~Shankar and E.~Witten,
  ``The S-matrix of the supersymmetric nonlinear sigma model'',
  Phys.\ Rev.\ D {\bf 17} (1978) 2134.

\bibitem{Parke}
  S.~J.~Parke,
  ``Absence of particle production and factorization of the S-matrix in (1+1)-dimensional models'',
  Nucl.\ Phys.\ B {\bf 174} (1980) 166.
  
\bibitem{Iagolnitzer:1977sw}
  D.~Iagolnitzer,
  ``Factorization of the multiparticle S-matrix in two-dimensional space-time models'',
  Phys.\ Rev.\ D {\bf 18} (1978) 1275.
  
\bibitem{Luscher:1977uq}
  M.~L\"uscher,
  ``Quantum nonlocal charges and absence of particle production in the two-dimensional nonlinear sigma model'',
  Nucl.\ Phys.\ B {\bf 135} (1978) 1.
  
\bibitem{Florian}
F.~Loebbert, ``Lectures on Yangian Symmetry'', in: ``An integrability primer for the gauge-gravity correspondence'', ed.: A.~Cagnazzo, R.~Frassek, A.~Sfondrini,
I.~M.~Sz\'ecs\'enyi and S.~J.~van Tongeren, J.\ Phys.\ A {\bf 49} (2016) no.32,  323002
  [arXiv:1606.02947 [hep-th]].
  
\bibitem{Yang-Baxter}
  C.~N.~Yang,
  ``Some exact results for the many-body problems in one dimension with repulsive delta-function interaction'',
  Phys.\ Rev.\ Lett.\  {\bf 19} (1967) 1312;
  R.~J.~Baxter,
  ``Partition function of the eight vertex lattice model'',
  Annals Phys.\  {\bf 70} (1972) 193,
   Annals Phys.\  {\bf 281} (2000) 187.
   
\bibitem{Miramontes:1999gd}
  J.~L.~Miramontes,
  ``Hermitian analyticity versus real analyticity in two-dimensional factorized S-matrix theories'',
  Phys.\ Lett.\ B {\bf 455} (1999) 231
  [hep-th/9901145].
   
\bibitem{Faddeev}
  L.~D.~Faddeev,
  ``Quantum completely integrable models of field theory'',
  Sov.\ Sci.\ Rev.\ C {\bf 1} (1980) 107.
    
\bibitem{CDD}
  L.~Castillejo, R.~H.~Dalitz and F.~J.~Dyson,
  ``Low's scattering equation for the charged and neutral scalar theories'',
  Phys.\ Rev.\  {\bf 101} (1956) 453.

\bibitem{Volin:2009uv}
  D.~Volin,
  ``Minimal solution of the AdS/CFT crossing equation'',
  J.\ Phys.\ A {\bf 42} (2009) 372001
  [arXiv:0904.4929 [hep-th]].

\bibitem{Klassen:1989ui}
  T.~R.~Klassen and E.~Melzer,
  ``Purely elastic scattering theories and their ultraviolet limits'',
  Nucl.\ Phys.\ B {\bf 338} (1990) 485.

\bibitem{Spill:2007bia}
  F.~Spill,
  ``Hopf algebras in the AdS/CFT correspondence'', Diploma thesis at Humboldt University, Berlin (2007).
  
\bibitem{Drinfeld:1986in}
  V.~G.~Drinfeld,
  ``Quantum groups'',
  J.\ Sov.\ Math.\  {\bf 41} (1988) 898,
   Zap.\ Nauchn.\ Semin.\  {\bf 155} (1986) 18.
  
\bibitem{Delius:1995he}
  G.~W.~Delius,
  ``Exact S-matrices with affine quantum group symmetry'',
  Nucl.\ Phys.\ B {\bf 451} (1995) 445
  [hep-th/9503079].
  
\bibitem{MacKay}
N.~J.~MacKay, ``On the algebraic structure of factorized S-matrices'',
Durham theses, Durham
University (1992), http://etheses.dur.ac.uk/5764/.

\bibitem{Spill:2012qe}
F.~Spill, ``Yangians in integrable field theories, spin chains and gauge-string dualities'', Rev.\ Math.\ Phys.\ {\bf 24} (2012) 1230001.

\bibitem{Wu:1975mw}
  T.~T.~Wu, B.~M.~McCoy, C.~A.~Tracy and E.~Barouch,
  ``Spin spin correlation functions for the two-dimensional Ising model: Exact theory in the scaling region'',
  Phys.\ Rev.\ B {\bf 13} (1976) 316;
  O.~Babelon and D.~Bernard,
  ``From form-factors to correlation functions: The Ising model'',
  Phys.\ Lett.\ B {\bf 288} (1992) 113
  [hep-th/9206003];
  O.~A.~Castro-Alvaredo and A.~Fring,
  ``Identifying the operator content, the homogeneous sine-Gordon models'',
  Nucl.\ Phys.\ B {\bf 604} (2001) 367
  [hep-th/0008044].
  
\bibitem{Orland:2014mya}
  P.~Orland,
  ``Seeing asymptotic freedom in an exact correlator of a large-$N$ matrix field theory'',
  Phys.\ Rev.\ D {\bf 90} (2014) no.12,  125038
  [arXiv:1410.2627 [hep-th]];
  A.~C.~Cubero,
  ``Yang-Mills Theories as Deformations of Massive Integrable Models'',
  [arXiv:1409.8341 [hep-th]].
  
\bibitem{Pozsgay}
B.~Pozsgay and G.~Takacs,
  ``Form factors in finite volume. II. Disconnected terms and finite temperature correlators'',
  Nucl.\ Phys.\ B {\bf 788} (2008) 209
  [arXiv:0706.3605 [hep-th]].
  
\bibitem{Lehmann:1954rq}
  H.~Lehmann, K.~Symanzik and W.~Zimmermann,
  ``On the formulation of quantized field theories'',
  Nuovo Cim.\  {\bf 1} (1955) 205.
      
\bibitem{Watson}
  K.~M.~Watson,
  ``Some general relations between the photoproduction and scattering of pi mesons'',
  Phys.\ Rev.\  {\bf 95} (1954) 228.
  
\bibitem{Faddeev:1973aj}
  L.~D.~Faddeev and L.~A.~Takhtajan,
  ``Essentially nonlinear one-dimensional model of the classical field theory'',
  Theor.\ Math.\ Phys.\  {\bf 21} (1975) 1046, 
  Teor.\ Mat.\ Fiz.\  {\bf 21} (1974) 160.
   
\bibitem{Kulish:1976xi}
  P.~P.~Kulish and E.~R.~Nissimov,
  ``Conservation laws in the quantum theory: $\cos\phi$ in two-dimensions and in the massive Thirring model'',
  JETP Lett.\  {\bf 24} (1976) 220,
   Pisma Zh.\ Eksp.\ Teor.\ Fiz.\  {\bf 24} (1976) 247.
   
\bibitem{Coleman:1974bu}
  S.~R.~Coleman,
  ``The quantum sine-Gordon equation as the massive Thirring model'',
  Phys.\ Rev.\ D {\bf 11} (1975) 2088.
  
\bibitem{Mandelstam:1975hb}
  S.~Mandelstam,
  ``Soliton operators for the quantized sine-Gordon equation'',
  Phys.\ Rev.\ D {\bf 11} (1975) 3026.
  
\bibitem{Klassen:1992eq}
  T.~R.~Klassen and E.~Melzer,
  ``Sine-Gordon not equal to massive Thirring, and related heresies'',
  Int.\ J.\ Mod.\ Phys.\ A {\bf 8} (1993) 413
  [hep-th/9206114]. 
  
\bibitem{Korepin:1975zu}
  V.~E.~Korepin and L.~D.~Faddeev,
  ``Quantization of solitons'',
  Theor.\ Math.\ Phys.\  {\bf 25} (1975) 1039,
   Teor.\ Mat.\ Fiz.\  {\bf 25} (1975) 147.


\bibitem{Korepin:1975vd}
  V.~E.~Korepin, P.~P.~Kulish and L.~D.~Faddeev,
  ``Soliton quantization'',
  JETP Lett.\  {\bf 21} (1975) 138,
   Pisma Zh.\ Eksp.\ Teor.\ Fiz.\  {\bf 21} (1975) 302.
   
\bibitem{Zamolodchikov:1976uc}
  A.~B.~Zamolodchikov,
  ``Exact S-matrix of quantum sine-Gordon solitons'',
  JETP Lett.\  {\bf 25} (1977) 468.
  
\bibitem{Jackiw:1975im}
  R.~Jackiw and G.~Woo,
  ``Semiclassical scattering of quantized nonlinear waves'',
  Phys.\ Rev.\ D {\bf 12} (1975) 1643.
  
\bibitem{Karowski:1977th}
  M.~Karowski, H.~J.~Thun, T.~T.~Truong and P.~H.~Weisz,
  ``On the uniqueness of a purely elastic S-matrix in (1+1)-dimensions'',
  Phys.\ Lett.\ B {\bf 67} (1977) 321.
  
\bibitem{Karowski:1977tv}
  M.~Karowski,
  ``An exact relativistic S-matrix in (1+1)-dimensions: The on-shell solution of the massive Thirring model and the quantum sine-Gordon equation'', Proc. 15th Erice School of Subnuclear Physics: The Why's of Subnuclear Physics, ed.: A.~Zichichi, Plenum Press, New York (1979).
  
\bibitem{Karowski:1977fu}
  M.~Karowski and H.~J.~Thun,
  ``Complete S-matrix of the massive Thirring model'',
  Nucl.\ Phys.\ B {\bf 130} (1977) 295.
  
\bibitem{Arefeva:1974bk}
  I.~Arefeva and V.~Korepin,
  ``Scattering in two-dimensional model with Lagrangian $1/\gamma ((d_{\mu}u)^2/2 + m^2 \cos(u-1))$'',
  Pisma Zh.\ Eksp.\ Teor.\ Fiz.\  {\bf 20} (1974) 680;
  S.~N.~Vergeles and V.~M.~Gryanik,
  ``Two-dimensional quantum field theories having exact solutions'',
  Sov.\ J.\ Nucl.\ Phys.\  {\bf 23} (1976) 704,
   Yad.\ Fiz.\  {\bf 23} (1976) 1324;
  B.~Schroer, T.~T.~Truong and P.~Weisz,
  ``Towards an explicit construction of the sine-Gordon field theory'',
  Phys.\ Lett.\ B {\bf 63} (1976) 422.
  
\bibitem{Weisz:1977ii}
  P.~H.~Weisz,
  ``Exact quantum sine-Gordon soliton form factors'',
  Phys.\ Lett.\ B {\bf 67} (1977) 179.
  
\bibitem{Gross:1974jv}
  D.~J.~Gross and A.~Neveu,
  ``Dynamical symmetry breaking in asymptotically free field theories'',
  Phys.\ Rev.\ D {\bf 10} (1974) 3235.
 
 
 \bibitem{Berg:1977dp}
  B.~Berg, M.~Karowski, P.~Weisz and V.~Kurak,
  ``Factorized U(n) symmetric S-matrices in two-dimensions'',
  Nucl.\ Phys.\ B {\bf 134} (1978) 125;
  B.~Berg and P.~Weisz,
  ``Exact S-matrix of the chiral invariant SU(N) Thirring model'',
  Nucl.\ Phys.\ B {\bf 146} (1978) 205;
  R.~Koberle, V.~Kurak and J.~A.~Swieca,
  ``Scattering theory and 1/N expansion in the chiral Gross-Neveu model'',
  Phys.\ Rev.\ D {\bf 20} (1979) 897;
  E.~Abdalla, B.~Berg and P.~Weisz,
  ``More about the S-matrix of the chiral SU(N) Thirring model'',
  Nucl.\ Phys.\ B {\bf 157} (1979) 387.
  
  \bibitem{Kurak:1978su}
  V.~Kurak and J.~A.~Swieca,
  ``Anti-particles as bound states of particles in the factorized S-matrix framework'',
  Phys.\ Lett.\ B {\bf 82} (1979) 289.  
  
\bibitem{Babujian:2006md}
  H.~M.~Babujian, A.~Foerster and M.~Karowski,
  ``The nested SU(N) off-shell Bethe Ansatz and exact form-factors'',
  J.\ Phys.\ A {\bf 41} (2008) 275202
  
\bibitem{Babujian:2009zg}
  H.~Babujian, A.~Foerster and M.~Karowski,
  ``Exact form factors of the SU(N) Gross-Neveu model and 1/N expansion'',
  Nucl.\ Phys.\ B {\bf 825} (2010) 396
  [arXiv:0907.0662 [hep-th]].
  
\bibitem{Alessandro}
A.~Torrielli, ``Lectures on Classical Integrability'', in: ``An integrability primer for the gauge-gravity correspondence'', ed.: A.~Cagnazzo, R.~Frassek, A.~Sfondrini,
I.~M.~Sz\'ecs\'enyi and S.~J.~van Tongeren, J.\ Phys.\ A {\bf 49} (2016) no.32,  323001
  [arXiv:1606.02946 [hep-th]].

\bibitem{Bena:2003wd}
  I.~Bena, J.~Polchinski and R.~Roiban,
  ``Hidden symmetries of the AdS$_{5} \times$S$^{5}$ superstring'',
  Phys.\ Rev.\ D {\bf 69} (2004) 046002
  [hep-th/0305116].
  
\bibitem{Minahan:2002ve}
  J.~A.~Minahan and K.~Zarembo,
  ``The Bethe Ansatz for $\mathcal{N}$=4 Super-Yang-Mills'',
  JHEP {\bf 0303} (2003) 013
  [hep-th/0212208].

\bibitem{Arutyunov:2009mi}
  G.~Arutyunov, M.~de Leeuw and A.~Torrielli,
  ``The bound state S-matrix for AdS$_{5} \times$S$^{5}$ superstring'',
  Nucl.\ Phys.\ B {\bf 819} (2009) 319
  [arXiv:0902.0183 [hep-th]].
  
\bibitem{Beisert:2015msa}
  N.~Beisert, M.~de Leeuw and P.~Nag,
  ``Fusion for the one-dimensional Hubbard model'',
  J.\ Phys.\ A {\bf 48} (2015) 32,  324002
  [arXiv:1503.04838 [math-ph]].  
  
\bibitem{Martins:2007hb}
  M.~J.~Martins and C.~S.~Melo,
  ``The Bethe Ansatz approach for factorizable centrally extended su($2|2$) S-matrices'',
  Nucl.\ Phys.\ B {\bf 785} (2007) 246
  [hep-th/0703086].
  
\bibitem{Beisert:2005fw}
  N.~Beisert and M.~Staudacher,
  ``Long-range psu(2,$2|4$) Bethe ans\"atze for gauge theory and strings'',
  Nucl.\ Phys.\ B {\bf 727} (2005) 1
  [hep-th/0504190].
  
\bibitem{Janik:2007wt}
  R.~A.~Janik and T.~Lukowski,
  ``Wrapping interactions at strong coupling: The giant magnon'',
  Phys.\ Rev.\ D {\bf 76} (2007) 126008
  [arXiv:0708.2208 [hep-th]];
  Z.~Bajnok and R.~A.~Janik,
  ``Four-loop perturbative Konishi from strings and finite size effects for multiparticle states'',
  Nucl.\ Phys.\ B {\bf 807} (2009) 625
  [arXiv:0807.0399 [hep-th]].
  
\bibitem{TBA}
  D.~Bombardelli, D.~Fioravanti and R.~Tateo,
  ``Thermodynamic Bethe Ansatz for planar AdS/CFT: A Proposal'',
  J.\ Phys.\ A {\bf 42}, 375401 (2009)
  [arXiv:0902.3930 [hep-th]];
  N.~Gromov, V.~Kazakov, A.~Kozak and P.~Vieira,
  ``Exact spectrum of anomalous dimensions of planar $\mathcal{N}$= 4 supersymmetric Yang-Mills theory: TBA and excited states'',
  Lett.\ Math.\ Phys.\  {\bf 91} (2010) 265
  [arXiv:0902.4458 [hep-th]];
  G.~Arutyunov and S.~Frolov,
  ``Thermodynamic Bethe Ansatz for the AdS$_{5} \times$S$^{5}$ mirror model'',
  JHEP {\bf 0905} (2009) 068
  [arXiv:0903.0141 [hep-th]].
  
\bibitem{Gromov:2013pga}
  N.~Gromov, V.~Kazakov, S.~Leurent and D.~Volin,
  ``Quantum Spectral Curve for planar $\mathcal{N}$=4 super-Yang-Mills theory'',
  Phys.\ Rev.\ Lett.\  {\bf 112} (2014) 1,  011602
  [arXiv:1305.1939 [hep-th]];
  N.~Gromov, V.~Kazakov, S.~Leurent and D.~Volin,
 ``Quantum spectral curve for arbitrary state/operator in AdS$_{5}$/CFT$_{4}$'',
  JHEP {\bf 1509} (2015) 187
  [arXiv:1405.4857 [hep-th]].
  
\bibitem{Ahn:2008aa}
   C.~Ahn and R.~I.~Nepomechie,
   ``$\mathcal{N}$=6 super Chern-Simons theory S-matrix and all-loop Bethe ansatz equations'',
   JHEP {\bf 0809} (2008) 010
   [arXiv:0807.1924 [hep-th]].
   
\bibitem{Gromov:2008qe}
  N.~Gromov and P.~Vieira,
  ``The all loop AdS$_{4}$/CFT$_{3}$ Bethe Ansatz'',
  JHEP {\bf 0901} (2009) 016
  [arXiv:0807.0777 [hep-th]].
 
\bibitem{Bombardelli:2008qd}
  D.~Bombardelli and D.~Fioravanti,
  ``Finite-size corrections of the CP$^{3}$ giant magnons: The L\"uscher terms'',
  JHEP {\bf 0907} (2009) 034
  [arXiv:0810.0704 [hep-th]];
   T.~Lukowski and O.~Ohlsson Sax,
  ``Finite size giant magnons in the SU(2)$\times$SU(2) sector of AdS$_{4}\times$CP$^{3}$'',
  JHEP {\bf 0812} (2008) 073
  [arXiv:0810.1246 [hep-th]];
   C.~Ahn and P.~Bozhilov,
  ``Finite-size effect of the dyonic giant magnons in $\mathcal{N}$=6 super Chern-Simons theory'',
  Phys.\ Rev.\ D {\bf 79} (2009) 046008
  [arXiv:0810.2079 [hep-th]];
  M.~C.~Abbott, I.~Aniceto and D.~Bombardelli,
  ``Real and virtual bound states in L\"{u}scher corrections for CP$^{3}$ magnons'',
  J.\ Phys.\ A {\bf 45} (2012) 335401
  [arXiv:1111.2839 [hep-th]].
  
\bibitem{Bombardelli:2009xz}
  D.~Bombardelli, D.~Fioravanti and R.~Tateo,
 ``TBA and Y-system for planar AdS$_{4}$/CFT$_{3}$'',
  Nucl.\ Phys.\ B {\bf 834} (2010) 543
  [arXiv:0912.4715 [hep-th]];
  N.~Gromov and F.~Levkovich-Maslyuk,
  ``Y-system, TBA and Quasi-Classical strings in AdS$_{4}\times$CP$^{3}$'',
  JHEP {\bf 1006} (2010) 088
  [arXiv:0912.4911 [hep-th]].
  
\bibitem{Klose:2010ki}
  T.~Klose,
  ``Review of AdS/CFT integrability, Chapter IV.3: $\mathcal{N}$=6 Chern-Simons and strings on AdS$_4\times$CP$^3$'',
  Lett.\ Math.\ Phys.\  {\bf 99} (2012) 401
  [arXiv:1012.3999 [hep-th]].
  
\bibitem{Cavaglia:2014exa}
  A.~Cavagli\`a, D.~Fioravanti, N.~Gromov and R.~Tateo,
  ``Quantum Spectral Curve of the $\mathcal N$=6 Supersymmetric Chern-Simons Theory'',
  Phys.\ Rev.\ Lett.\  {\bf 113} (2014) 2,  021601
  [arXiv:1403.1859 [hep-th]]. 
  
\bibitem{Ahn:2012hw} 
   C.~Ahn and D.~Bombardelli,
   ``Exact S-matrices for AdS$_3$/CFT$_2$'',
   Int.\ J.\ Mod.\ Phys.\ A {\bf 28} (2013) 1350168
   [arXiv:1211.4512 [hep-th]].

\bibitem{Babichenko:2009dk}
  A.~Babichenko, B.~Stefanski, Jr. and K.~Zarembo,
  ``Integrability and the AdS$_3$/CFT$_2$ correspondence'',
  JHEP {\bf 1003} (2010) 058
  [arXiv:0912.1723 [hep-th]].
  
\bibitem{Borsato:2012ud}
  R.~Borsato, O.~Ohlsson Sax and A.~Sfondrini,
  ``A dynamic su$(1|1)^2$ S-matrix for AdS$_3$/CFT$_2$'',
  JHEP {\bf 1304} (2013) 113
  [arXiv:1211.5119 [hep-th]].
  
\bibitem{Borsato:2012ss}
  R.~Borsato, O.~Ohlsson Sax and A.~Sfondrini,
  ``All-loop Bethe Ansatz equations for AdS$_3$/CFT$_2$'',
  JHEP {\bf 1304} (2013) 116
  [arXiv:1212.0505 [hep-th]]. 
  
  \bibitem{Borsato:2015mma}
  R.~Borsato, O.~Ohlsson Sax, A.~Sfondrini and B.~Stefanski,
   ``The $\mathrm{AdS}_3\times \mathrm{S}^3\times \mathrm{S}^3\times\mathrm{S}^1$ worldsheet S-matrix'',
  J.\ Phys.\ A {\bf 48} (2015) 41,  415401
   [arXiv:1506.00218 [hep-th]].
  
\bibitem{Borsato:2014exa}
  R.~Borsato, O.~Ohlsson Sax, A.~Sfondrini and B.~Stefanski,
  ``Towards the all-loop worldsheet S-matrix for AdS$_3\times$S$^3\times$T$^4$'',
  Phys.\ Rev.\ Lett.\  {\bf 113} (2014) 13,  131601
  [arXiv:1403.4543 [hep-th]];
   R.~Borsato, O.~Ohlsson Sax, A.~Sfondrini and B.~Stefanski,
   ``The complete AdS$_{3} \times$S$^3 \times$T$^4$ worldsheet S-matrix'',
   JHEP {\bf 1410} (2014) 66
   [arXiv:1406.0453 [hep-th]];
   

\bibitem{Borsato:2013hoa}
  R.~Borsato, O.~Ohlsson Sax, A.~Sfondrini, B.~Stefanski, Jr. and A.~Torrielli,
  ``Dressing phases of AdS$_3$/CFT$_2$'',
  Phys.\ Rev.\ D {\bf 88} (2013) 066004
    [arXiv:1306.2512 [hep-th]].
    
\bibitem{Borsato:2016kbm}
  R.~Borsato, O.~O.~Sax, A.~Sfondrini and B.~Stefanski,
  ``On the spectrum of AdS$_3\times$ S$^3\times$ T$^4$ strings with Ramond-Ramond flux'',
  arXiv:1605.00518 [hep-th].
  
\bibitem{Abbott:2015pps}
  M.~C.~Abbott and I.~Aniceto,
  ``Massless L\"uscher terms and the limitations of the AdS$_3$ asymptotic Bethe Ansatz'', Phys.\ Rev.\ D {\bf 93} (2016) no.10,  106006
  [arXiv:1512.08761 [hep-th]].
  
\bibitem{Sfondrini:2014via}
  A.~Sfondrini,
  ``Towards integrability for ${\rm Ad}{{{\rm S}}_{{ 3}}}/{\rm CF}{{{\rm T}}_{{ 2}}}$'',
  J.\ Phys.\ A {\bf 48} (2015) 2,  023001
  [arXiv:1406.2971 [hep-th]].
   

\bibitem{Hoare:2014kma}
  B.~Hoare, A.~Pittelli and A.~Torrielli,
   ``Integrable S-matrices, massive and massless modes and the AdS$_{2}\times$S$^{2}$ superstring'',
   JHEP {\bf 1411} (2014) 051
   [arXiv:1407.0303 [hep-th]];
    B.~Hoare, A.~Pittelli and A.~Torrielli,
  ``The S-matrix algebra of the AdS$_2\times$S$^2$ superstring'', Phys.\ Rev.\ D {\bf 93} (2016) no.6,  066006
  [arXiv:1509.07587 [hep-th]].
  
  
\end{thebibliography}
\end{document}